\documentclass{article}

\usepackage{arxiv}

\usepackage[utf8]{inputenc} 
\usepackage[T1]{fontenc}    
\usepackage{hyperref}       
\usepackage{url}            
\usepackage{booktabs}       
\usepackage{amsfonts}       
\usepackage{nicefrac}       
\usepackage{microtype}      
\usepackage{lipsum}
\usepackage{graphicx}
\usepackage{array,multirow}
\usepackage{colortbl}
\usepackage{hhline}
\usepackage{graphics}
\usepackage{subcaption}
\usepackage{float}
\usepackage{placeins}
\usepackage{amsmath}
\usepackage{wrapfig}
\usepackage[english]{babel}
\usepackage[
    backend=biber,
    style=ieee,
  ]{biblatex}
\title{CheckSoft : A Scalable Event-Driven Software Architecture for Keeping Track of People and Things in People-Centric Spaces}

\author{Rohan Sarkar}

\author{Avinash C. Kak}

\author{
 Rohan Sarkar \\
  School of Electrical and Computer Engineering\\
  Purdue University\\
  West Lafayette, IN 47907 \\
  \texttt{sarkarr@purdue.edu} \\
   \And
 Avinash C. Kak \\
  School of Electrical and Computer Engineering\\
  Purdue University\\
  West Lafayette, IN 47907 \\
  \texttt{kak@purdue.edu} \\
}

\begin{document}
\maketitle
\begin{abstract}
We present CheckSoft, a scalable
  event-driven software architecture for keeping track of
  people-object interactions in people-centric applications
  such as airport checkpoint security areas, automated
  retail stores, smart libraries, and so on. The
  architecture works off the video data generated in real
  time by a network of surveillance cameras. Although there
  are many different aspects to automating these
  applications, the most difficult part of the overall
  problem is keeping track of the interactions between the
  people and the objects.  CheckSoft uses
  finite-state-machine (FSM) based logic for keeping track
  of such interactions which allows the system to quickly
  reject any false detections of the interactions by the
  video cameras.  CheckSoft is easily scalable since the
  architecture is based on multi-processing in which a
  separate process is assigned to each human and to each
  ``storage container'' for the objects. A storage container
  may be a shelf on which the objects are displayed or a bin
  in which the objects are stored, depending on the specific
  application in which CheckSoft is deployed.
\end{abstract}



\section{Introduction}\label{sec1}

This paper presents a scalable software architecture for
automating video monitoring of people-centric spaces.  We
must mention at the very outset that the purpose of this
paper is {\em not} to address the computer vision issues
related to detecting people, objects, and their
interactions.\footnote{Even though the computer vision
  aspects of the overall problem are not the focus of this
  paper, for the sake of validating the software
  architecture presented here, we will show results using
  actual and simulated video streams.}

On the other hand, our aim in this paper is solely to
present an error-tolerant and scalable software design that
uses finite-state machine (FSM) logic capable of rejecting
false detections reported by the underlying video cameras.
As we show in this paper, with FSM the system can quickly
catch moment-to-moment discrepancies and inconsistencies in
the detections reported by the sensors.

Whereas FSM allows for quick checks on the validity of the
detected human-object interactions, the multi-processing
design we have used for the architecture allows it to be
scalable.  Scalability means that the architecture would
automatically allow for an arbitrary number of people and
objects to be present in the space being monitored with
latencies limited only by the computational power of the
underlying hardware platform.

With its high-level attributes as described above, a
software architecture of the type presented in this paper
can be expected to find applications in the new-age retail
stores with no sales clerks and cash registers, smart
libraries that allow a customer to simply walk out with the
book desired, airport checkpoint security areas where it is
important to keep track of the passengers and their
belongings, and so on.  Additionally, since CheckSoft is
capable of recording on a continuous basis the history of
people-object interactions in the monitored space, another
important application of such a software system is that it
lends itself to an easy post-facto analysis of the log files
for gaining insights into people's reaction to the displayed
objects.  The results of such analysis may be used for a
more productive arrangement of the objects vis-a-vis the
people.

The application scenarios presented above might cause a
reader to think that our work is similar to or overlaps
significantly with a rather popular research area: smart (or
intelligent) spaces in which wireless sensors attached to
the objects and cameras are used to keep track of {\em just}
the people or {\em just} the objects.  The goal of the work
presented in this paper is different: {\em Keeping track of
  people-object interactions \cite{sarkar2020CheckSoft}, in the sense that we want our
  system to track ownership and possession relationships as
  people interact with the objects, exchange them, leave
  them behind in monitored spaces when they shouldn't, and
  so on.}\footnote{By ownership of an object, we mean the
  first individual identified as possessing the object.}

A generic software architecture for keeping track of
people-object interactions should be able to accommodate the
fact that the objects may present themselves differently in
different applications.  For example, in a retail store, the
objects are likely to be placed on the shelves where either
each object is directly accessible to a customer or, when
the objects are stacked, the topmost object is directly
accessible.  On the other hand, in an airport checkpoint
security area, the objects will be the personal belongings
that the passengers divest in the bins that are then placed
on the conveyor belt.  In this case, the objects will be in
a heap in the bins or placed directly on the conveyor belt.
In the software system presented in this report, we use the
notion of {\tt storage} to address at a generic level these
differences between the different applications. For retail
store application, an instance of {\tt storage} could be a
shelf containing items.  On the other hand, an instance of
the same for airport checkpoint security would be a bin in
which the passengers divest their belongings.

In addition to the differences in how the objects may
present themselves to the system in different applications,
a system like ours must also be mindful of the fact that the
acceptable rules for people-object interactions may be
different in different applications.  For example, in an
airport checkpoint security application, passengers should
be allowed to collect their own items only at the end of the
screening process. In a museum, visitors should not be
allowed to touch any expensive exhibits, and, in a library,
readers should not be allowed to misplace books in wrong
shelves at the time of returning the books.

The software architecture we present in this paper meets the
challenges described above and is also scalable at the same
time. We achieve scalability by using multiprocessing and
concurrency.  A unique process is associated with each
entity, that is with each individual and with each storage
unit containing objects in the space being monitored.  This
decentralizes the bookkeeping for keeping track of the
different states of each individual and each storage
container --- the process assigned to each entity takes care
of all such details.  The decentralization also eliminates
the risk of deadlock that may arise due to contention for
the entity state information.  The communication between the
processes is implemented using Message Passing Interface
(MPI).


Here is a summary of the main contributions of this paper:

\begin{list}{\labelitemi}{\leftmargin=1em}

\item CheckSoft can keep track of concurrent person-object
  interactions involving an arbitrary number of people and
  also an arbitrary number of objects in a monitored space
  in real-time using an architecture based on
  multi-processing and inter-module communications based on
  message passing.  Each significant entity in the system is
  assigned a separate process.  The overall design achieved
  in this manner eliminates the risk of any deadlocks that
  may arise due to contention for the entity state
  information.

\item CheckSoft allows for an arbitrary number of video
  trackers to be plugged into the system on a plug-n-play
  basis. It is event-driven and can operate asynchronously
  in real-time vis-a-vis the events detected from the
  video-feeds of any arbitrary number of video cameras.

\item The architecture adheres to the time-honored
  principles of good object-oriented design and uses
  finite-state-machine based logic for fault tolerance
  vis-a-vis any temporal discrepancies in the events
  detected by the sensors.

\item The architecture can be applied to a variety of
  applications that require tracking the interactions
  between people and objects by only making minor
  modifications to the application-specific FSM logic and
  without any major architectural changes.
\end{list}

CheckSoft was validated with data from actual video cameras.
The scalability of CheckSoft was validated with a simulator.

In the rest of this report, we start in Section
\ref{sec:lit_review} with a quick review of the literature
that has two parts to it: the first part deals with the
software engineering literature that has guided the design
of CheckSoft, and second part with the literature related to
keeping track of people and objects in video monitored
spaces.  Subsequently, in Section \ref{sec:principal_vars},
we then introduce the principal data structures of the
system for representing the relational information in the
system. The definitions presented for these data structures
should give the reader a sense of the generality of the
software architecture. Finally, the overall system
architecture is summarized in Section
\ref{sec:overall_architecture} and presented in details in
Section \ref{sec:details_architecture}.  In Section
\ref{sec:verification}, we verify scalability and
deadlock-free operation for the proposed architecture. In
Section 7, we validate the operation of CheckSoft with
actual video trackers and test the scalability and
robustness of CheckSoft using a simulator.

%

\section{Literature Review}
\label{sec:lit_review}

We will first present the literature that guided our
software design in Section \ref{sec:lit_principles} and then
compare related software systems in Section
\ref{sec:comparison}.

\subsection{Literature that Guided Our Software Design }
\label{sec:lit_principles}

Ning et al. \cite{Ning1997} states that the two primary
aspects of a complex software system are the {\em
	components}, which would be the basic building blocks of
the system, and the {\em architecture}, which describes how
the individual components interact so that the overall
system possesses the desired behavior.  There is a close 
relationship between {\em Component-Based Software Engineering} (CBSE)
\cite{Ning1997} and {\em Modular programming} \cite{Parnas1972} 
that emphasizes the importance of dividing
the overall functionality of a complex software system into
individual functional components that can be developed
separately such that each module contains all that
is needed to execute a particular aspect of the desired
overall functionality. CheckSoft is comprised
of different modules/components, each responsible for a
distinct functionality as explained in Section \ref{sec:eda}. 

CheckSoft components can produce or consume events and,
through the events, work collaboratively to achieve the
desired overall behavior, as we explain in Section
\ref{sec:overall_architecture}.  The resulting architecture
is what may be referred to as an {\em event-driven
  architecture} (EDA). In a wide ranging survey of event
based systems by Hinze et al \cite{Hinze2009}, the authors
discuss the role and the modalities for event processing in
reactive applications. Etzion et al. \cite{Etzion2005}
discuss the necessity of incorporating event-driven
functionality in software systems that must exhibit
on-demand and just-in-time behavior.  Since in EDA, the
event producers are unaware of the nature and the number of
event consumers, there is a low coupling between them which
contributes to \textit{extensibility} as new consumers can
be added as and when required as well as the
\textit{scalability} of such systems \cite{Roda2016}.

In such systems, an event is defined as a significant change
in state \cite{Chandy2006} that, in general, must result in
the execution of a certain functionality by what is commonly
referred to as an {\em event handler}.  Anvur et
al. \cite{Anvur1990} and Wagner et al. \cite{Wagner2006}
discuss various finite-state-machine (FSM) based approaches
for designing event handlers for real-time software systems.
The event handler modules in CheckSoft are also based on FSM
based logic that is capable of fast rejection of
``illegal''state changes reported by the video monitoring
system.

Of the various applications where the EDA approach to
software architecture design has been found to be effective,
the two that are closest to CheckSoft are IoT and smart
environments \cite{Sciphor2019}. The networks used in IoT
generally involve heterogeneous devices capable of
generating a large number of events and it is necessary to
integrate, process, and react to the events on the fly.
Almeida et al. \cite{Almeida2019} have proposed a
distributed hierarchical architectural model based on
Situational Awareness that can support scalability,
flexibility, autonomy and heterogeneity demands in
distributed IoT environments. With regard to EDA for smart
environments, Roda et al. \cite{Roda2016} have proposed an
architecture for a scalable and collaborative ambient
intelligence environment designed for applications such as
smart homes, hospitals, health monitoring and for daily life
assistance.

That brings us to the previous work in which researchers
have addressed the concurrency issues that arise when there
is a need to process in real-time a large number of events
simultaneously \cite{Ben-Ari1990}, as is the case with
CheckSoft.  There are generally two important aspects to
such software systems: the ability to detect and respond to
events occurring in any random order, and ensuring that the
software responds to these events within some required time
interval. One of the most notable features of CheckSoft is
its ability to process concurrent events in a scalable
fashion by using multiprocessing with a distributed memory
model as discussed in Section \ref{sec:concur}.


\subsection{Survey of the Literature in Related Areas}
\label{sec:comparison}


In this subsection, we briefly review other software systems
that are related to our work in the sense that these systems
involve software architectures for making real-time
inferences from video streams of surveillance data.  As an
illustration, Vezzani et al. \cite{Vezzani2010} have
proposed a Service Oriented Architecture (SOA) that uses
event-driven communications to analyze video feeds from multiple cameras
for detecting and classifying faces,
postures, behaviors, etc.  In that sense, this system is
mostly for monitoring people, as opposed to monitoring
people-object interactions as we do in CheckSoft.  There
also exists a commercial framework for video surveillance,
the IBM Smart Surveillance Engine \cite{IBM2008} that is
capable of generating real-time alerts for events triggered
by changes in the locations of objects. This system again is
not about tracking human-object interactions as we do in
CheckSoft.


More closely related to CheckSoft --- more closely from the
standpoint of the end purpose of the software architectures
--- are the works described in \cite{Radke2011} and
\cite{Bhargava2007} for video surveillance meant for airport
checkpoint security, and in \cite{Frontoni2013},
\cite{Singh2016}, and \cite{Gyori2018} for retail store
automation.  The architecture for airport checkpoint
security reported in \cite{Radke2011} attempts to detect the
associations between the bags and the passengers using a
rudimentary FSM that is specific to passengers divesting
objects and reclaiming them at airport checkpoint security.
The manner in which this logic is implemented and, also,
lack of multi-processing significantly impacts the
scalability of that work.  The goal of the work reported in
\cite{Bhargava2007} is even more limited --- it only seeks
to detect abandoned bags.  To compare, CheckSoft addresses
system scalability with regard to both the number of people
and the number of objects with multi-processing and message
passing --- this was one of the most important
considerations in the design of the CheckSoft architecture.
Equally important in CheckSoft is the extendibility of the
system with regard to the types of interactions between
people and objects.

With regard to the previous work on retail store automation,
the work reported in \cite{Frontoni2013} and
\cite{Singh2016} is about just categorizing actions, as
opposed to {\em tracking} human-object interactions in a
fault-tolerant and scalable framework as accomplished by
CheckSoft.
Another contribution worth mentioning in the context of
retail store automation is the work reported in
\cite{Gyori2018} that deals with using cameras and weight
sensors for cashierless grocery shopping. The main focus of
such a system is
to identify object transfers from the shelves to the
customer baskets and vice versa.  On the other hand, the
design we have used for CheckSoft allows the same
architecture to be used for different applications (to name
just two for highlighting the variety: automatic libraries
and airport checkpoint security) with just a tweak of the
FSM states and state transitions.

Another area of research that is tangentially related to
CheckSoft is video-based security monitoring of what are
known as cyber-physical spaces.  By cyber-physical space is
meant a monitored space with access control. In addition to
physical assets, such spaces also may contain cyber assets
that need to be protected from unauthorized intrusions by
people. Greaves et al. \cite{Greaves2018} have shown how a
virtual perimeter based on different types of sensors,
including video cameras, can be used for access
control. Another relevant publication in this area is by
Tsigkanos et al. \cite{Tsigkanos2018} where the authors have
shown how the physical layout and the topology of the space
involved can be analyzed for the level of protection offered
by the sensors used against potential threats. Their
proposed analysis techniques also includes the dynamics of
people movement in the spaces.  The approaches used in such
systems for detecting unauthorized behaviors by people are
generally based on violations of access control and on
spatial modeling of people movement.  These concepts are
not applicable to the goals for which CheckSoft is designed.

\section{The Principal Data Structures}
\label{sec:principal_vars}

\begin{figure}[htb]
\centering
\includegraphics[width=0.6\textwidth]{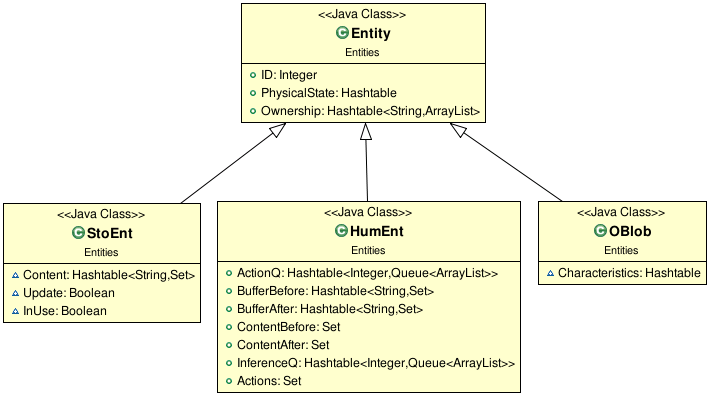}
\caption{\em The class diagram depicting the inheritance hierarchy for the different entities in CheckSoft. }
\label{fig:CD_Entity}
\end{figure}

CheckSoft is about keeping track of the different entities
that are present in a space that is being monitored, and,
even more importantly, about keeping track of the
interactions between those entities.  We refer to the
interactions --- such as when a human picks up an object ---
as events.  Therefore, we need data structures for the
entities and for the events.  In keeping with the best
practices in modern programming, we represent these data
structures by class hierarchies. This organization of the
classes as shown in Figs. \ref{fig:CD_Entity} and \ref{fig:CD_Event} allows the
functionality that is common to all the classes to be placed
in the root class, which makes it more efficient to maintain
and extend the code for different applications.

In the two subsections that follows, Subsection
\ref{sec:Entities} presents the {\tt Entity} class hierarchy
and Subsection \ref{sec:Events} the {\tt Event} class
hierarchy.

\subsection{The Entity Class Hierarchy}
\label{sec:Entities}

As shown by the inheritance hierarchy in
Fig. \ref{fig:CD_Entity}, we use the Java class {\tt Entity} as
the root class for all the entities that CheckSoft is
required to keep track of, these being the individuals, the
objects, the storage units, etc.  


Here is a brief description of the contract of each class shown
in Fig. \ref{fig:CD_Entity}:


\noindent 
\textbf{Entity} (\texttt{Entity}): As mentioned previously,
this is the parent class of all the entity related classes
in CheckSoft.

\noindent 
\textbf{Human Entity} (\texttt{HumEnt}): The subclass
\texttt{HumEnt} serves as a base class for different types
of human entities that may be present in the space being
monitored with the video cameras.  As to what these
different types of human entities would be depends on the
application.  In a retail application, the different
subclasses of \texttt{HumEnt} would represent the customers
and the different categories of employees in the store.  For
an automated library, the same would be the users and the
librarians, and so on.  For airport checkpoint security, the
different subclasses of \texttt{HumEnt} would be the
passengers and the TSA agents.

%

\noindent
\textbf{Storage Entity} (\texttt{StoEnt}): The subclass {\tt
  StoEnt} can be used as the parent class of other storage
related subclasses that tell us how the objects in the space
being monitored present themselves to the humans.  The
objects could be on shelves, in bins, on the floor, etc.  An
instance of {\tt StoEnt} may be real or virtual.  An example
of a virtual instance of {\tt StoEnt} would a heap of
objects created by a passenger dumping his/her belongings
directly on a conveyor belt.



\noindent
\textbf{Object Blobs} (\texttt{OBlob}): This subclass in the
       {\tt Entity} hierarchy can be used as the parent
       class for representing different types of objects in
       the space being monitored.  Ordinarily, one would
       expect an {\tt OBlob} instance to represent a single
       object in the space being monitored.  However, it is
       not always possible to discern objects individually.
       For example, for the airport checkpoint security
       case, when a passenger divests all his or her belongings
       on, say, the conveyor belt, all that the cameras
       would be able to see would be a blob of pixels
       occupied by the heap. So an instance of {\tt OBlob}
       represents what the system ``thinks'' is an object.
       It is possible for such an instance to implicitly
       represent a collection of objects that are not
       visible separately.

\begin{figure}[ht]
	\centering
	\includegraphics[width=0.3\textwidth]{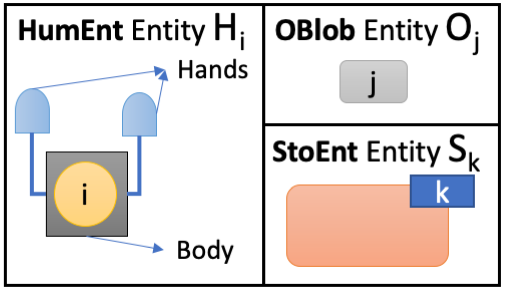}
	\caption{\em Iconic representation of the different entities in CheckSoft. }
	\label{fig:Icons}
\end{figure}

The reader will notice the attribute {\tt ID} of the {\tt
  Entity} root class in Fig. \ref{fig:CD_Entity}.  This
attribute, inherited by all the subclasses, is a unique
integer that is assigned to each instance of type {\tt
  Entity}. Our explanation of CheckSoft in this paper uses
the notation {\tt H$_i$} to refer to that {\tt HumEnt} whose
{\tt ID} attribute has value {\tt i}.  Along the same lines,
the notation {\tt O$_j$} will stand for an {\tt OBlob}
instance that was assigned the {\tt ID} value {\tt j}.  And
the $k^{th}$ instance of {\tt StoEnt} in an explanation will
be referred to as {\tt S$_k$}.  In our diagrams, we will use
the iconic representations of the different entities as
shown in Fig. \ref{fig:Icons}.

Table \ref{tbl:ds} in the Appendix elaborates on the
attributes for the classes in the inheritance hierarchy.
Obviously, for each child class, what is shown for it
specifically in the table is in addition to what it inherits
from the base class {\tt Entity}.

\subsection{The Class Hierarchy for the Events}
\label{sec:Events}
\begin{figure}[htb]	
	\centering
	\includegraphics[width=0.4\textwidth]{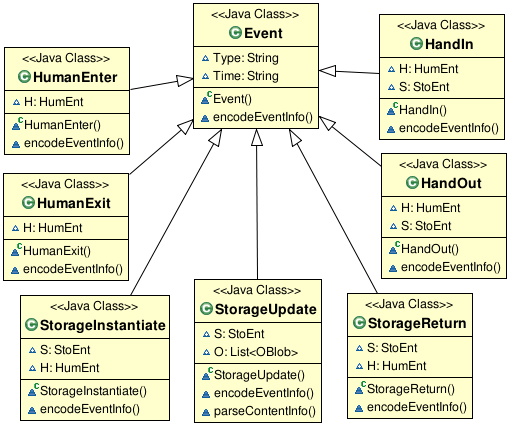}
	\caption{\em The class diagram depicting the inheritance hierarchy for the different events in CheckSoft.}
	\label{fig:CD_Event}
\end{figure}
As mentioned at the beginning of Section
\ref{sec:principal_vars}, CheckSoft is about keeping track
of the entities and the events that are generated when there
is any object related interaction between the entities.
This subsection will describe the class hierarchy for the
events.

However, before presenting the hierarchy of classes for the
events, it is important to mention that we assume that a
video-camera client of CheckSoft can track the individuals
and identify the object blobs that the individuals are
interacting with and do so on a continuous basis.  We assume
that the events pertaining to humans interacting with the
objects are all hand-based.  We also assume that an
interaction between a human entity and an object entity that
is in a storage entity starts with a {\tt HandIn} event and
ends with a {\tt HandOut} event as explained below.

More specifically, we assume that the video-camera clients
that are used to monitor the space have continuously running
processes that can detect the following events:

\begin{enumerate}
	
\item A human entering the monitored area triggers the {\tt
  HumanEnter} event and the human exiting the area triggers the {\tt
  HumanExit} event.
	
\item When a human instantiates a new storage entity (such
  as a cart) or ``returns'' a storage entity, either the
  {\tt StorageInstantiate} or the {\tt StorageReturn} event
  is triggered as the case may be.  A storage entity is
  considered returned when it is empty and its user has
  exited the monitored area.

\item When an object is placed in a storage entity or taken
  out of it, that triggers a {\tt StorageUpdate} event which
  updates the content list of the storage entity. If the
  space being monitored involves shelves for storing the
  objects, each shelf would require its own camera for
  detecting such events.

\item A human hand reaching inside a storage entity triggers
  a {\tt HandIn} event and the hand being pulled back
  triggers a {\tt HandOut} event.  In
  Fig. \ref{fig:Problem}, the direction of the arrow on the
  hand extension of the HumEnt icon indicates the direction
  of the hand movement with respect to the storage
  container. The direction of the arrow should help the
  reader figure out whether the corresponding event is {\tt
    HandIn} or {\tt HandOut}.

\end{enumerate}

The {\tt Event} class hierarchy is shown in
Fig. \ref{fig:CD_Event}. As shown there, we use the Java
Class {\tt Event} as the root class for all events in {\tt
  CheckSoft} that are detected by video-camera clients.
Different applications of {\tt CheckSoft} will differ with
regard to the types of events and the entities involved in
the events.  Table \ref{tbl:events} in the Appendix
elaborates on the different classes in the inheritance
hierarchy shown in Fig. \ref{fig:CD_Event}.  This
organization of classes allows a user to introduce new types
of events in {\tt CheckSoft} with ease as and when required.

\subsection{An Illustration of How the State of the Entities is Stored}

\begin{figure}[htb]
	\centering
        \includegraphics[width=0.4\textwidth]{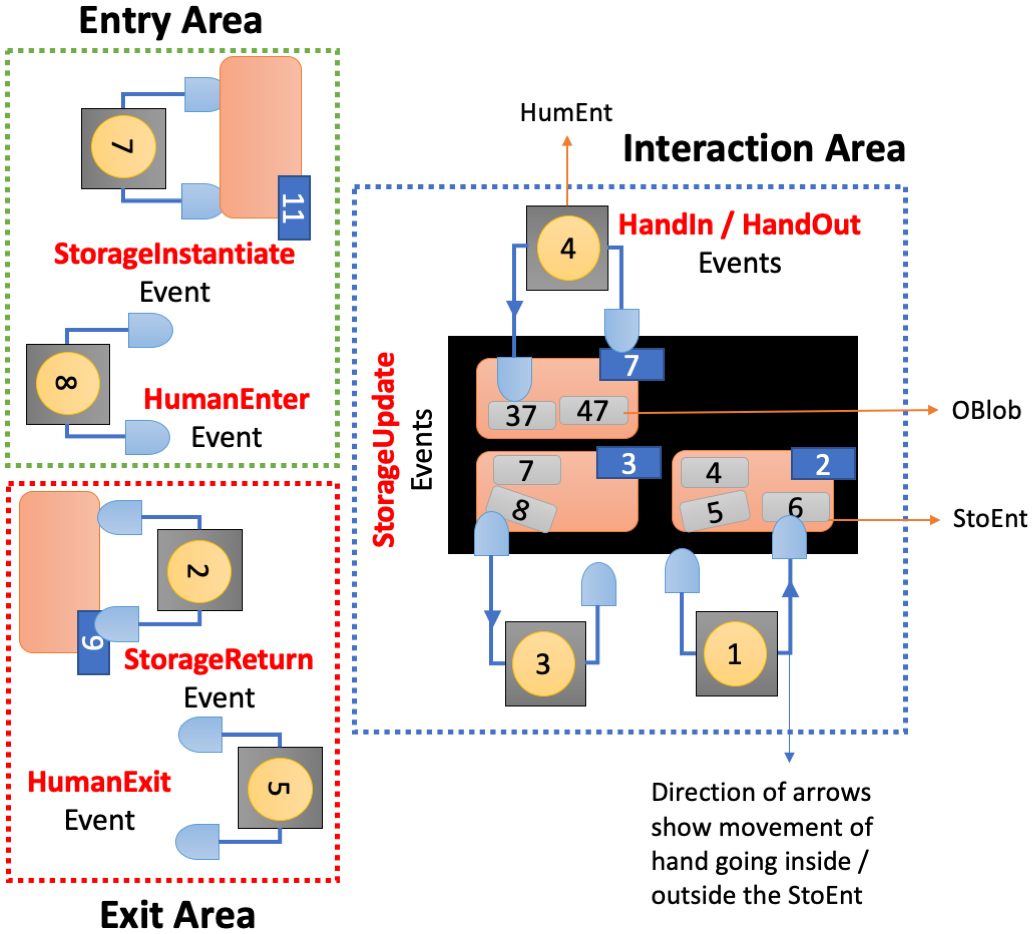}
	\caption{\em An illustration of the different entities and the events detected in the monitored space. }
	\label{fig:Problem}
	\vspace{-0.2in}
\end{figure}

Fig. \ref{fig:Problem} illustrates an example of the
different entities involved and the events detected in a
people-centric space. The space being monitored can be
divided into three main regions: the entry area, the exit
area, and the interaction area where people interact with
objects. The three different types of entities -- {\tt
  HumEnt}, {\tt StoEnt} and {\tt OBlobs} --- are shown, each
with its own unique integer.  The icons used for the
entities are as indicated in Fig. \ref{fig:Icons}.

The example shown in Fig. \ref{fig:Problem} will generate
several different events simultaneously.  In the entry and
exit areas, the entry of HumEnt $H_8$ and the exit of HumEnt
$H_5$ will generate the {\tt HumanEnter} and {\tt HumanExit}
events, respectively. Similarly, HumEnt $H_7$ instantiating
StoEnt $S_{11}$ and HumEnt $H_2$ returning StoEnt $S_{9}$
will result in the {\tt StorageInstantiate} and {\tt
  StorageReturn} events being detected. In the interaction
area of Fig. \ref{fig:Problem}, many individuals could be
interacting at the same time and hence several {\tt HandIn},
{\tt HandOut} and {\tt StorageUpdate} events would be
generated simultaneously. It could be that HumEnt $H_3$ has
previously placed the object represented by OBlob $O_{8}$ in
the StoEnt $S_{3}$ and is now retrieving his/her hand, which
would trigger the {\tt HandOut} event.  Along the same
lines, HumEnt $H_4$ appears in the process of placing an
object in StoEnt $S_{7}$, which would trigger the {\tt
  HandIn} event; and so on. The {\tt StorageUpdate} event is
generated whenever the content information of any StoEnt
needs to be updated.

\begin{figure*}[b]
	\begin{minipage}[t]{0.35\textwidth}
		\centering
		\includegraphics[width = \textwidth]{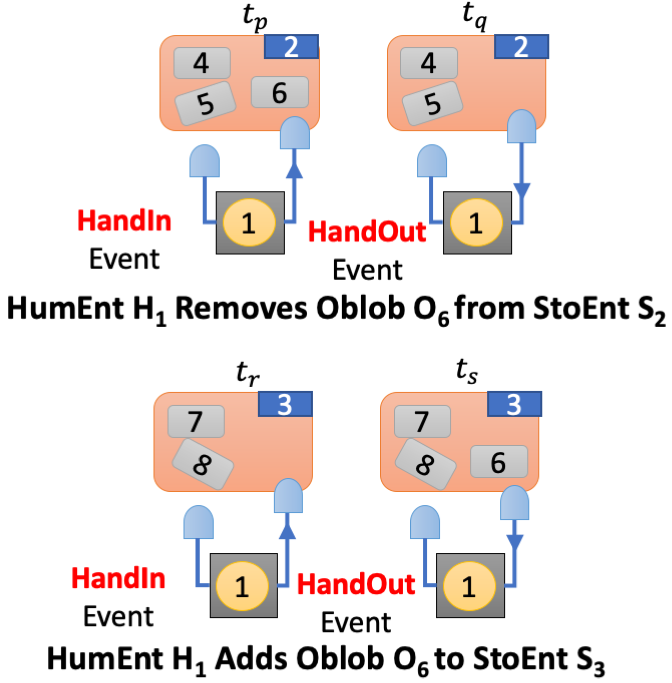}
		\caption*{(a)}
	\end{minipage}
	\hfill
	\begin{minipage}[t]{0.64\textwidth}
		\centering
		\includegraphics[width = \textwidth]{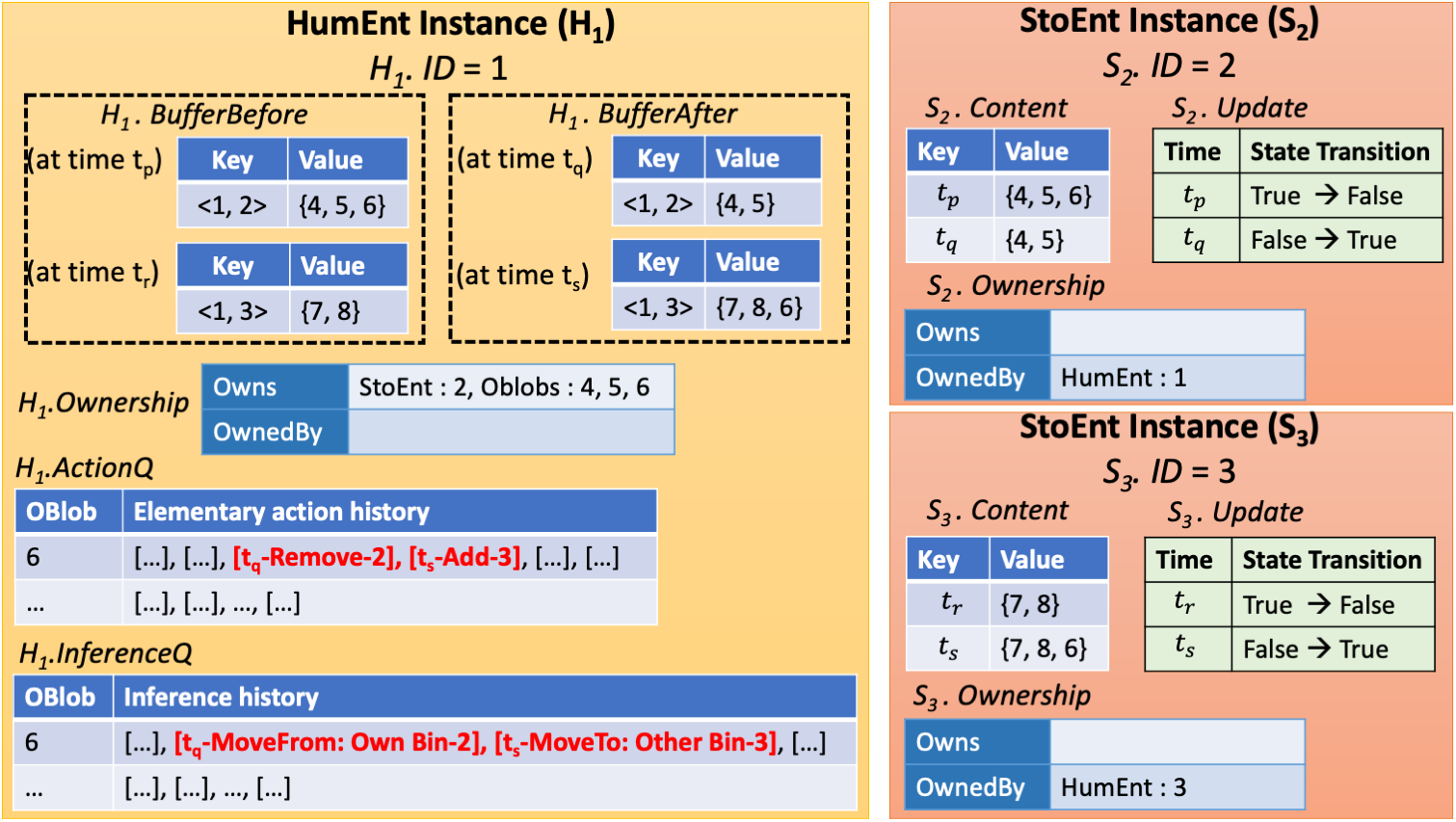}
		\caption*{(b)}
	\end{minipage}
	\caption{\em A pictorial description of HumEnt $H_1$
		transferring OBlob $O_6$ from StoEnt $S_2$ to StoEnt $S_3$
		is shown in (a). An illustration of how the state
		information of the different entities due to the
		interactions is stored is shown in (b)}
	\label{fig:PDSExample}
\end{figure*}

Focusing on a specific subset of the interactions in
Fig. \ref{fig:Problem}, Fig. \ref{fig:PDSExample}
illustrates how the state information for the entities is
stored in the different data structures.  We will consider
the interactions involved in HumEnt $H_1$ transferring OBlob
$O_6$ from StoEnt $S_2$ to StoEnt $S_3$
in Fig. \ref{fig:Problem}.

There are two interactions relevant here: HumEnt $H_1$
removing OBlob $O_6$ from StoEnt $S_2$ between time $t_p$ to
$t_q$, followed by another interaction in which the same
HumEnt places the same OBlob in StoEnt $S_3$ between time
$t_r$ to $t_s$. Each interaction starts with a {\tt HandIn}
event and ends with a {\tt HandOut} event as shown in
Fig. \ref{fig:PDSExample}(a). Fig. \ref{fig:PDSExample}(b)
shows all the entity instances relevant to the two
interactions.

As shown in Fig. \ref{fig:PDSExample}(b), the StoEnt
instances maintain the timestamped information related to
the OBlobs instances they contain in the {\tt Content} data
attribute. In Fig. \ref{fig:PDSExample}(b), before the interaction starts
with StoEnt $S_2$ at $t_p$ , $S_2.{\tt Content} = \{4, 5, 6\}$ as
StoEnt $S_2$, contains the OBlobs $O_4, O_5$ and $O_6$ and
similarly after the interaction ends at $t_q$, we have
$S_2.{\tt Content} = \{4, 5\}$. On a similar note, before the
interaction starts with StoEnt $S_3$ at $t_r$, we have
$S_3.{\tt Content} = \{7, 8\}$ as can be seen in
Fig. \ref{fig:PDSExample}(b). Similarly, after the interaction
ends at $t_s$, we have $S_3.{\tt Content} = \{7, 8, 6\}$. 

One important point to note here is that during the time of
the interaction, when the hands are inside the storage area,
the hands may occlude the objects from the
camera. Therefore, if the video-trackers generate a {\tt
  StorageUpdate} event during the interaction, the content
information reported would be inaccurate. It must be ensured
that the state of StoEnt $S_k$ is not corrupted when the
data reported by the video-trackers is unreliable.The
$S_k$.{\tt Update} attribute shown in
Fig. \ref{fig:PDSExample}(b) allows updates to the
$S_k$.{\tt Content} attribute only when it is set to
True. To prevent the state of $S_k$ from being updated
during the interaction when a {\tt HandIn} event involving
$S_k$ occurs, $S_k$.{\tt Update} transitions from True to
False and, subsequently, when a {\tt HandOut} event occurs,
$S_k$.{\tt Update} transitions back to True.

Each HumEnt instance stores locally the `before' and `after'
state of the storage containers that it is interacting with.
These are stored in the data attributes {\tt BufferBefore}
and {\tt BufferAfter} of the HumEnt instance.  In our
example, the HumEnt instance $H_1$ will store the `before'
and `after' contents of the StoEnts $S_2$ and $S_3$ in the
variables $H_1.{\tt BufferBefore}$ and $H_1.{\tt
  BufferAfter}$.  A unique key is associated with each such
stored state that depends on the IDs of both the HumEnt and
the StoEnt.  In our example, the key associated with what is
stored in $H_1.{\tt BufferBefore}$ and $H_1.{\tt
  BufferAfter}$ would be <$1, 2$> for the first interaction
and <$1, 3$> for the second interaction.  In these two keys,
$1$ is the HumEnt ID and $2$ and $3$ are the StoEnt IDs
involved in the interactions.

An attribute common to all the entities is {\tt Ownership}
that consists of two lists: the {\tt OwnedBy} list in which
the information regarding who owns the corresponding entity
is stored, and {\tt Owns} in which the information regarding
what entities the corresponding entity owns is stored. For
example, as shown in Fig. \ref{fig:PDSExample}(b), HumEnt
$H_1$ owns StoEnt $S_2$ and hence HumEnt $H_1$ appears in
the {\tt OwnedBy} list of $S_2.{\tt Ownership}$ and StoEnt
$S_2$ appears in the ${\tt Owns}$ list of $H_1.{\tt
  Ownership}$. Similarly, the reader can see that the StoEnt
$S_3$ is owned by some other HumEnt $H_3$. HumEnt $H_1$ owns
Oblobs $O_4, O_5$ and $O_6$ and therefore this appears in
the ${\tt Owns}$ list of $H_1.{\tt Ownership}$.

The data attributes $H_1.{\tt ActionQ}$ and $H_1.{\tt
  InferenceQ}$ shown in the left panel in
Fig. \ref{fig:PDSExample}(b) are hash tables of queues
storing the interaction history for each OBlob that HumEnt
$H_1$ interacted with. The interaction history, in the exact
temporal sequence of occurrence, is stored for each OBlob in
a separate queue. Each interaction is recorded in the
following form:
\begin{center}[$t$ - {\it action/inference} - $k$]\end{center}
where $t$ is the time of the interaction and $S_k$ the
StoEnt involved in the interaction.

Regarding the two queues mentioned above, in $H_1.{\tt
  ActionQ}$, we store the elementary actions such as adding
or removing OBlobs. On the other hand, in $H_1.{\tt
  InferenceQ}$, we store the inferences and anomalies
detected by the logic of CheckSoft. To illustrate this, for
both $H_1. {\tt ActionQ}$ and $H_1.{\tt InferenceQ}$, we
will consider the entries for key $6$ in
Fig. \ref{fig:PDSExample}(b), which is the $OBlob$ involved
in the interaction. Since OBlob $O_6$ was transferred from
StoEnt $S_2$ to $S_3$, the elementary actions in this case
are {\it Remove} and {\it Add} and these are stored in
$H_1. {\tt ActionQ}$ as the following string:
\begin{center}[$t_q$ - {\it Remove} -2], [$t_s$ - {\it Add} -3]\end{center} The inferences are stored in $H_1.{\tt InferenceQ}$ as:\begin{center}[$t_q$ - {\it MoveFrom: Own Bin} - 2], [$t_s$ -  {\it MoveTo: Other Bin} - 3]\end{center}  The ownership relationships {\it Own Bin} and {\it Other Bin} are derived from the fact that StoEnt $S_2$ is owned by HumEnt $H_1$ but StoEnt $S_3$ is not, as shown in $H_1.{\tt Ownership}$. This interaction in which the OBlob $O_6$ was transferred to StoEnt $S_3$ which belongs to some other HumEnt is an example of what will be detected as an anomalous interaction by CheckSoft.

\section{Overall System Architecture}
\label{sec:overall_architecture}

The goal of this section is to give the reader a top-level
view of the architecture of CheckSoft and the rationale
underlying the architectural design. Subsequently, a more
detailed presentation of the architecture that describes the
roles of the different components will be presented in
Section \ref{sec:details_architecture}.

With regard to the top-level view presented in the next
three subsections, we first describe in Section
\ref{sec:concur} how concurrency is exploited for handling
the interactions that may occur simultaneously, and,
subsequently, in Section \ref{sec:eda} we present the major
modules of the event-driven software architecture. Finally,
we provide a high-level overview of how Checksoft's finite
state machine based logic provides significant immunity
against noisy data reported by video-trackers in Section
\ref{sec:handlenoise}.

\begin{figure*}[h]
	\centering
	\begin{subfigure}{\textwidth}
		\centering
		\includegraphics[width = 0.5\textwidth]{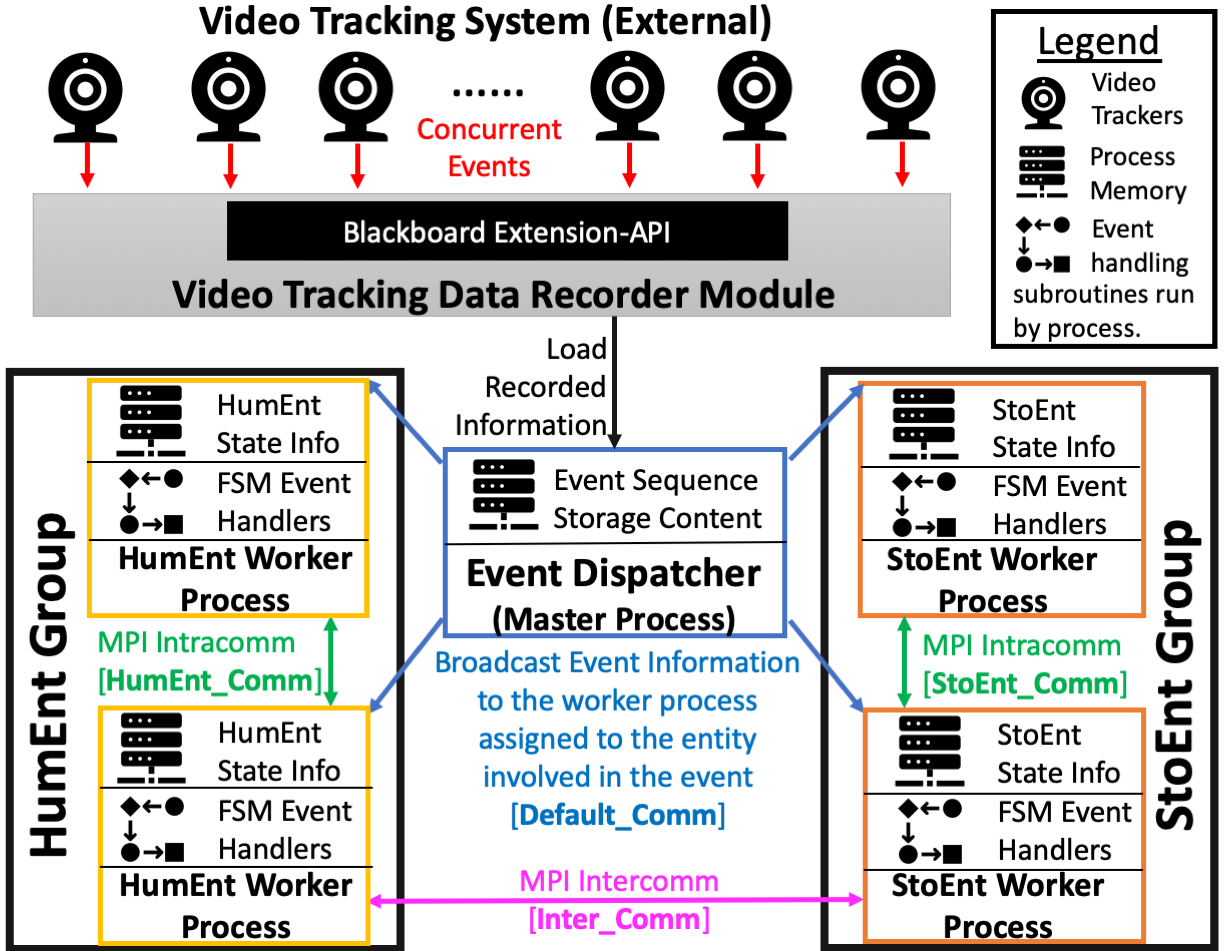}
		\caption*{\normalsize (a) {\em Each {\tt
                      HumEnt} and {\tt StoEnt} entity is
                    allocated a separate worker process. The
                    master process dispatches events
                    detected by the video-trackers to the
                    different worker processes. The worker
                    processes then run the event handling
                    modules shown below and communicate
                    using MPI as and when required.}}
	\end{subfigure}
	\newline
	\begin{subfigure}{\textwidth}
		\centering
		\includegraphics[width = 0.75\textwidth]{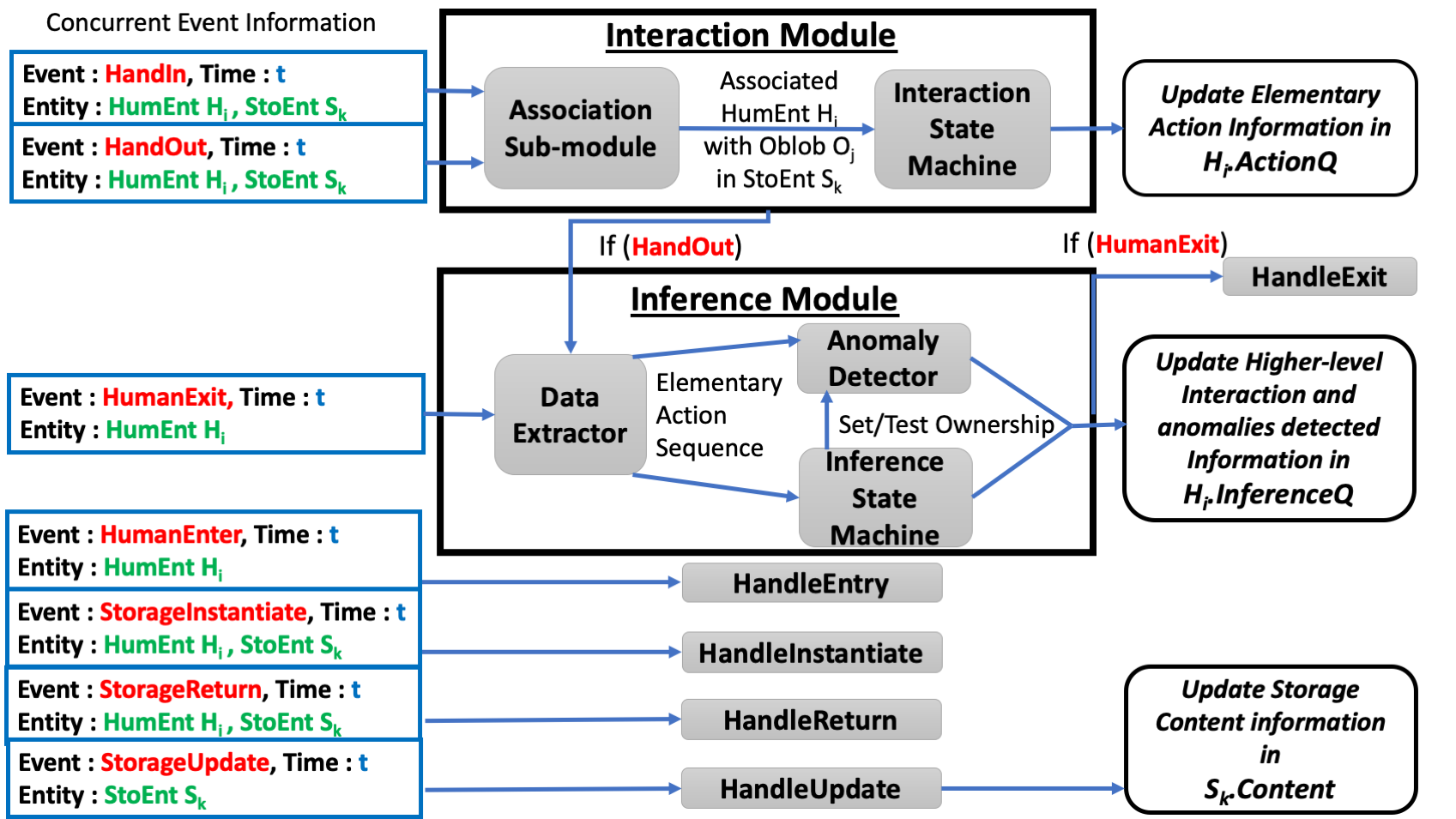}
		\caption*{\normalsize (b) {\em The different
                    CheckSoft modules that handle concurrent
                    events are shown in grey boxes. (The
                    subscripted notation {\tt H$_i$} refers
                    to the $i^{th}$ {\tt HumEnt} instance,
                    {\tt O$_j$} refers to the $j^{th}$ {\tt
                      OBlob} instance and {\tt S$_k$} refers
                    to the $k^{th}$ {\tt StoEnt} instance
                    involved in the corresponding events.)}}

	\end{subfigure}
	\caption{ \em The Overall Architectural Diagram.  }
	\label{fig:architecture}
\end{figure*}

\subsection{Exploiting Data Parallelism with Concurrency and Communications}
\label{sec:concur}
%

As illustrated in Fig. \ref{fig:architecture}(a), the data
produced by the video camera clients (displayed by the 6
icons in the top row in the figure) is inherently parallel.
Most applications of CheckSoft will involve several video
cameras monitoring different sections of the space as shown
in Fig. \ref {fig:Problem}. Additionally, any number of
individuals may be interacting with an arbitrary number of
objects at any given time. Therefore, one can expect an
arbitrary number of events to be generated concurrently. In
order to not introduce undesirable latencies in reacting to
these events, they would need to be processed in parallel.
Obviously, any attempt at exploiting this parallelism must
not result in memory corruption and deadlock conditions
should the computations being carried out in parallel access
the same portions of the memory.


The two basic approaches to achieving concurrency in
software are multi-threading and multi-processing.  Since
threads can be thought of as lightweight processes, one main
advantage of multithreading is that it is faster to create
and destroy the threads and faster to access the objects in
the memory that are meant to be shared by the threads.
However, through synchronization and semaphores,
multithreading requires greater care in ensuring that the
memory meant to be shared by the different objects is not
corrupted by the threads accessing the same object
simultaneously.

In CheckSoft, we have chosen to use multiprocessing instead.
As each {\tt HumEnt} or {\tt StoEnt} instance is created, it
is handed over to a separate process by the master process
that runs CheckSoft. This decentralizes the bookkeeping for
keeping track of the different states of each individual and
each storage container, as the process assigned to each
entity takes care of all such details.  This is one of the
important reasons for why our software can easily be scaled
up.  As is to be expected, a collection of processes for
similar entities, such as all the HumEnt entities, are
clones of the same basic process. We refer to such a
collection as a {\em process group}. For the purpose of
explanation, and as shown in Fig. \ref{fig:architecture}(a),
we use the notation {\tt HumEnt\_Group} to refer to the
HumEnt processes and the notation {\tt StoEnt\_Group} to
refer to the StoEnt processes.

CheckSoft uses multiprocessing with a distributed memory
model in which every process in the system has its own
private memory. If an event involves any of the HumEnt
and/or StoEnt instances, the corresponding worker process
executes the appropriate event handling module described in
Section \ref{sec:eda} and updates the corresponding entity
state information in its own private memory as shown in
Fig. \ref{fig:architecture}(a). Since all of the
computations carried out by a process only involve the
private memory of the process, it follows that the processes
that are in charge of analyzing the human-object
interactions must somehow become aware of both the humans
and the objects involved.  {\em As opposed to using a
  shared-memory model, we take care of such inter-process
  interactions through the communication facilities provided
  by MPI (Message Passing Interface).}

CheckSoft uses the standard MPI for intra-communication
within each process group and for inter-communication
between different process groups.  MPI provides us with what
are known as {\tt communicators} for those purposes.  The
messaging achieved with intra-communicators within each
process group can work in both point-to-point mode, in which
each process sends a message to one other specific process,
and in broadcast mode, in which a process sends a message to
all other sibling processes. The messaging achieved with
inter-communicators can work only in point-to-point mode in
which a process of a particular group sends a message to a
chosen process in another process group.

For an example of within-group communications with
intra-communicators, let's say that some aspect of the state
of all \texttt{StoEnt} entities needs to be updated at the
same time, this would require using an intra-communicator in
the broadcast mode.  And for an example that requires an
intra-communicator in a point-to-point mode, consider an
object that has been transferred from one {\tt StoEnt}
instance to another {\tt StoEnt} instance.  In this case,
while the receiving StoEnt would know directly from the
video cameras that it was in possession of a new object, it
would nonetheless need to hear directly from the sending
{\tt StoEnt} instance for confirmation.

The master {\tt CheckSoft} process uses the MPI's default
communicator, {\tt Default\_comm}, to dispatch events to all
worker process groups, meaning the processes in {\tt
  HumEnt\_Group} and in {\tt StoEnt\_Group} as shown in
Fig. \ref{fig:architecture}(a).  We denote the
intra-communicators within the process groups {\tt
  HumEnt\_Group} and {\tt StoEnt\_Group} by
\texttt{HumEnt\_comm} and \texttt{StoEnt\_comm},
respectively.  As mentioned earlier, in addition to
facilitating within-group communications, the
intra-communicators are also needed for the functioning of
the inter-communicators.

A particularly useful application of the inter-communicator
is in fetching content data from a process in the
\texttt{StoEnt\_Group} when a {\tt HandIn} or a {\tt
  HandOut} event is recorded (because, say, a human divested
an object in a bin). The data and the entities involved can
subsequently trigger the downstream finite-state based logic
for checking the legality of the event and the legality of
the consequences of the event.

As to how the communicators are actually used for passing
messages between the processes and fetching data and results
from the processes, that is accomplished with an MPI based
function {\tt gptwc()}, whose name stands for: ``General
Purpose Two-way Communication''.  This function is general
purpose in the sense that it can be used to establish a
two-way communication link between any two processes.  The
process that wants to establish a communication link is
called the {\em initiator process} and the other endpoint of
the communication link the {\em target process}. An
initiator process sends data to a target process and
indicates what operation needs to carried out on the sent
data vis-a-vis the data that resides in the private memory
of the target process.  The target process acts accordingly
and sends the result back to the initiator process.  The
implementational details and how this function is invoked is
presented in Appendix \ref{appendix:GPTWC} .

\subsection{Components of the Event Driven Architecture}
\label{sec:eda}

As mentioned in Section \ref{sec:lit_principles}, CheckSoft
is comprised of different modules, each responsible for a
distinct functionality, such as extracting person-object
ownership information from the data; establishing and
updating associational information related to the possibly
changing relationships between the objects and the storage
units; monitoring various interactions in the environment;
detecting anomalies; and so on.

The highly modular architecture is event-driven as shown in
Figure \ref{fig:architecture}(b). The events listed on the
left side of the figure trigger specific event-handler
modules shown on the right side. The most critical
components of CheckSoft are the Interaction and Inference
modules which are responsible for analyzing the events and
drawing inferences regarding the outcome of each
person-object interaction. Each submodule within the
Interaction and Inference Module is a finite state machine
(FSM) that works independently of the FSMs in all other
submodules --- independently in the sense that the FSM in a
submodule cares only about the changes in the state
information of the entities involved in the event that
triggered the particular FSM.

The use of FSM logic allows us to endow the modules with
event-driven behaviors that possess efficient (with
polynomial-time guarantees) implementations. As a result of
employing an event-driven state machine model, the event
handlers remain idle while waiting for an event, and then,
when an event occurs, the relevant event handler reacts
immediately. This approach can be thought of as a set of
computational agents collaborating asynchronously using
event-based triggers. Additionally, an FSM based
implementation also allows for easy scalability with regard
to all the variables that require an arbitrary number of
instantiations and additionally provides flexibility to
easily extend software functionalities as and when required.

CheckSoft consists of the following four modules:

\noindent
\textbf{Video Tracker Data Recorder (VTDR) Module:} It is
this module that the video camera clients talk to directly.
As mentioned earlier, CheckSoft is meant to sit on top of a
one or more video camera clients installed in the space
being monitored. As we will present in Section
\ref{sec:vtdr}, the VTDR module provides a Java RMI (Remote
Method Invocation) \cite{JavaRMI} based E-API
(Extension-API) that a video camera client can use for
directly sending its event detection results to CheckSoft.
VTDR associates a time stamp with every detection reported
by a video camera client, and, as the reader will see
later, that plays an important role in the temporal
sequencing of the events that may be received simultaneously
from multiple video cameras.\footnote{It is implicitly
  assumed that the video camera clients are running the NTP
  protocol as a background process for staying globally
  synchronized} The VTDR module records the event
and storage content information from multiple video-trackers in an encoded format into a concurrent unbounded persisted queue named {\tt EventSeq} with micro-second latencies.

\noindent
\textbf{Event Dispatcher Module:}  This module dispatches the
encoded event information in the {\tt EventSeq} queue sequentially in the temporal order of occurrence for further processing. This triggers the different event
handlers (and thus the modules shown on the right side of
Fig. \ref{fig:architecture}(b)).

\noindent
\textbf{Interaction Module: } This module is triggered by
the {\tt HandIn} and {\tt HandOut} events involving
\texttt{HumEnt} and \texttt{StoEnt} instances and detects
the elementary interactions such as the addition and the
removal of \texttt{OBlobs} to and from \texttt{StoEnt}
instances. This module has two submodules:

\begin{enumerate}
	
	\item \textbf{Association Submodule:} This module processes
	the {\tt HandIn} and {\tt HandOut} events and, by
	monitoring changes in the \texttt{StoEnt} content due to
	an interaction, establishes associations between the
	\texttt{OBlob}, \texttt{StoEnt} and the \texttt{HumEnt}
	instances.
	
	\item \textbf{Interaction State Machine Submodule:} This
	module implements the finite-state machine based logic to
	keep track of the elementary interactions between the
	\texttt{HumEnt} instances and the \texttt{OBlob} instances
	that are present in the different \texttt{StoEnt}
	instances.  The finite-state machine logic associates
	different states with the entities and checks on the
	legality of the state transitions --- and, thus, adds
	robustness to the operation of CheckSoft.
	
\end{enumerate}

\noindent
\textbf{Inference Module:} This module specializes in making
inferences from the elementary interactions of the
\texttt{HumEnt} instances with each of the \texttt{OBlob}
instances that the \texttt{HumEnt} interacted with. After each interaction, this module is triggered by a {\tt HandOut} event (after the elementary action involving the {\tt HandOut} event is recorded by the Interaction Module). This module is also triggered by the {\tt HumanExit} events. When the inference module is triggered, that causes  the \textbf{Data Extractor} 
submodule to extract the interaction
history recorded by the Interaction module in a
time-sequential manner for each {\tt OBlob} instance the  \texttt{HumEnt} interacted with and sends this data
to two concurrently running submodules:

\begin{enumerate}
	\item \textbf{Inference State Machine Submodule:} This
	submodule implements the finite-state machine based logic
	to infer higher level interactions from the elementary
	interactions involving the \texttt{HumEnt}, \texttt{OBlob}
	and the \texttt{StoEnt} instances.
	
	\item \textbf{Anomaly Detector:} This submodule detects
	anomalous interactions (such as when a non-owner
	\texttt{HumEnt} interacts with an \texttt{OBlob} and/or
	\texttt{StoEnt} instance) and raises appropriate alarms
	and warnings.
\end{enumerate}

Since the higher level interactions as well as the type of
anomalies vary with the specific application that our
architecture is applied to, we may need to tweak the
finite-state machine logic in this module. However, it is
important to note that these changes in the logic are
expected to be minor and should not require any changes in
the overall architectural framework of CheckSoft.

\begin{figure*}[b]
	\begin{minipage}[t]{0.45\textwidth}
		\centering
		\includegraphics[width = \textwidth]{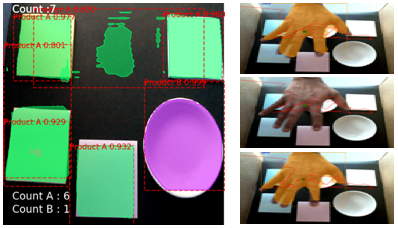}
		\caption*{(a)}
	\end{minipage}
	\hfill
	\begin{minipage}[t]{0.52\textwidth}
		\centering
		\includegraphics[width = \textwidth]{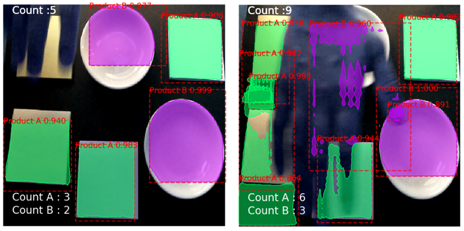}
		\caption*{(b)}
	\end{minipage}
	\caption{\em Inconsistencies in object detection due to sudden changes in ambient conditions and inconsistencies in hand detection due to motion artifacts and sudden motion are illustrated in Fig. (a). Inconsistencies arising when objects are partially or completely occluded by hand(s) are illustrated in Fig. (b)}
	\label{fig:cvprob}
\end{figure*}

\subsection{Tolerance to noisy data from video-trackers}
\label{sec:handlenoise}
The events listed on the left side of Fig.
\ref{fig:architecture}(b) are generated by the video-trackers.  Such events generally are noisy due to occlusion, missed and false detections, inaccurate localization of objects \cite{Greaves2018}, and ineffective tracking under noisy conditions \cite{Mustanar2019}. Hence, CheckSoft needs to be designed such that it can provide some immunity against noisy events detected by the video trackers. This section just qualitatively describes how the software handles different sources of noise and a quantitative analysis is shown in Section \ref{sec:validation}. The former part of this section briefly describes how CheckSoft handles false and missed detections and the latter part describes how CheckSoft ensures that the entity state is not corrupted by erroneous events reported by video trackers.

%
CheckSoft is designed to be robust with respect
to missed detections by the video cameras and also with
respect to any false detections. 
To give the reader a sense of what we mean by missed and
false detections,
note that all the objects in a {\tt StoEnt} instance may not
be detected accurately in every image frame, as can be seen in Figure
\ref{fig:cvprob}(a). These variations in the detected objects
might lead to inconsistencies.  An additional source of
difficulty arises from the fact
that when a hand that is trying to reach an object in a
{\tt StoEnt} moves too suddenly, it may not be detected in one
or more frames until the image of the hand stabilizes. CheckSoft 
provides some protection against such effects by filtering out 
momentary fluctuations in the detections reported by the video trackers.


Yet another source of difficulty is that when a
human hand reaches into a {\tt StoEnt} instance it will
block the objects in the container from the camera.
This is illustrated in Figure \ref{fig:cvprob}(b).
Therefore, to eliminate the inconsistencies introduced due to
occlusion, the {\tt Interaction
	Module} in Section \ref{sec:IntM} only considers the object content before a hand has
entered the storage area and after the hand has left.

\begin{figure*}[hbt!]
	\begin{minipage}[t]{0.23\textwidth}
		\centering
		\includegraphics[width = \textwidth]{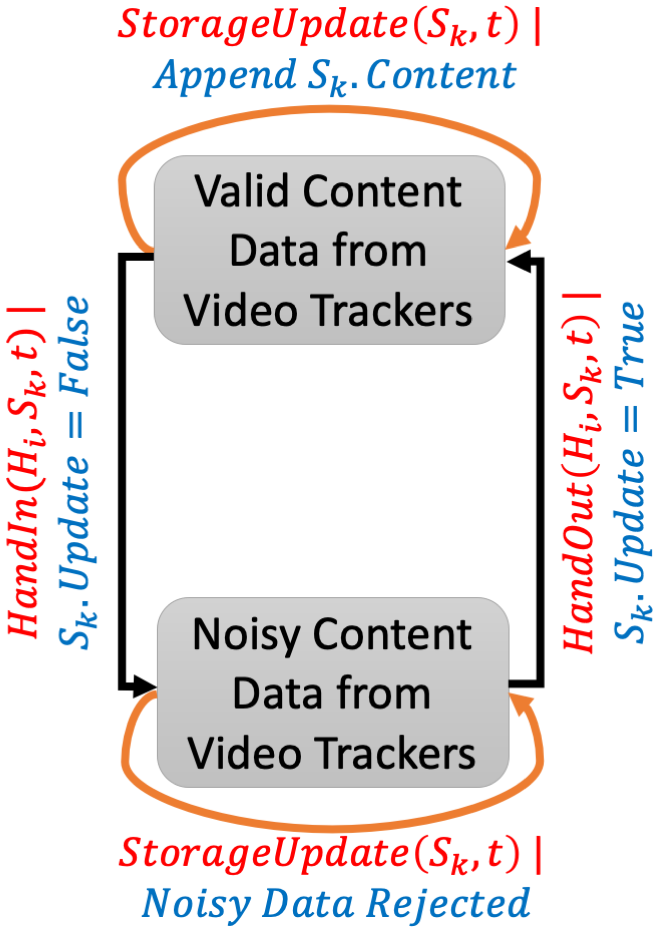}
		\caption*{(a)}
	\end{minipage}
    \vline  
	\hfill
	\begin{minipage}[t]{0.76\textwidth}
		\centering
		\includegraphics[width = \textwidth]{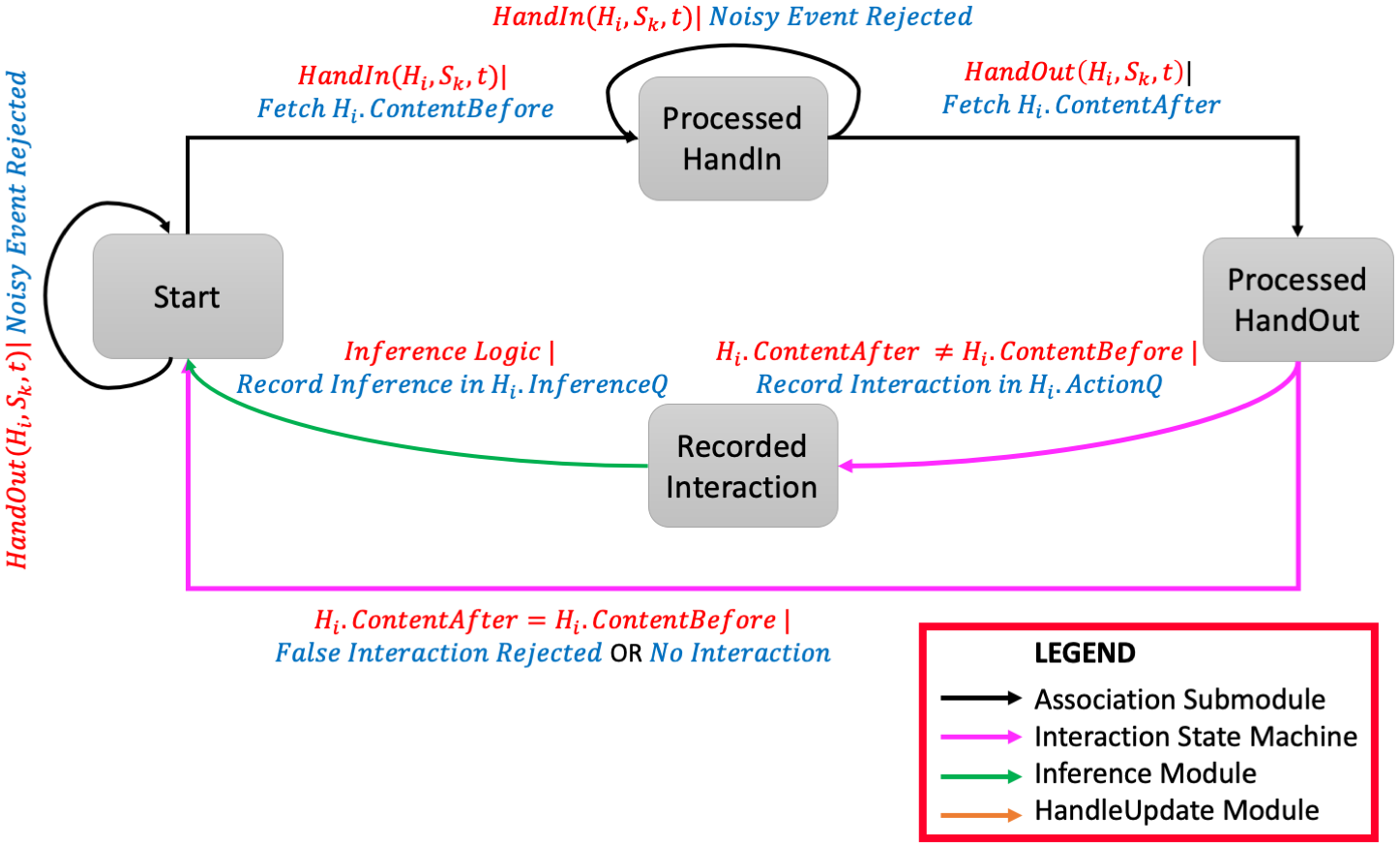}
		\caption*{(b)}
	\end{minipage}
	\caption{\em State Diagrams showing how CheckSoft is designed to be robust to noisy data reported by video-trackers. The state diagram for rejecting noisy storage content information is shown in (a). The state diagram for rejecting noisy HandIn, HandOut events and false interactions is shown in (b)}
	\label{fig:NoiseFSM}
\end{figure*}

The state of the entities are updated by the frequently occurring {\tt HandIn}, {\tt HandOut} and {\tt StorageUpdate} events. It is critical to ensure that the state of the entities don't get changed by erroneous events reported by the video trackers so that the higher-level reasoning logic of CheckSoft functions as desired. This is done by enforcing consistency in the finite-state logic between the different events related to the same overall person-object interaction as shown by the state diagrams in Fig. \ref{fig:NoiseFSM}. In Fig. \ref{fig:NoiseFSM}, a state is represented by the grey boxes, the event or condition that needs to be satisfied for a state transition is shown in red and the corresponding output as a result of the transition is shown in blue alongside the arrows.

The state diagram in Fig. \ref{fig:NoiseFSM}(a) illustrates how CheckSoft rejects noisy content information reported by video-trackers. The {\tt HandleUpdate} module in Fig. \ref{fig:architecture}(b), appends the {\tt Content} data attribute with the latest content of the StoEnt $S_k$ with the corresponding time-stamp $t$  when a {\tt StorageUpdate($S_k, t$)} event occurs only if the data attribute {\tt $S_k$.Update} is set to {\tt True}. This ensures that the content information for any StoEnt is updated only when reliable data is reported by video-trackers. As a case in point, the video-trackers might report erroneous content information during the time of an interaction when objects inside the storage area are occluded by hands. To provide protection against this, when any {\tt HandIn} event involving StoEnt $S_k$ occurs, $S_k$.{\tt update} is set to False thereby blocking any updates to state of $S_k$ during the interaction. When a {\tt HandOut} event occurs involving the same StoEnt, $S_k$.{\tt Update} is set to True thereby allowing updates to the state of $S_k$ after the interaction is over. 

The state diagram in Fig. \ref{fig:NoiseFSM}(b) illustrates how noisy {\tt HandIn}, {\tt HandOut} events as well as false interactions reported by video-trackers are filtered by the {\tt Interaction Module}. 
    As mentioned before, every interaction starts with a {\tt
	HandIn} event and ends with a {\tt HandOut} event. Typically a video-tracker monitoring any storage area should report a {\tt HandIn} event when motion is detected and subsequently report a {\tt HandOut} event when the motion has completely subdued in storage area.  {\tt CheckSoft} considers an interaction to be valid only if both of these events involving the same HumEnt and StoEnt have been reported in the correct temporal sequence by the video trackers. The {\tt Association Submodule} is designed to filter out the noisy hand-based events as shown by the black arrows in Fig. \ref{fig:NoiseFSM}(b). The Association Module thereby only processes the hand-based events for a valid interaction and consequently fetches the content before and after the interaction from StoEnt $S_k$ and finally stores it in $H_i$.{\tt ContentBefore} and $H_i$.{\tt ContentAfter} respectively. 

Also, it is very common for the video-trackers to report false interactions, where an interaction didn't actually happen but the video trackers report false {\tt HandIn} and {\tt HandOut} events indicating an interaction has happened. For such cases, there is no change in the content since an actual interaction has not happened. The {\tt Interaction State Machine} (shown by the magenta arrows in \ref{fig:NoiseFSM}(b)) analyses the content before and after the interaction and records the interaction in $H_i$. {\tt ActionQ} for further analysis only if there is some change in the content. Otherwise, the interaction is not recorded and thereby false interactions are filtered out. Only the interactions recorded are further considered for analyzing inferences and detecting anomalies by the {\tt Inference Module} and recorded in $H_i$.{\tt InferenceQ} (shown by the green arrow).


\section{The CheckSoft Modules}
\label{sec:details_architecture}

The previous section, Section
\ref{sec:overall_architecture}, provided a high-level
summary of the modular architecture of CheckSoft and also
talked about how data parallelism is exploited through
concurrent processing as made possible by the MPI standard.
The goal of this section is to present a more detailed look at
each of the CheckSoft modules.

\subsection{Video Tracker Data Recorder Module}
\label{sec:vtdr}

We start by reminding the reader that video tracking per se
is outside the scope of the system architecture we present
in this paper. That is, we assume that a video tracker unit
is implemented externally and provides our system with the information related to the
occurrence of various kinds of events mentioned in Section \ref{sec:Events}.
As mentioned before, any arbitrary number of such video-trackers might be employed to monitor a space and thereby it is critical to design an interface that would allow CheckSoft to operate vis-a-vis any arbitrary number of video cameras. 
\begin{figure*}[hbt!]
	\begin{minipage}[t]{0.68\textwidth}
		\centering
		\includegraphics[width = 0.9\textwidth]{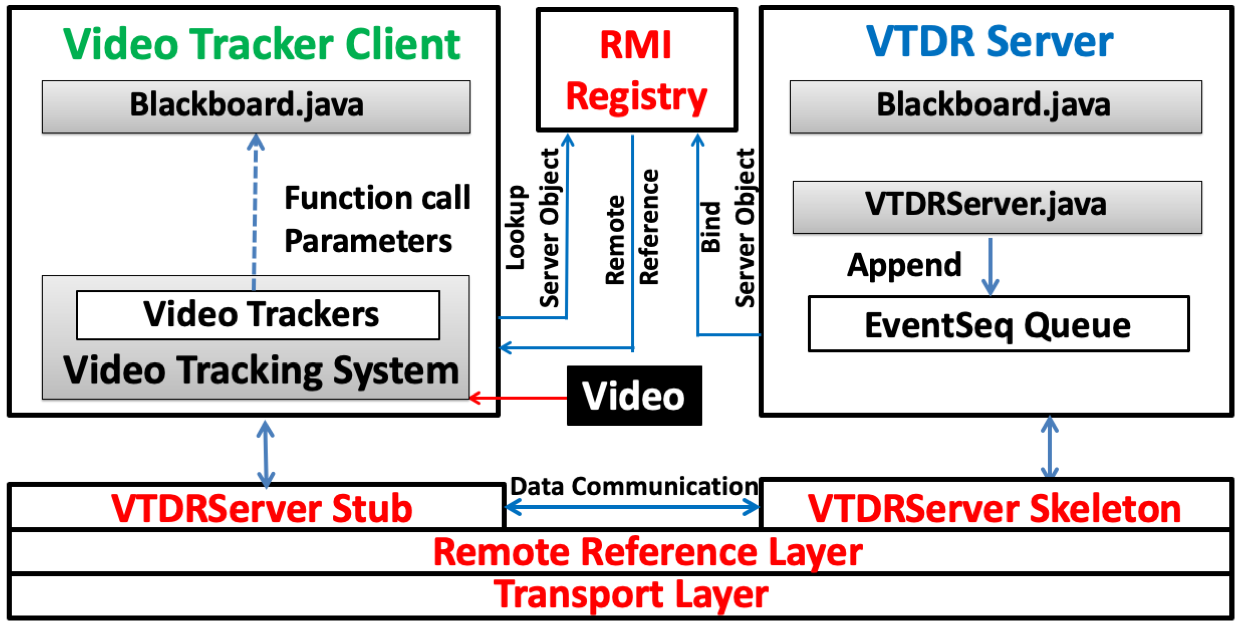}
		\caption*{(a)}
	\end{minipage}
	\vline  
	\hfill
	\begin{minipage}[t]{0.31\textwidth}
		\centering
		\includegraphics[width = \textwidth]{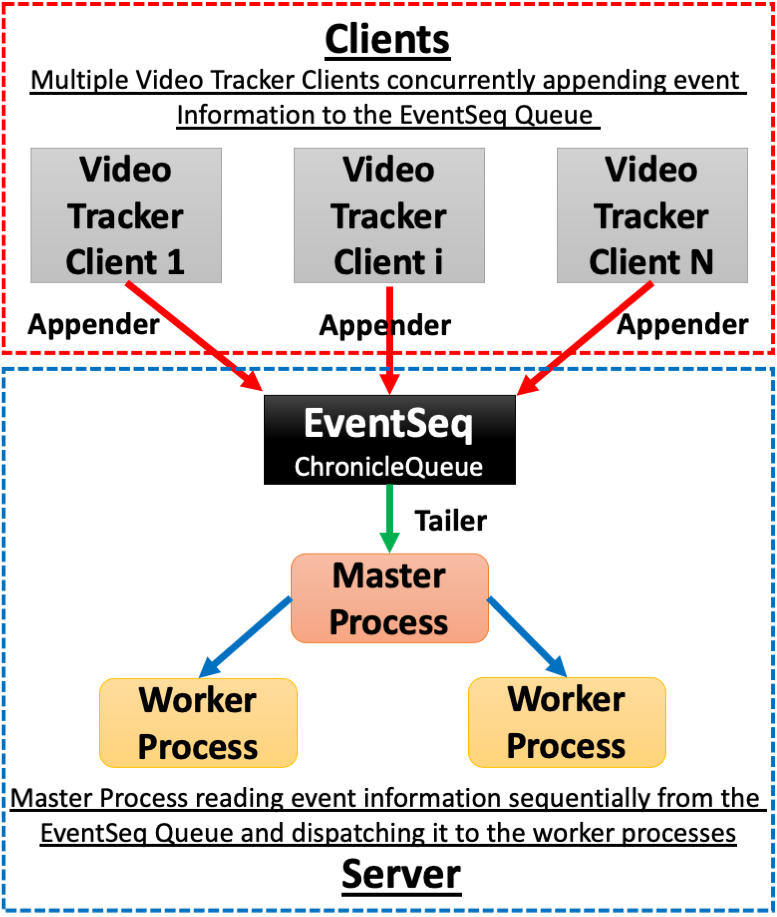}
		\caption*{(b)}
	\end{minipage}
	\caption{\em The plug and play architecture for incorporation with video trackers is shown in (a). A high-level overview illustrating the mechanism of recording and dispatching event information using the EventSeq Queue is shown in (b).}
	\label{fig:ERD}
\end{figure*}

The main purpose of the VTDR module shown in Figure
\ref{fig:ERD}(a) is to provide a Java's RMI (Remote
Method Invocation) \cite{JavaRMI} based plug-n-play
interface for the software clients running video cameras. In the language
	of RMI, a {\em client} refers to the machine that wants to
	invoke a certain functionality in the namespace of another
	machine, which is referred to as the {\em server}.  The
	distributed object programming made possible by RMI is
	founded on the tenet that to a client a remote object should
	{\em appear} as if it were available locally.  
	
	In other words, after a client has located the remote object on the
	server, the client's invocation of the methods of the server
	object should have the same syntax as the methods of an
	object local to the client. 
That is, as long as the client implements the functions
declared in the E-API (Extension Application Programming
Interface) of the VTDR module, the VTDR module and the
client would be able to exchange information seamlessly.  By
calling on its RMI stub classes, the client would be able to
write information to VTDR server's memory.


The E-API\footnote{When a software system defines
		an E-API, that makes it much easier to write code for
		plug-n-play modules for that system.  Technically
		speaking, an E-API for a software system is just like a
		regular API, except that the E-API's functions are meant
		to be implemented by an external entity.} made available by the VTDR Server is shown in
	Figure \ref{fig:ERD}(a). The {\tt Blackboard} interface is a
	pure interface with the function declaration of the
	function that is implemented in the Java
	class {\tt VTDRServer}. The function signature is as follows : 
	\fontsize{9}{4}
	\begin{verbatim}
	void recordEventSequence(Event E)
	\end{verbatim}
	\normalsize where {\tt Event} is a superclass of all events
	that are detected by the video tracker client.  Therefore,
	at runtime, an argument of type {\tt Event} will actually be
	a derived instance of the specific event depending on the
	application.  
	
	As each event of the type listed in Section \ref{sec:Events}
	occurs, the video tracker client software invokes the {\tt recordEventSequence()} function on the
	stub of the {\tt VTDRServer} class.
	VTDR's E-API sits on top of Java RMI, which allows for
	asynchronous interactions between the VTDR module and
	multiple video tracker clients. This design allows CheckSoft to
	operate vis-a-vis any arbitrary number of video cameras monitoring the space.

{ 	
	The information provided by multiple video-tracker clients
	through the E-API is encoded and appended concurrently to a queue named {\tt EventSeq}. This provides a buffer between the video-tracker clients (producers) and the downstream CheckSoft event handler modules (consumers) which facilitates CheckSoft to process events independent of the incidence rate of events. 
	
	In our implementation, we use the Chronicle-Queue\footnote{\url{https://github.com/OpenHFT/Chronicle-Queue}} (CQ) which is a distributed unbounded persisted queue used for high performance and latency-critical applications. It uses `appenders' that write data to the end of the queue and `tailers' that read the next available data in the queue without deleting any data. In our design shown in Fig. \ref{fig:ERD}(b), each video-tracker client records the event information to the EventSeq queue concurrently using an appender. Then the master process of CheckSoft reads the event data using a tailer and dispatches the information to different worker processes. 
	
	The use of CQ in the design of the VTDR module offers the following main advantages:
	\begin{enumerate}
		\item It allows fast communication between the process that runs VTDRServer (writes event data) and the master process of CheckSoft (reads and dispatches event data to other worker processes).
		
		\item It provides data persistence through memory-mapped files with no data loss and additionally storing the data to disk periodically for post-facto analysis. 
		\item It is highly suitable for real-time applications that demand high throughput involving a large number of events with low latency because it uses off-heap memory which is not affected by garbage collection overheads.
		\item It supports concurrent read and write operations and guarantees total ordering of messages. For us this ensures that each video-tracker client appends the event data to the queue in the exact temporal sequence of occurrence. 
	\end{enumerate}}

	It is important to note that a client downloads only a stub
	version of the {\tt VTDRServer} class and the client only knows
	about the signature of the function of this class that
	is declared in the {\tt Blackboard} interface. The client only has to program to the interfaces of the root classes in our software system and has no
	access to any of the implementation code in the server
	class and that creates maximum separation between our software architecture and the code in the video-tracker module. Additionally, the VTDR server and the video tracker clients could be running on different machines in a computer network.

\subsection{Event Dispatcher Module}

The concurrent event information from the video trackers is
recorded in the following encoded format :
\[\textit{[Event type, Event time, Entity Information]}\]
\noindent The master process for {\tt CheckSoft} reads the
encoded event information from the {\tt EventSeq} queue
sequentially in the temporal order of occurrence and broadcasts it to the {\tt HumEnt} and {\tt StoEnt} worker
processes using the {\tt Default\_comm} communicator as
shown in Figure \ref{fig:architecture}(a).

The worker processes decode the event data received from the master process. From the {\it Entity
	Information}, the entities involved in each event are determined and the worker processes assigned to the corresponding entities involved then call the appropriate event handlers to perform a variety of
functions/computations based on the {\it Event Type}, as
shown in the Table \ref{tbl:evhandler} { in the Appendix}.

\subsection{Interaction Module}
\label{sec:IntM}

This module is responsible for detecting the elementary
interactions such as the addition and the removal of
\texttt{OBlob} instances to/from the \texttt{StoEnt}
instances and associating these entities with the {\tt
	HumEnt} entity involved in the interaction. This module is
triggered for every interaction between \texttt{HumEnt} and
\texttt{OBlob} instances contained within \texttt{StoEnt}
instances. Since these interactions can happen simultaneously
in the real-world, this module is implemented such that it
can handle concurrent events.

	Additionally, the Interaction Module is designed to be robust to various sources of noisy event data reported by the video trackers such that the higher-level inferences made by the modules downstream are less prone to errors. We had provided a brief overview of the approach in Section \ref{sec:handlenoise}. In this section we provide a detailed discussion of the same.

	More specifically, the Interaction Module provides immunity against the following : 
	\begin{enumerate}
\item noisy hand-based events and false interactions by enforcing consistency in the finite-state logic between
the different events related to the same overall person-object interaction.
\item inconsistent storage content information as shown in Fig. \ref{fig:cvprob}, due to the following: 
\begin{enumerate}
	\item false and missed object detections in the storage area by filtering out momentary fluctuations in the detections reported by the video trackers. 
	\item occlusion of objects by hands during an interaction by determining the object content before a hand has entered the {\tt StoEnt} area and after the hand has left.
\end{enumerate}
\end{enumerate}

The Interaction module has two submodules:

\subsubsection{Association Submodule}

\begin{figure}[htbp]
	\centering
	\includegraphics[width = 0.5\textwidth]{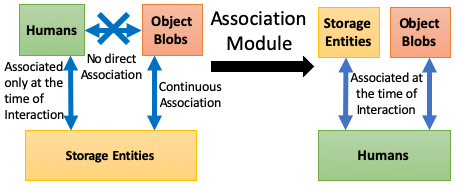}
	\caption{\em{Establishing associations between the HumEnt, StoEnt and OBlobs at the time of interaction.}}
	\label{fig:AMIdea}
\end{figure}

This submodule processes the {\tt HandIn} and {\tt HandOut} events and by monitoring changes in the \texttt{StoEnt} content
due to an interaction, establishes associations between the \texttt{OBlob}, \texttt{StoEnt}
and the \texttt{HumEnt} instances. To elaborate further on this, let us refer to Fig. \ref{fig:AMIdea}. There is a continuous association between {\tt OBlob} and {\tt StoEnt} instances since we know the content information at any given timestamp reported by the video-trackers. An association between a {\tt HumEnt} instance and a {\tt StoEnt} instance is formed at the time of interaction. However, there is no direct association between {\tt HumEnt} and {\tt OBlob} instances. This submodule monitors the change in the content of the {\tt StoEnt} before and after an interaction and figures out the change in the \texttt{OBlob} instances as a result of the interaction and passes this information of the \texttt{OBlob}, \texttt{StoEnt}
and the \texttt{HumEnt} instances involved in any interaction to the Interaction State Machine Submodule. 

\begin{figure*}[t]
	\centering
	\includegraphics[width = 0.8\textwidth]{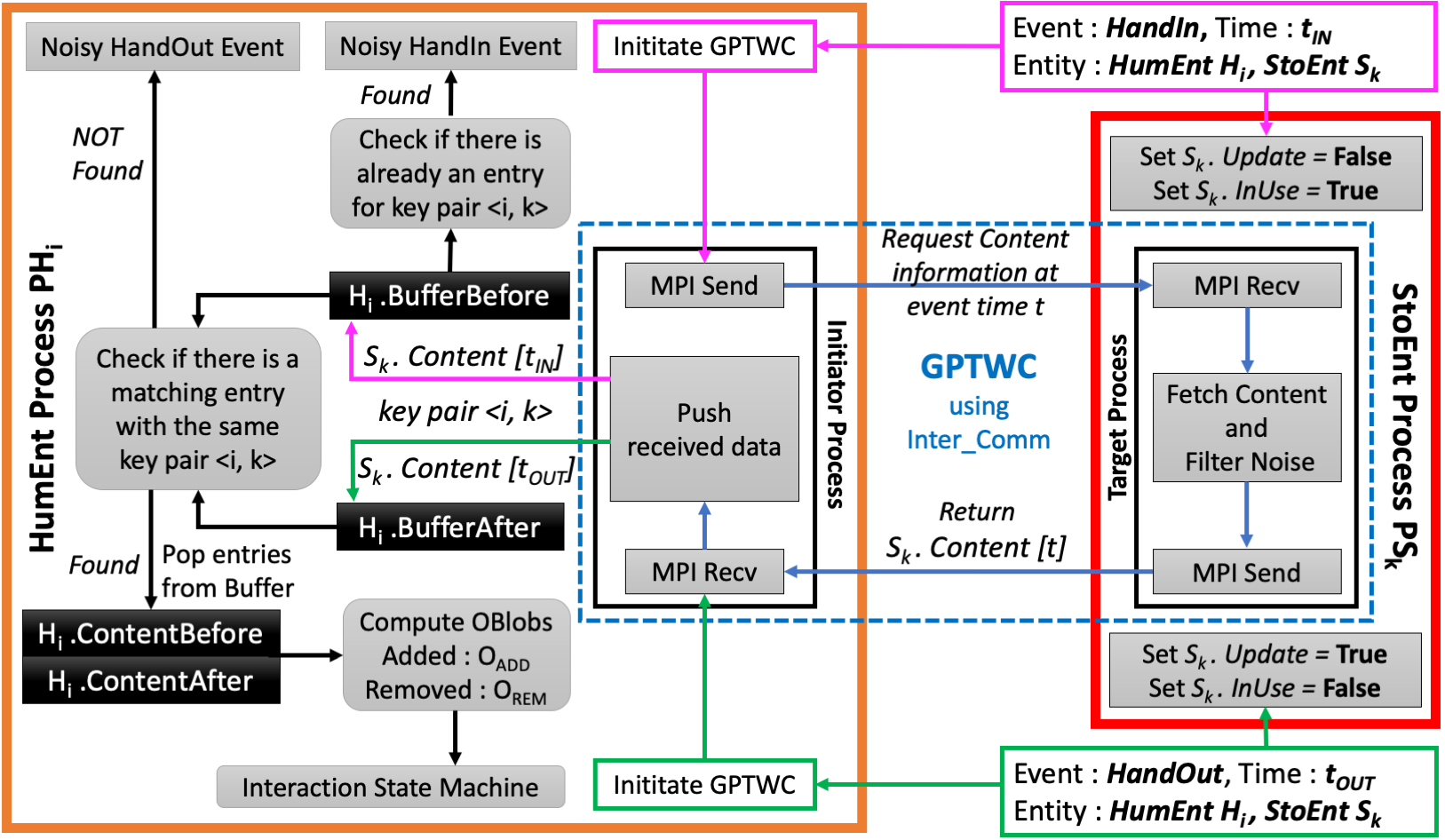}
	\caption{\em{MPI Communication and computations in the Association Module}}
	\label{fig:AM}
\end{figure*}

Let us now refer to the Figure \ref{fig:AM} to understand how the Association Submodule is designed to handle concurrent events. Let us consider an interaction between {\tt HumEnt} $H_i$ and {\tt StoEnt} $S_k$, that in general starts with a {\tt HandIn} event at time $t_{in}$ and ends with a {\tt HandOut} event at time $t_{out}$. Let us denote the process assigned for the $H_i$ instance as $PH_i$ and the process assigned for the $S_k$ instance as $PS_k$. 

When a {\tt HandIn} event occurs, $PS_k$ sets $S_k$.{\tt Update} to False which prevents any erroneous updates to $S_k.${\tt Content} during the interaction when objects might be occluded from the camera by hands. The $S_k$.{\tt Update} value is set to True when the {\tt HandOut} event involving the same HumEnt $H_i$ and StoEnt $S_k$ occurs. On the other hand, $S_k$.{\tt InUse} is set to True when the interaction begins and subsequently set to False at the end of the interaction. During both the {\tt HandIn} and {\tt HandOut} events, there would be a gptwc() function call 
using the inter-communicator {\tt Inter\_comm} between $PH_i$ and $PS_k$, where the process $PH_i$ would be the  initiator process and the $PS_k$ would be the target process.

The \texttt{HumEnt} instances have a specialized buffer data structure that is designed to store information pertaining to concurrently occurring interactions temporarily as well as filter out noisy invalid events effectively as will be explained later. Each entry in this buffer stores the content information ($S_k$.\texttt{Content}($t$)), at interaction time $t$ returned by a gptwc() function call between $PH_i$ and $PS_k$ with an associated unique key pair (<$i,k$>), which denotes the {\sc ID}s of the interacting {\tt HumEnt} $H_i$ and {\tt StoEnt} $S_k$ entities. 
Each {\tt HumEnt} instance has two such buffer data attributes {\tt BufferBefore} and {\tt BufferAfter}, for storing the results returned by a gptwc() function call for {\tt HandIn} and {\tt HandOut} events respectively. \\

During a {\tt HandIn} event, $PH_i$ initiates a request to $PS_k$ and $PS_k$ would be responsible for fetching the content before time $t_{in}$ and filtering the noise across multiple time-stamps before that and send the noise-filtered content information ($S_k$.\texttt{Content}($t_{in}$)) back to $PH_i$ and this would be pushed into $H_i$.{\tt BufferBefore} as:
\[S_k.\texttt{Content}(t_{in}) \xrightarrow[<i,k>]{push} H_i.\texttt{BufferBefore}\]

Similarly during a {\tt HandOut} event calling the gptwc() function would provide the noise-filtered content information ($S_k$.\texttt{Content}($t_{out}$)) after time $t_{out}$. The result would be pushed into $H_i$.{\tt BufferAfter} as :

\[S_k.\texttt{Content}(t_{out}) \xrightarrow[<i,k>]{push} H_i.\texttt{BufferAfter}\]

Now, it will be checked if there is a matching entry with the same key pair (<$i,k$>) in $H_i$.{\tt BufferBefore} and $H_i$.{\tt BufferAfter}, which would denote that this corresponds to a valid interaction between the  unique pair of {\tt HumEnt} $H_i$ and {\tt StoEnt} $S_k$. If a matching entry is found, these entries are popped from $H_i$.{\tt BufferBefore} and $H_i$.{\tt BufferAfter} as :

\[H_i.\texttt{BufferBefore} \xrightarrow[<i,k>]{pop} H_i.\texttt{ContentBefore} = S_k.\texttt{Content}(t_{in})\] 
\[H_i.\texttt{BufferAfter} \xrightarrow[<i,k>]{pop} H_i.\texttt{ContentAfter} = S_k.\texttt{Content}(t_{out})\]

The Process $PH_i$ then computes the objects added ($O_{ADD}$) and removed ($O_{REM}$) by computing the set difference as follows: 

\[O_{ADD} = H_i.\texttt{ContentAfter} - H_i.\texttt{ContentBefore}\]
\[O_{REM} = H_i.\texttt{ContentBefore} - H_i.\texttt{ContentAfter}\]

The extracted information is then passed to the Interaction State Machine Submodule.\\

Now, if there is a {\tt HandOut} event {\em without a matching} {\tt HandIn} event between {\tt HumEnt} $H_i$ and {\tt StoEnt} $S_k$, then there would be no entry in $H_i$.{\tt BufferBefore} with the key pair (<$i,k$>) and the system would detect a {\em Noisy} {\tt HandOut} event and exit with a operation to flush out the erroneous entry in $H_i$.{\tt BufferAfter}. The Interaction State Machine will not be triggered in this case.

Similarly, if there is a previous {\tt HandIn} event {\em without a matching} {\tt HandOut} event, between {\tt HumEnt} $H_i$ and {\tt StoEnt} $S_k$, then there would be a previous entry in $H_i$.{\tt BufferBefore} with the key pair (<$i,k$>) however this error will not be propagated any further because the Interaction State Machine is only triggered if there is a matching {\tt HandOut} event.

This validates the fact that only {\em a matching} {\tt HandIn} and {\tt HandOut} event triggers the Interaction State Machine and all other {\em noisy invalid} {\tt HandIn} and {\tt HandOut} events would be filtered out. 
This design provides immunity against various types of noisy event data as well as handles concurrent interactions reported by video-trackers. \footnote{In some cases, an end user might want to keep track of these interactions for a group comprising of multiple individuals together. For example, in an airport checkpoint security it would be beneficial to keep track of passengers traveling together as a group so that there are no false alarms when members of the same group divest/collect common items. Another possible scenario is keeping track of families shopping together in a retail store such that they can be billed to a single account. In such cases we can associate all the members of the group to a single {\tt HumEnt} instance with a single ID. The design of the Association Module can also effectively handle complicated situations where multiple members in a group associated with the same {\tt HumEnt} instance interacts with multiple {\tt StoEnt} instances at the same time, without any additional changes. This is because each such interaction will have different key pairs (<$i,k$>), where $i$ will remain same but $k$ will be different for different {\tt StoEnt} instances.}

\subsubsection{Interaction State Machine Submodule}
\label{sec:intsm}
\begin{figure*}[h]
	\begin{minipage}[t]{0.55\textwidth}
		\centering
		\includegraphics[width = \textwidth]{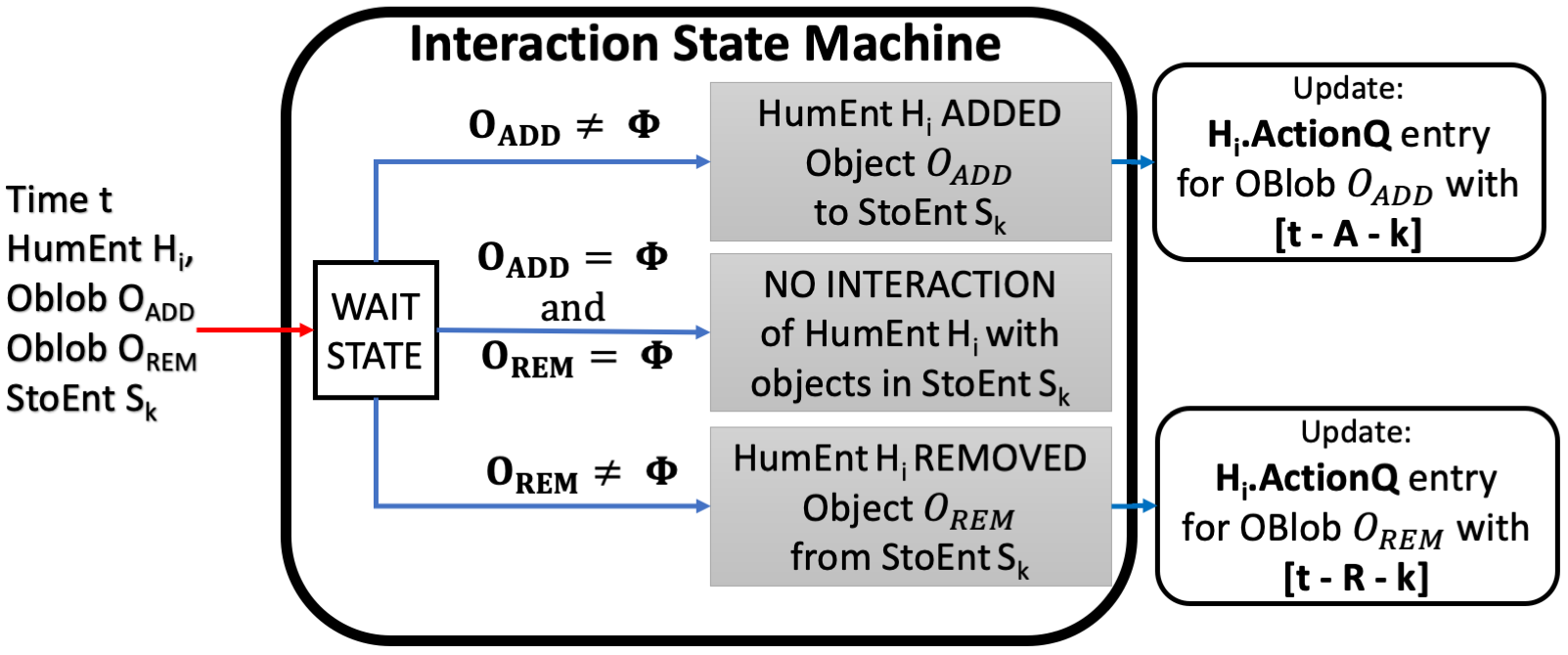}
		\caption*{(a)}
	\end{minipage}
	\hfill 
	\begin{minipage}[t]{0.45\textwidth}
		\centering
		\includegraphics[width = \textwidth]{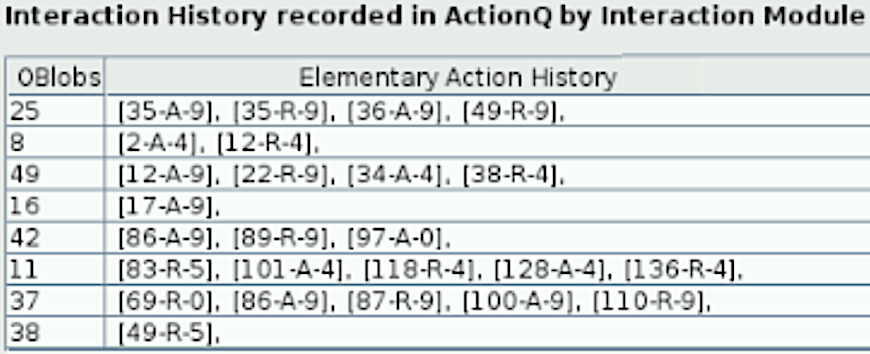}
		\caption*{(b)}
	\end{minipage}
	\caption{\em The Interaction State Machine diagram for each Interaction between \texttt{HumEnt$_i$} (\texttt{$H_i$}) and \texttt{StoEnt$_k$} (\texttt{$S_k$}) at time t is shown in Fig. (a). An example of the elementary action history stored in $H_i.ActionQ$ for each OBlob (stored in a different row) that the particular HumEnt $H_i$ interacted with is shown in Fig. (b)}
	\label{fig:IntSM}
\end{figure*}
This module implements the finite-state machine based logic
to keep track of the elementary interactions between the
\texttt{HumEnt} instances and the \texttt{OBlob} instances
that are present in the different \texttt{StoEnt}
instances. By monitoring the changes in the contents in the
particular {\tt StoEnt} instance the Interaction State
Machine can determine if any {\tt OBlobs} were added and
removed as a result of the interaction.

The states shown in Figure \ref{fig:IntSM}(a) represent the
elementary interactions between the three instances,
\texttt{HumEnt} $H_i$, \texttt{StoEnt} $S_k$ and
\texttt{OBlob} $O$, that would typically be involved in any
interaction. { Let us consider the following three cases:}
\begin{enumerate}
	\item If an object $O_{ADD}$ was added to $S_k$
	($O_{ADD} \neq \phi$) at time $t$, then the string [$t$ - A - $k$]
	is appended to the queue corresponding to \texttt{OBlob}
	$O_{ADD}$ in the data attribute $H_i$.{\tt
		ActionQ}. 
	\item Similarly, if an object $O_{REM}$ was removed
	($O_{REM} \neq \phi$) from $S_k$ at time $t$, then the string  
	[$t$ - R - $k$] is appended to the queue corresponding to
	\texttt{OBlob} $O_{REM}$ in the data attribute $H_i$.{\tt
		ActionQ}. 
	\item However, if no objects were added to/removed
	from $S_k$ ($O_{ADD} = \phi$ and $O_{REM} = \phi$), it could
	either mean that this is a false interaction reported by the video trackers or the {\tt HumEnt} $H_i$ interacted
	with {\tt StoEnt} $S_k$, but did not displace any objects
	and thus no changes are made to $H_i$.{\tt ActionQ}. 
\end{enumerate}
An example of the elementary action history stored in $H_i$.{\tt ActionQ} for each OBlob that the particular HumEnt $H_i$ interacted with can be seen in Fig. \ref{fig:IntSM}(b).

\subsection{Inference Module}
\label{sec:InfM}

This module specializes in
making inferences from the elementary interactions between the \texttt{HumEnt} instances and the \texttt{OBlob} instances. After each interaction, this module is triggered by a {\tt HandOut} event (after the elementary action involving the {\tt HandOut} event is recorded by the Interaction Module). This module is also triggered by the {\tt HumanExit} event. The input to this module is the data attribute $H_i$.{\tt ActionQ}, where $H_i$ is the \texttt{HumEnt} who is either involved in the interaction or is exiting the monitored area. 

Since the higher level interactions as well as the anomalous interactions vary with the specific application that our architecture is used in, the  finite-state machine logic in this module needs to be tweaked. Hence, it is particularly important to design this module such that this does not involve any changes in the overall architectural framework and so that these changes in the logic are minor and can be easily updated, specific to the requirements of the application.

\begin{figure}[h]
	\centering
	\includegraphics[width = 0.4\textwidth]{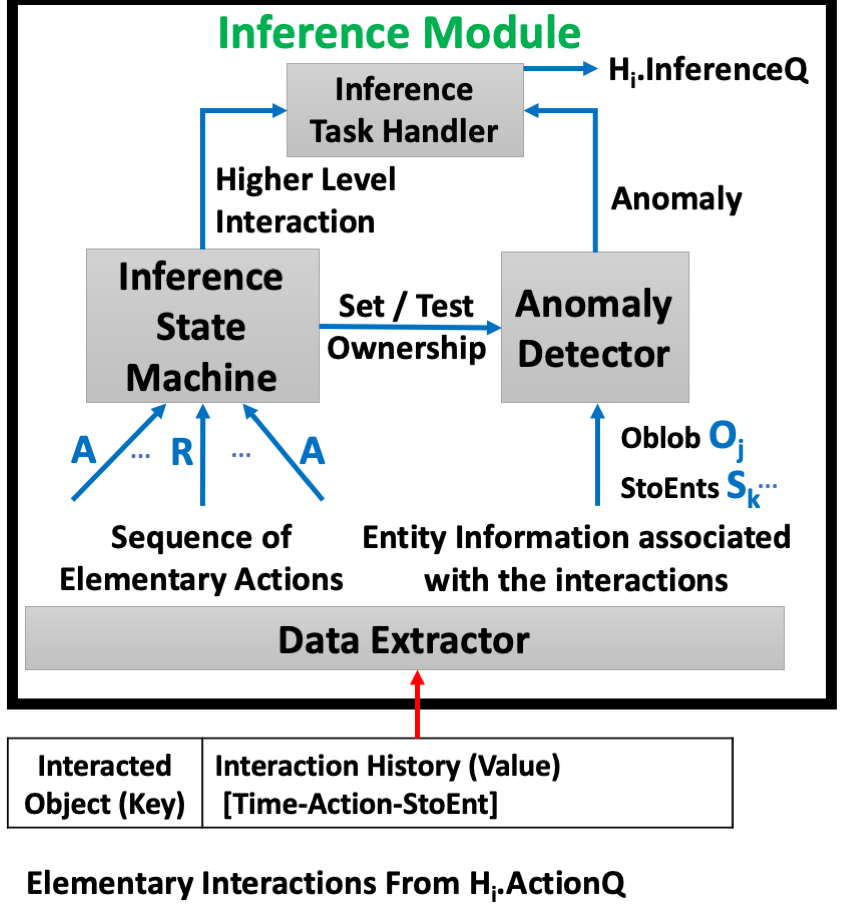}
	\caption{\em The Inference Module architectural diagram with concurrently running Inference State Machine and Anomaly Detector submodules analyzing each entry for \texttt{HumEnt} \textsc{$H_i$} and \texttt{OBlob} \textsc{$O_j$}}
	\label{fig:InfM}
\end{figure}

\subsubsection{Data Extractor} This submodule extracts the interaction history from the {\tt ActionQ} of the {\tt HumEnt} for each {\tt OBlob} that the {\tt HumEnt} has interacted with. 
If the Inference Module is triggered after an interaction (by the {\tt HandOut} event), then only the information for the {\tt OBlob} that was involved in the interaction is extracted. However, if the Inference Module is triggered by the {\tt HumanExit} event then the Data Extractor extracts the information for every {\tt OBlob} that the exiting {\tt HumEnt} interacted with. 

For example according to Figure \ref{fig:IntSM}(b), the latest interaction of the HumEnt was at time 136 with OBlob $O_{11}$.  So the HandOut event at this time triggers the Inference module and the Data Extractor would fetch the following string:
\[ [83-R-5], [101-A-4], [118-R-4], [128-A-4], [136-R-4] \]

The sequence of elementary actions\footnote{An elementary action can either be {\tt Add} or {\tt Remove}, which is represented as {\sc `A'} or {\sc `R'}} for the particular {\tt OBlob} is sent to the Inference State Machine submodule and the information pertaining to the {\tt StoEnt} and {\tt OBlob} instances involved in the interactions are sent to the Anomaly Detector, as shown in Fig. \ref{fig:InfM}. For OBlob $O_{11}$ in Fig \ref{fig:IntSM}(b), the sequence of elementary actions (\{R, A, R, A, R\}) is sent to the Inference State Machine and the information for OBlob $O_{11}$ and StoEnts $S_4$ and $S_5$ is sent to the Anomaly Detector. 
\subsubsection{Inference State Machine Submodule} This submodule implements the finite-state machine based logic to infer the higher level interactions and the ownership relationships between the \texttt{HumEnt}, \texttt{OBlob} and the \texttt{StoEnt} instances. More specifically, this submodule processes the causal relationship between the elementary actions to understand higher level interactions based on the rules of interaction specific to the application. The application-specific inference logic can be customized based on the requirements, as discussed in {Appendix \ref{appendix:InfLogic}.}

If the higher-level interaction inferred alters the ownership relationship then a control signal to {\em Set Ownership} is sent to the Anomaly Detector. Otherwise, the control signal to {\em Test Ownership} is sent to the Anomaly Detector. Typically for the applications we are considering, the first interaction is what determines the Ownership relationship between {\tt HumEnt} instances and {\tt OBlob} instances. 

\subsubsection{Anomaly Detector} This submodule
detects anomalous interactions based on ownership relationships
and raises appropriate alarms when anomalies are
detected. This submodule runs in parallel vis-a-vis the 
Inference State Machine, that dictates its mode of operation.

When the Inference State Machine infers any change in ownership relationships,
it indicates that the Anomaly Detector should set the ownership information and
remember it for detecting anomalies in successive interactions. For example in an airport checkpoint security application, this is done by appending the 
{\tt OBlob} $O_j$ and {\tt StoEnt} $S_k$ information in the {\tt Owns} list of the data attribute $H_i$.{\tt Ownership}. 

On the contrary, when any other type of higher-level interaction is inferred that does not change the ownership relationship, the 
Inference State Machine indicates that the Anomaly Detector should test if 
the entities involved in the interaction belongs to {\tt HumEnt} $H_i$. For testing the ownership of the entities {\tt OBlob} and {\tt StoEnt} we would check if it exists in $H_i$.{\tt Ownership}. If it does not exist in  $H_i$.{\tt Ownership}, then $H_i$ is not the owner and then an appropriate alarm or warning message is issued.\footnote{The actual owner can be determined by a gptwc() function call using {\tt Inter\_comm} to fetch the ownership information of the corresponding entity. }
Each of the latter two submodules have their own handlers to handle application-specific tasks based on the outcome of the inferences and anomaly detections. 

{ In this section, we are only showing the architectural highlights of the Inference Module to demonstrate that CheckSoft can be used in different applications. Obviously, the rules of person-object interactions are application-specific and hence the FSM based inference logic must be tweaked for each application. The inference logic for the Airport Checkpoint Security and Automated Retail Store applications is described in Tables \ref{tbl:InfCheck} and \ref{tbl:InfStore} respectively in Appendix \ref{appendix:InfLogic}. To show examples of higher-level interactions inferred and anomalies detected by the Inference Module from the elementary action history in these applications, we refer the reader to Fig. \ref{fig:GUI}. }



\section{Scalability and Deadlock-Free Operation}
In this section we
present the main features of our software design that
guarantee scalability and deadlock-free operation. The supplemental material includes a Petri Net based modeling of the software for verification of deadlock-free operation and liveness properties. 

\begin{figure}[h]
	\centering
	\includegraphics[width = 0.5\textwidth]{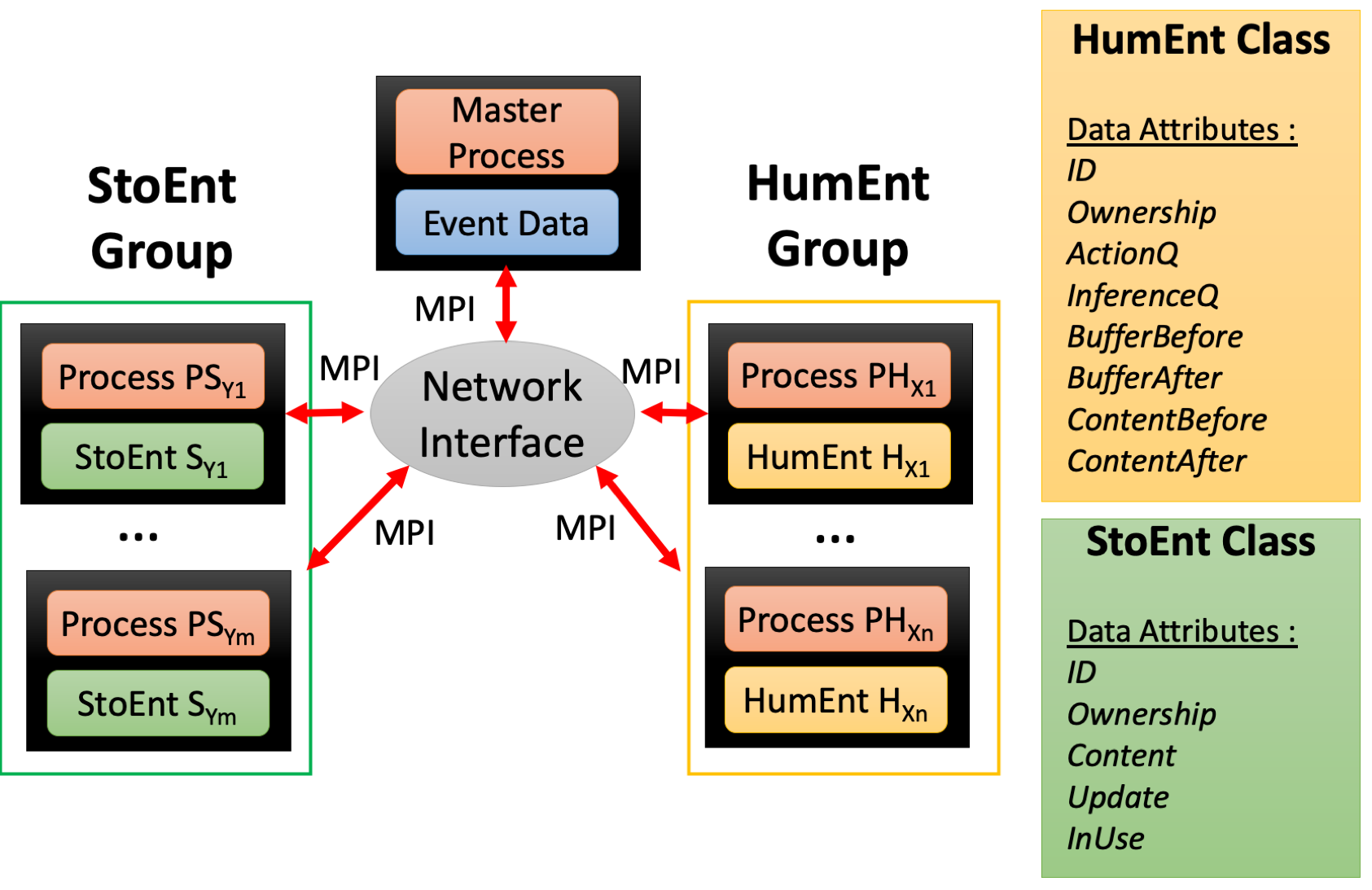}
	\caption{\em CheckSoft architecture using multiprocessing with a distributed memory model.}
	\label{fig:DMM}
\end{figure}

CheckSoft uses multiprocessing with a distributed memory
model in which every process in the system has its own
private memory and all of the computations carried out by a
process only involve the private memory.  Obviously, the
processes that are in charge of analyzing human-object
interactions must somehow become aware of both the humans
involved and the objects in the storage containers. As
opposed to using a shared-memory model, we take care of such
inter-process interactions through the communication
facilities provided by MPI, as shown in Fig \ref{fig:DMM}.
The isolation between the processes achieved in this manner
eliminates any possibility of the processes stepping on
one-another for accessing shared resources, which is a major
reason for deadlock in shared memory systems.


\label{sec:verification}

The master process reads the encoded event information
recorded in the \texttt{EventSeq} queue mentioned earlier
in Section \ref{sec:vtdr} and loads the next available entry into its local memory.  This information is
then broadcast through MPI communications to the different
{\tt HumEnt} and {\tt StoEnt} worker processes, as shown in
Fig \ref{fig:DMM}.  When there is a {\tt HumanEnter} or a
{\tt StorageInstantiate} event, a previously created worker
process is launched and the corresponding entity information
is stored in the local versions of the principal data
structure derived from the base class {\tt Entity}. For
example, the worker processes in the HumEnt group ($PH_{i1}$
to $PH_{in}$) each has a {\tt HumEnt} instance ($H_{i1}$ to
$H_{in}$ respectively) in its own local memory to store the
corresponding human entity information. Similarly, the
worker processes in the StoEnt group ($PS_{k1}$ to
$PS_{km}$) each has a {\tt StoEnt} instance ($S_{k1}$ to
$S_{km}$ respectively) in its own local memory to store the
corresponding storage entity information. When there is a
{\tt HumanExit} or {\tt StorageReturn} event, these worker
processes are freed up and made available for new entities.

We will now use an example to illustrate the fact that each
worker process only needs to work with its own local memory.
Assume that a {\tt StorageUpdate} event has just been
recorded for a {\tt StoEnt} instance $S_k$.  This would
cause the attribute $S_k.{\tt Content}$ of the $S_k$ instance to
be updated by the worker process $PS_k$ if $S_k.{\tt Update}$ is True. To avoid erroneous updates during an interaction involving $S_k$, the worker process $PS_k$ sets $S_k.{\tt Update}$ to False at the beginning of the interaction and subsequently sets it back to True at the end of the interaction. 

It is important to
note that only the process $PS_k$ has access to this content
information of $S_k$. In general, when a process needs some
information from another entity, it fetches the information
using MPI's communication primitives. In our example, when
there is a {\tt HandIn} or {\tt HandOut} event due to an
interaction between a {\tt HumEnt} $H_i$ and a {\tt StoEnt}
$S_k$, the {\em Interaction Module} is triggered in order to
figure out what object has either been placed in the {\tt
	StoEnt} instance or taken out of it. For that, the {\tt
	HumEnt} worker process $PH_i$ needs the storage content of
the {\tt StoEnt} entity $S_k$ before the {\tt HandIn} or
after the {\tt HandOut} event. Towards that end, the process
$PH_i$ initiates a {\tt gptwc()} function call that results
in $PS_k$ fetching the $S_k$.{\tt Content} before/after the
event. Subsequently, $PH_i$ stores this information
temporarily in the $H_i.{\tt BufferBefore}$ buffer or the
$H_i.{\tt BufferAfter}$ buffer, depending on whether the primary
triggering event was {\tt HandIn} or {\tt HandOut}.  The
difference between the content before and after a particular
interaction is then analyzed to figure out what object was
involved in the interaction and the elementary actions
associated with the interaction is then recorded by the process
$PH_i$ in $H_i.{\tt ActionQ}$.  In this manner, only the worker
process $PH_i$ is in charge of recording all the
interactions related to the {\tt HumEnt} $H_i$.   All
the elementary interactions involving any {\tt HumEnt} $H_i$ can be found in the data
structure $H_i.{\tt ActionQ}$.  As a consequence,  at the end of every interaction involving a {\tt
	HumEnt} $H_i$ or when a {\tt
	HumEnt} $H_i$ exits the area being monitored by CheckSoft,
this aspect of our software design allows for all high-level
inferences related to the {\tt HumEnt} $H_i$ to be made
immediately and without any resource contention.

%
%


This design makes CheckSoft scalable to any number of {\tt
	HumEnt}, {\tt StoEnt} and {\tt OBlob} entities. The level
of concurrency, which is the number of active processes or
entities at any given point of time, is, of course, limited
by the computational resources available on the hardware
platform running CheckSoft. The event handlers of {\tt
	CheckSoft} are non-preemptive and hence for handling
certain events, a task may need to wait if the required
process is busy handling a previous task or when
communication between processes is required. In general, a
task would wait until all the required processes are
available and this might incur unwanted latency in handling
events, which should be within reasonable limits of
tolerance. The scalability related aspects have
been analyzed in Section \ref{sec:scalabilityval}.

\section{Performance and Validation}
\label{sec:validation}
For validation, we have adopted a dual approach in
which we use a simulator to study the scalability and
robustness of CheckSoft and use actual video trackers to
analyze other aspects of the system. This is because it
would be highly non-trivial to also analyze the scalability
and robustness issues with real video data in a laboratory
setting.

In this section, we first report in Section \ref{sec:simval}
on the scalability results that we have obtained with the
help of the simulated data involving a large number of {\tt
	HumEnt}, {\tt StoEnt} and {\tt OBlob} entities being monitored by 
a large number of  video trackers. The
scalability study involves investigating two performance
parameters: {\em level of concurrency} and {\em latency}
with simulated data for two different types of applications of CheckSoft: {\em airport checkpoint security} and {\em automated retail store}.  The
simulated data, after noise is added to it, is also used to
validate the robustness of CheckSoft with respect to errors
made by event detectors.

Subsequently, in Section \ref{sec:vidval}, we demonstrate
with real-time data from several cameras that CheckSoft can
indeed process feeds simultaneously from multiple video
trackers.

\subsection{A Simulation Based Large-scale Testing of CheckSoft}
\label{sec:simval}
\begin{figure}[hbtp]
	\centering
	\includegraphics[width = 0.6\textwidth]{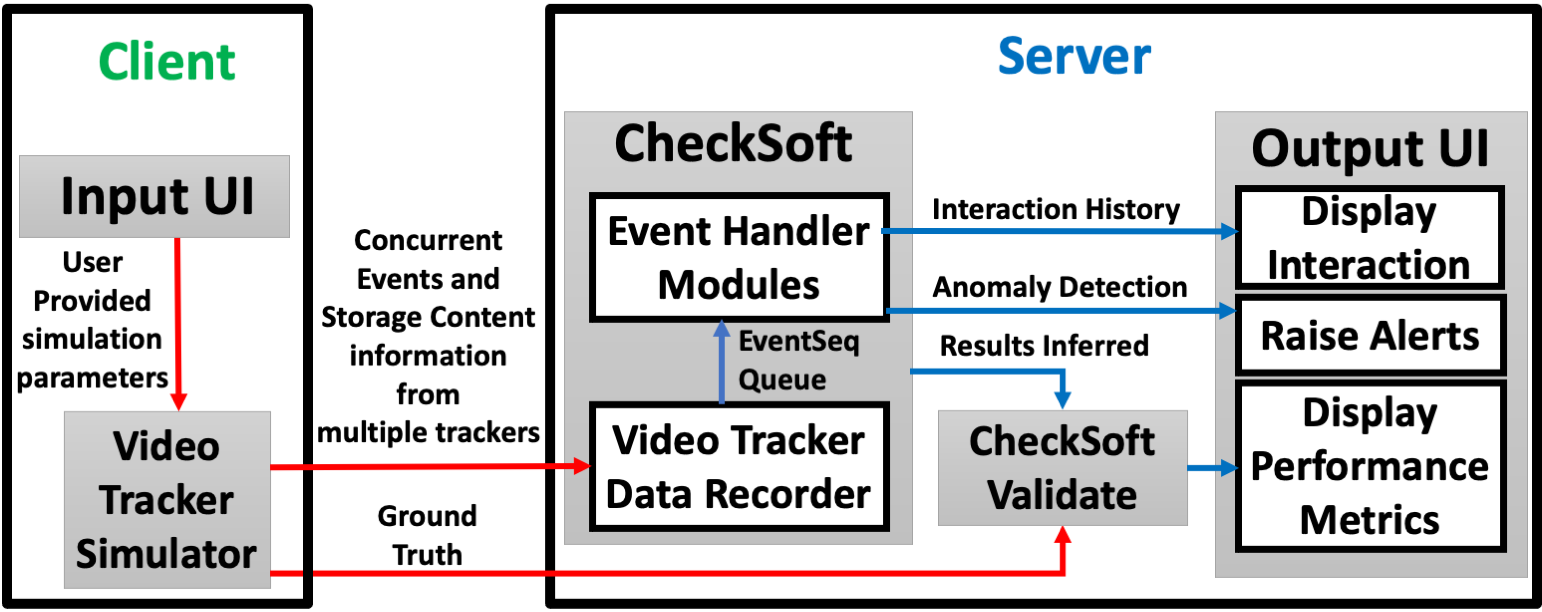}
	\caption{\em Validation Framework for {\tt CheckSoft}.}
	\label{fig:CSVal}
\end{figure}

\begin{figure}[htb]
	\centering
	\includegraphics[width = 0.48\textwidth]{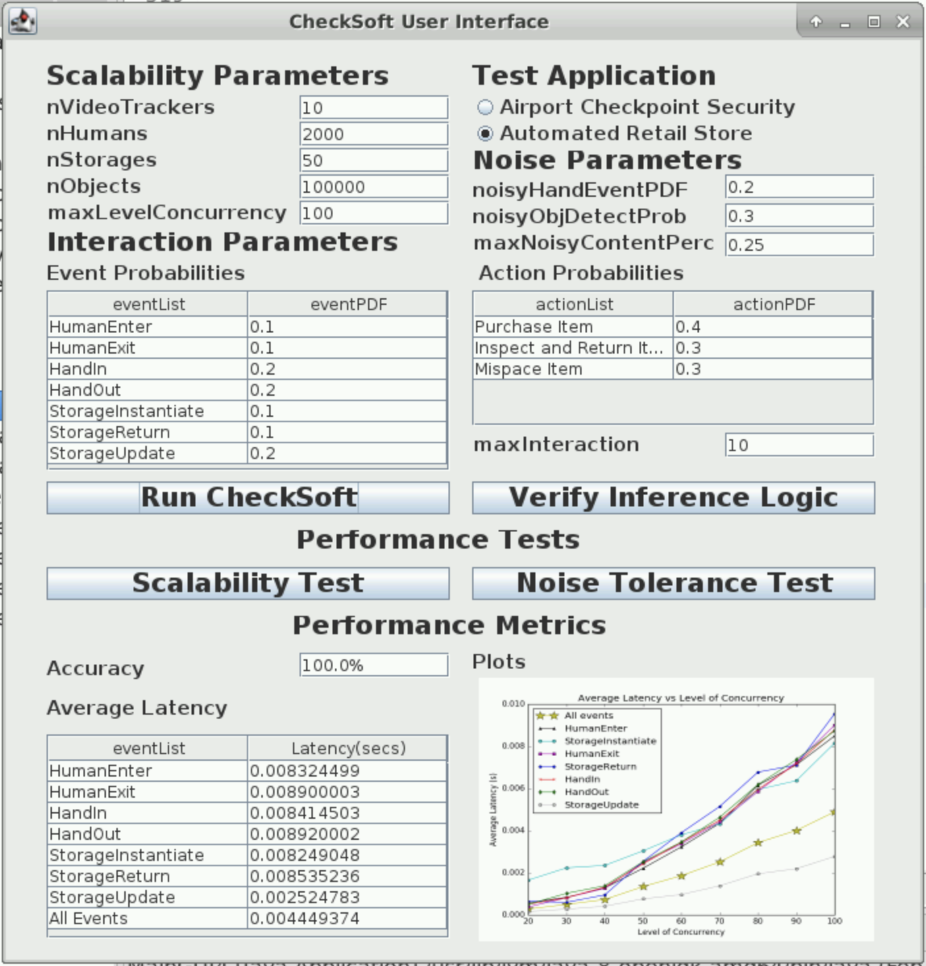}
	\caption{\em The main GUI of {\tt CheckSoft}.}
	\label{fig:MainGUI}
\end{figure}

We have tested the logic of CheckSoft with a
simulation-based validation framework\footnote { The verification framework is available at the following url: 
	\begin{center}https://github.com/sarkar-rohan/CheckSoft \end{center} 
	where a user can run the CheckSoft simulator and investigate its performance with respect to all the design parameters. }, called
CheckSoftValidate, whose ``architecture'' is shown in Fig
\ref{fig:CSVal}.  The validation framework can be used for
testing and evaluating the rules used for associating human
entities with the objects they interact with and for
analyzing the outcomes of these interactions. The front-end
to the validation framework is a UI on the client machine 
that can be used to set the different parameters of a simulated environment
as shown in Fig. \ref{fig:MainGUI}. 
The UI on the server machine displays the complete
state of the monitored area and any detected anomalies, as
inferred by the logic of CheckSoft. 

  \begin{table*}[t]\centering
	\resizebox{0.8\textwidth}{!}{%
		\begin{tabular}{@{}p{0.23\linewidth}|p{0.75\linewidth}@{}}\hline
			{\cellcolor[rgb]{0.792,0.792,0.792}}{\bf \large Parameter} & {\cellcolor[rgb]{0.792,0.792,0.792}}{\bf \large Description} \\ \hline \hline 
			{\cellcolor[rgb]{1,0.906,0.741}} {\bf \normalsize (a)} & {\bf \normalsize Parameters that control the number of entities and video-trackers monitoring the simulated environment}\\
			\hline
			{\cellcolor[rgb]{1,0.906,0.741}}\texttt{nVideoTrackers} & {Number of video-trackers recording event information in server memory simultaneously using RMI based E-API. This represents the number of threads running the Video Tracker Simulator.}\\
			\hline
			{\cellcolor[rgb]{1,0.906,0.741}} \texttt{nHumans}    &  Number of {\tt HumEnt} entities in total. \\
			\hline
			{\cellcolor[rgb]{1,0.906,0.741}}\texttt{nStorages}  &  Number of {\tt StoEnt} entities in total. \\
			\hline
			{\cellcolor[rgb]{1,0.906,0.741}}\texttt{nObjects}   & Number of {\tt OBlob} entities in total. The {\tt OBlob} entities are distributed in the different {\tt StoEnt} entities based on the event sequences generated depending on the application.\\
			\hline
			{\cellcolor[rgb]{1,0.906,0.741}}\texttt{maxLevelConcurrency}   & Maximum number of allowed active {\tt HumEnt} and {\tt StoEnt} entities at any given time instant.\\
			\hline \hline 
			{\cellcolor[rgb]{0.835,1,0.835}} {\bf \normalsize (b)}& {\bf \normalsize Parameters that control the frequency of events and different types of interactions}\\
			\hline
			{\cellcolor[rgb]{0.835,1,0.835}}\texttt{eventList} & List of all possible events \texttt{{[}HumanEnter, HumanExit, HandIn, HandOut, StorageInstantiate, StorageReturn, StorageUpdate{]}} \\
			\hline
			{\cellcolor[rgb]{0.835,1,0.835}}\texttt{eventPDF}  & Probability of the corresponding event in \texttt{eventList} to be drawn at any time step (basically a PDF). The probabilities determine the frequency of the events and control the rate at which humans enter/exit the simulated environment, instantiate/return storage units and interact with objects etc.\\
			\hline
			{\cellcolor[rgb]{0.835,1,0.835}}\texttt{actionList} & List of all possible higher level interactions depending on the specific type of application. This list includes the anomalous interactions as well.\\
			\hline
			{\cellcolor[rgb]{0.835,1,0.835}}\texttt{actionPDF}  & Probability of the corresponding action in \texttt{actionList} to be drawn at any time step (basically a PDF). The probabilities determine the randomized sequence of events for the different interactions that emulate application-specific behavior of any {\tt HumEnt} entity.\\
			\hline
			{\cellcolor[rgb]{0.835,1,0.835}}\texttt{maxInteraction}   & Maximum number of interactions allowed for each {\tt HumEnt} entity.\\
			\hline \hline 
			{\cellcolor[rgb]{1,1,0.675}} {\bf \normalsize (c)}&{\bf \normalsize Parameters that control the noise affecting the simulated data}\\
			\hline
			{\cellcolor[rgb]{1,1,0.675}}\texttt{noisyHandEventPDF}  & Probability of the hand-based events (\texttt{HandIn} and \texttt{HandOut}) to be corrupted by noise. The probabilities determines the effect of noise in the sequence of hand-based events generated. \\
			\hline
			{\cellcolor[rgb]{1,1,0.675}}\texttt{noisyObjDetectProb} & The probability that a particular type of object will be detected wrongly (due to false or missed detections) and the storage content would be corrupted by noise.\\
			\hline
			{\cellcolor[rgb]{1,1,0.675}}\texttt{maxNoisyContentPerc} & The maximum extent that the content of any {\tt StoEnt} will be affected by noise. This determines the maximum percentage of image frames that are corrupted by noise before and after interactions. \\
			\hline

	\end{tabular}}
	\caption{\em Input parameters for simulator to generate stochastic event sequences.}
	\label{tbl:simulator}
\end{table*}

\begin{figure*}[htb!]
	\begin{minipage}[b]{0.49\textwidth}
		\centering
		\includegraphics[width = \textwidth]{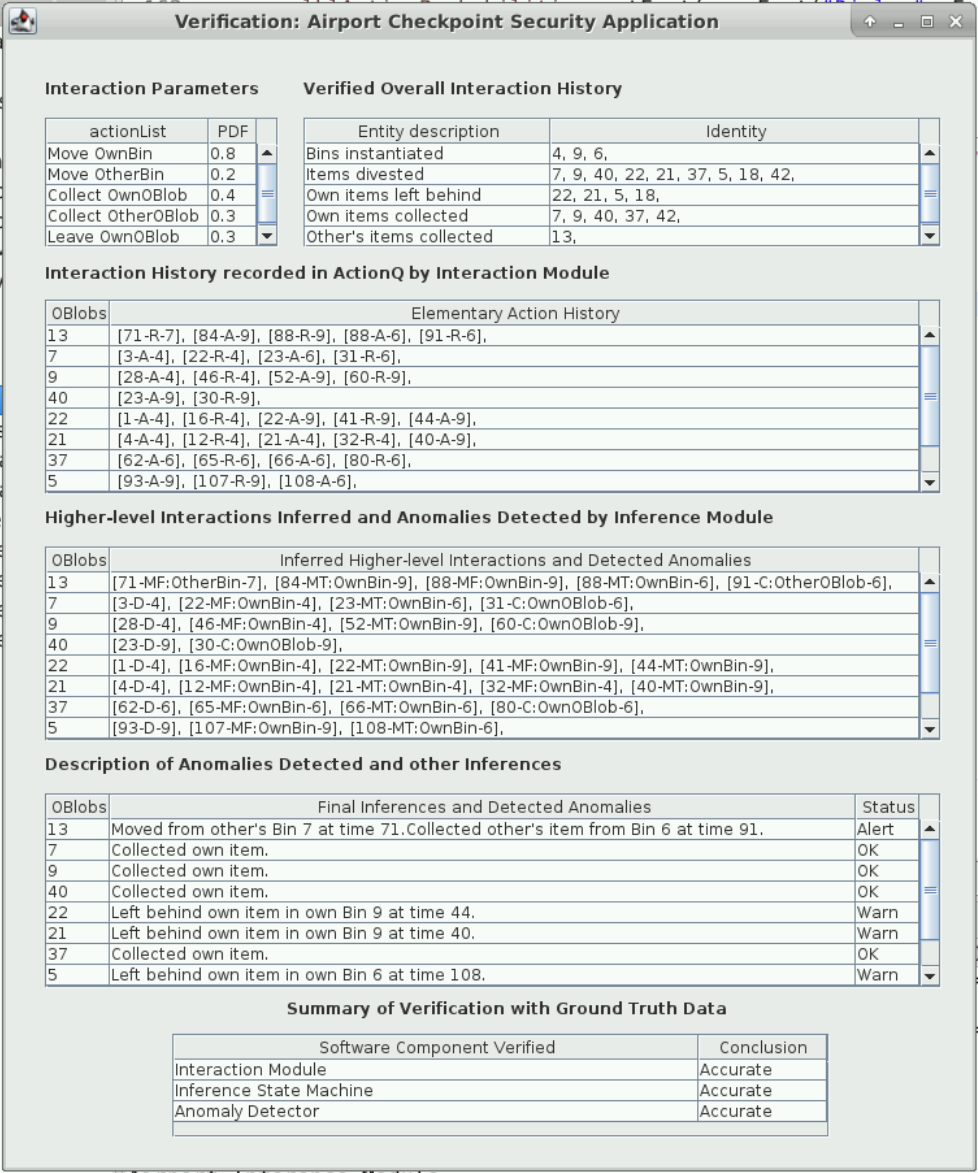}
		\caption*{(a)}
	\end{minipage}
	\hfill
	\begin{minipage}[b]{0.49\textwidth}
		\centering
		\includegraphics[width = \textwidth]{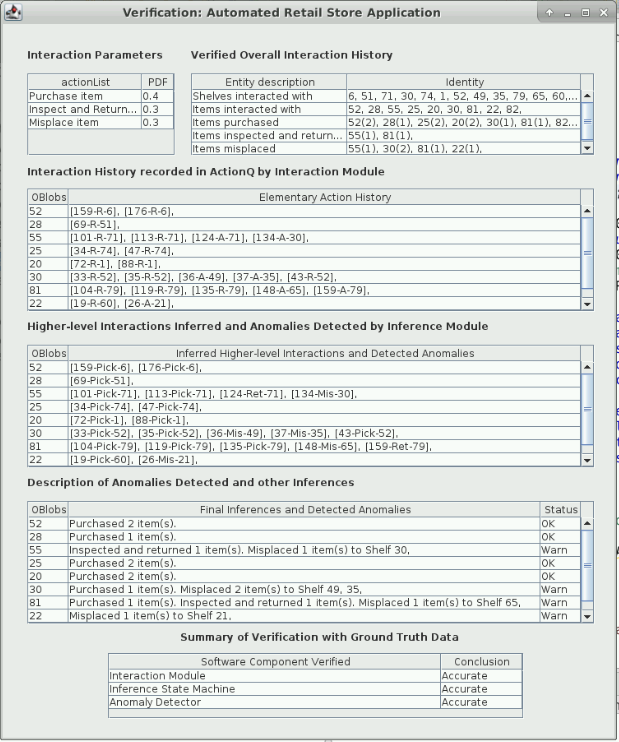}
		\caption*{(b)}
	\end{minipage}
	\caption{\em GUI that shows the interaction history, detected anomalies for an exiting human and verifies the different software modules of {\tt CheckSoft} for Airport Checkpoint Security and Automated Retail Store applications respectively.}
	\label{fig:GUI}
\end{figure*}

As shown in Fig \ref{fig:CSVal}, an important component of
CheckSoftValidate is a video-tracker simulator module running on the 
client machine that
generates the event and storage content information which is
input to CheckSoft. The video-tracker simulator models the
monitored area symbolically and emulates
application-specific human behavior as closely to the real
world as possible. The simulator is multi-threaded 
where each thread symbolically represents a video-tracker detecting concurrent events. The threads upload the
event information parallely to the server memory.

The number of 
video trackers monitoring the simulated area as well as the number of 
human entities, storage entities used, 
objects in the simulated environment and the number 
of human and storage entities that are active at any moment of time 
are controlled by values for the relevant UI parameters mentioned 
in Table \ref{tbl:simulator}(a).
The rate at which the humans enter/exit 
the simulated environment and storage entities are instantiated/returned
as well as the frequency of interactions with objects is
controlled by the values for the relevant UI parameters mentioned 
in Table \ref{tbl:simulator}(b). 
We can also simulate noise in the
environment which generates false hand-based events as well
as false and missed detections of objects in the storage
entities, which are controlled by the values for the relevant 
UI parameters mentioned in Table \ref{tbl:simulator}(c). 
The values of the parameters set in the UI
specifically in Table \ref{tbl:simulator}(b) and (c),
that control the interactions and events as well as the effect of 
noise in the simulated data are
merely the mean values for what are otherwise random
numbers.

CheckSoftValidate uses the data generated by the video
tracker simulator for testing the scalability of the system
as the number of people, the storage units, and the objects
are increased. Besides, the validation framework verifies the operation of the overall decentralized system by testing
the E-API that allows CheckSoft and
the Video Tracker Simulator to work in two different
machines in a network.

As shown in Fig \ref{fig:CSVal}, CheckSoftValidate compares
the results inferred by the logic of CheckSoft with the
ground truth generated by the simulator and computes the
accuracy of CheckSoft under different test scenarios,  generated by varying the 
different parameters in Table \ref{tbl:simulator}(a), (b) and (c). The
results of this comparison are shown in the GUI
displayed in Fig. \ref{fig:GUI} 
for {\em Airport Checkpoint Security}
and {\em Automated Retail Store} applications respectively. 

\subsubsection{Verification of FSM based Inference Logic for Different Applications}
{
	The inference logic presented in Tables \ref{tbl:InfCheck} and \ref{tbl:InfStore} is tested using the CheckSoftValidate framework. In this subsection we are going to briefly discuss the verification results and refer the reader to Appendix \ref{appendix:InfLogic} for a more detailed discussion regarding the same. }

\noindent
{\bf \em Airport Checkpoint Security Application } \\
Fig. \ref{fig:GUI}(a) shows how {\tt CheckSoft} draws inferences about each item that a particular passenger interacted with and raises appropriate warnings or alerts.
	
It can be easily seen
that CheckSoft can correctly verify if passengers collected their 
own items in which case no anomalies are reported. It can also 
raise warnings when passengers leave behind their items 
and raise security alerts when any passenger either moves an item 
from or to a tray that is not their own or collects an item that does 
not belong to them.

\noindent
{\bf \em Automated Retail Store Application}\\
Fig. \ref{fig:GUI}(b) shows how {\tt CheckSoft} draws inferences about each item that the customer interacted with  and raises appropriate warnings if some item is misplaced.
	
It can be easily seen
that CheckSoft can accurately draw inferences regarding how many items of each type were purchased, inspected and returned as well as infer which shelves any particular item has been misplaced to.

\subsubsection{Scalability}
\label{sec:scalabilityval}

Scalability refers to CheckSoft's ability to track
person-object interactions on a continuing basis when it has
to deal with a large number of person entities, storage
containers, and objects.  The number of these entities in 
the simulated environment is controlled by the {\tt nHumans}, 
{\tt nStorages} and {\tt nObjects} parameters in Table \ref{tbl:simulator}(a).
We test for scalability through {\em latency} at a given
{\em concurrency level}.
The {\em concurrency level} is defined as the total number
of worker processes that can simultaneously be active.  As
the reader would recall, a worker process is assigned to
each new {\tt HumEnt} instance and to each new {\tt StoEnt}
instance. This is controlled by the {\tt maxLevelConcurrency} parameter
in Table \ref{tbl:simulator}(a).
The maximum level of concurrency depends obviously on the
computational resources available to run the CheckSoft
software.  As a case in point, in our simulations with
CheckSoftValidate, we have no problems running CheckSoft at
a concurrency level of { 200} on a VM with 24 vCPU and 8 GB RAM as shown in Fig. \ref{fig:MainGUI}.
We are able to do so for both the application domains
mentioned in the introduction to Section 7 -- the {\em airport
	checkpoint security domain} and the {\em automated retail domain}.

The number of video-trackers monitoring the simulated environment is controlled by setting the {\tt nVideoTrackers} parameter in Table \ref{tbl:simulator}(a). The simulated data in the previous paragraph was generated using 50 threads uploading
	event information parallely on the server memory. 
	This validates that multiple video-trackers can asynchronously and simultaneously
	record data using the Blackboard interface of CheckSoft. Therefore, CheckSoft can operate vis-a-vis any number of video-trackers that connect with it on a plug and play basis through the E-API. We have also validated this with actual video-trackers in Section \ref{sec:vidval}.
	
	%

\begin{figure}[htbp]
		\centering
		\includegraphics[width=0.5\textwidth]{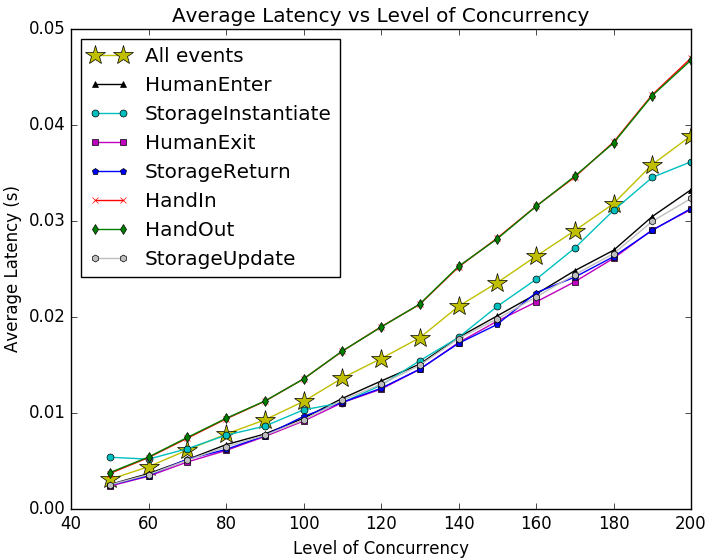} 
		\caption{\em Average Latency for the different event types}
		\label{fig:latency}
	\end{figure}

	By {\em latency} at any given level of concurrency we refer
	to the time taken by CheckSoft event handlers to process any
	event from the time it was first detected.  This metric
	helps us understand the responsiveness of CheckSoft to
	different types of events for a specific level of
	concurrency. Fig \ref{fig:latency} shows the result of an
	experiment in which the level of concurrency was varied from
	{ 50 to 200} and the average latency for each event type was
	then averaged over 10 experiments.
	
	As can be seen in Fig \ref{fig:latency}, the average latency
	of CheckSoft at a concurrency level of 200 is well within
	the reasonable limits considering the fact that the time
	constants associated with typical human-object interactions
	are of the order of a second if not longer.  Fig
	\ref{fig:latency} also shows that the average latency
	increases at a slow rate as the number of active processes
	increases and therefore CheckSoft would scale well to even
	higher levels of concurrency.

	  The {\em HandIn} and {\em HandOut} events involve MPI communication between one of the {\tt HumEnt} and one of the {\tt StoEnt} worker processes and trigger additional FSM based event handling subroutines that filters out noisy events and draws inferences at the end of every interaction and hence has the highest response time. The {\em HumanEnter}, {\em HumanExit}, {\em StorageInstantiate} and {\em StorageReturn} events require collective operations within the {\tt HumEnt\_group} or {\tt StoEnt\_group} and have roughly similar response time because of which the curves overlap. The {\em StorageUpdate} event involves only one of the {\tt StoEnt} worker processes and only updates the latest storage content information for the corresponding {\tt StoEnt}. Since the event handlers for these events do not involve any computation and MPI communication, these events have the lowest response time.
	
	
	
	
	\subsubsection{Tolerance to Noise}
	\label{sec:noiseval}
	The video-tracker simulator part of CheckSoftValidate was
	designed specifically to generate randomized data that would
	correspond to noisy hand-based events as well as erroneous
	storage content for testing the tolerance of CheckSoft to
	measurement and event uncertainties. 
	\begin{figure}[htbp]
		\centering
		\includegraphics[width=0.6\textwidth]{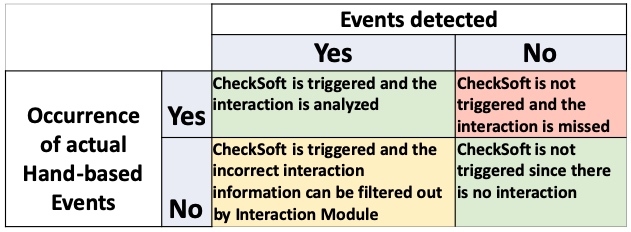} 
		\caption{\em Tolerance of CheckSoft to different types of missed or false hand-based events}
		\label{fig:falseevents}
	\end{figure}

	The probability supplied through the simulation parameter
	{\tt noisyHandEventPDF} in Table 4(c) is responsible for
	generating the noisy hand-based event data. Fig
	\ref{fig:falseevents} describes the effect of missed or
	false hand-based events. While {\tt CheckSoft} can filter
	out false event detections to some extent, a missed
	hand-based event would not even trigger CheckSoft and hence
	it is important that the {\em hand-based event detectors
		have a low false negative rate} as explained in Fig
	\ref{fig:falseevents}.
	
	\begin{figure}[htbp]
		\centering
		\includegraphics[width = 0.5\textwidth]{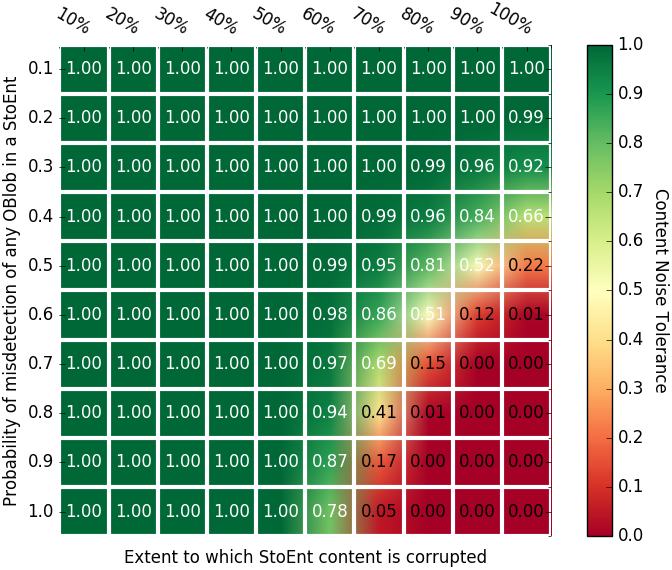}
		\caption{\em Tolerance of {\tt CheckSoft} to Storage Content Noise due to false and missed object detections. }
		\label{fig:noisetol}
	\end{figure}
	
	The simulation parameter {\tt noisyObjDetectProb} in Table 4(c)
	controls the probability of missed and false detections of
	any {\tt OBlob} in a {\tt StoEnt}. At the same time, the
	parameter {\tt maxNoisyContentPerc} shown in the same table
	controls the extent to which the content in any {\tt StoEnt}
	is corrupted by noise. CheckSoft uses a polling based noise
	filtering algorithm to filter out the noise before and after
	interactions across the content information for multiple
	time-stamps. Fig \ref{fig:noisetol} shows how the accuracy
	of CheckSoft decreases as the values supplied for the
	parameters listed here increase  because there is greater sensory 
	noise in recognizing the objects that naturally leads to 
	reduced overall accuracy.
	It gives us a rough
	estimate of the accuracy needed from the video trackers
	detecting objects in the storage units. 
	The probability of 
	missed and false detections and the extent to which the content 
	information is corrupted by noise should be such that the accuracy 
	of CheckSoft remains within the region that is marked by 
	dark green color. The red color in Fig \ref{fig:noisetol}
	indicates low tolerance to noisy content information.


	\subsection{Validation with video-based data from multiple video-trackers}
	\label{sec:vidval}
	\begin{figure}[t]
		\centering
		\includegraphics[width=0.6\textwidth]{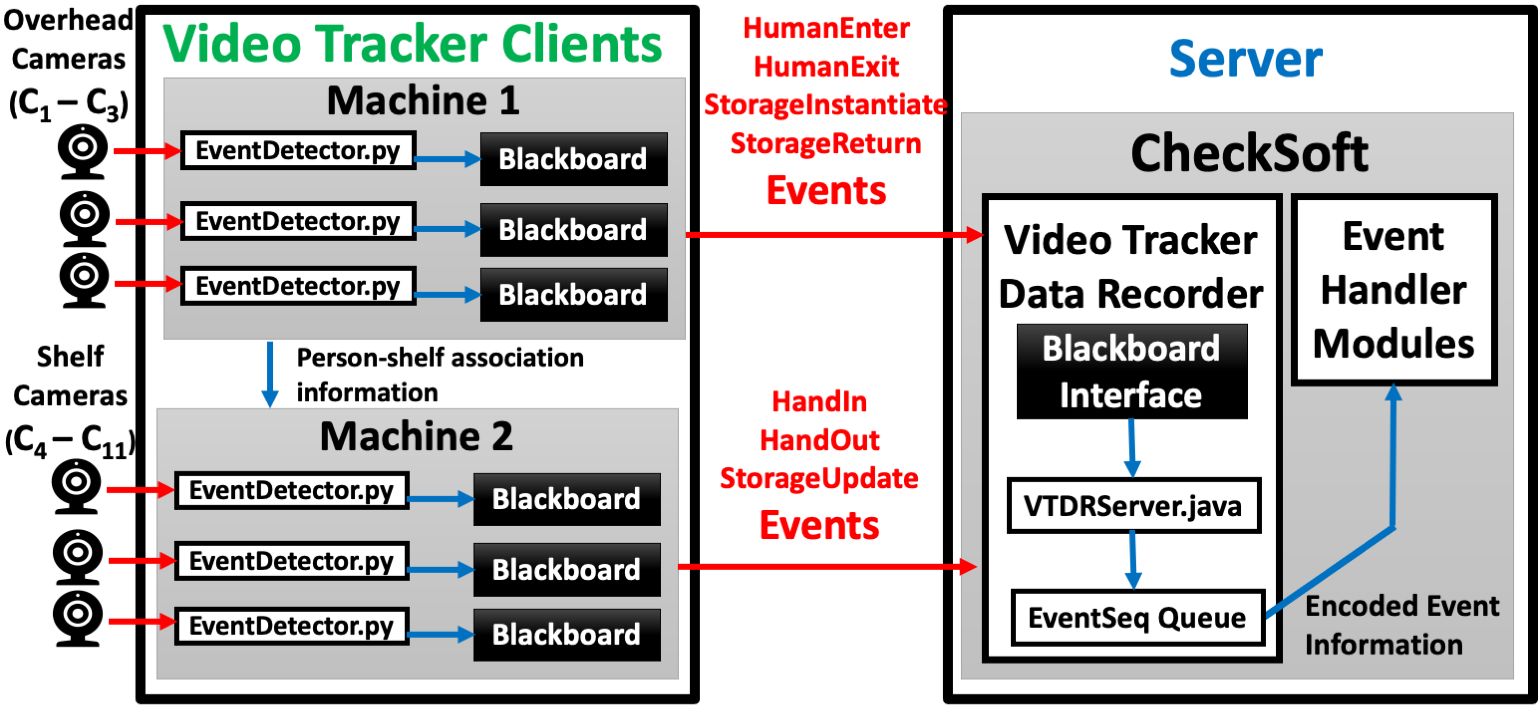} 
		\caption{\em Block Diagram of the Video Trackers implemented to test {\tt CheckSoft}}
		\label{fig:VTBlock}
	\end{figure}

			
	
	While we have tested the scalability of CheckSoft and its
	robustness to noise using simulators, we have used real
	video trackers to establish that CheckSoft can indeed
	process feeds simultaneously from multiple video
	trackers operating in real-time.\footnote{It would be highly non-trivial to also
		analyze the scalability and robustness issues with real
		video data in a laboratory setting.  Hence our dual
		approach in which we use a simulator to study the
		scalability and robustness of CheckSoft and actual video
		trackers to analyze other aspects of the system.} 
	Our experimental setup is a simple retail-store application 
	with shelves storing different types of objects.

	\begin{figure*}[hb]
		\begin{minipage}[b]{0.49\textwidth}
			\centering
			\includegraphics[width = \textwidth]{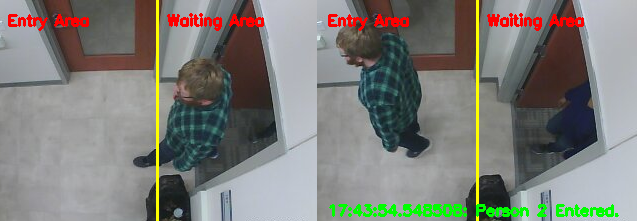}
			\caption*{(a)}
		\end{minipage}
		\hfill
		\begin{minipage}[b]{0.49\textwidth}
			\centering
			\includegraphics[width = \textwidth]{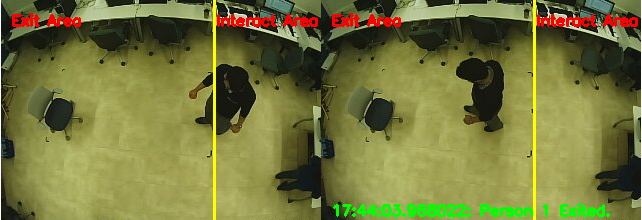}
			\caption*{(b)}
		\end{minipage}
		\caption{\em The video-trackers track people and detect human entry and exit events from the video-feed of overhead cameras.}
		\label{fig:entryexit}
		\centering
		\vspace{0.1in}
		\includegraphics[width=0.8\textwidth]{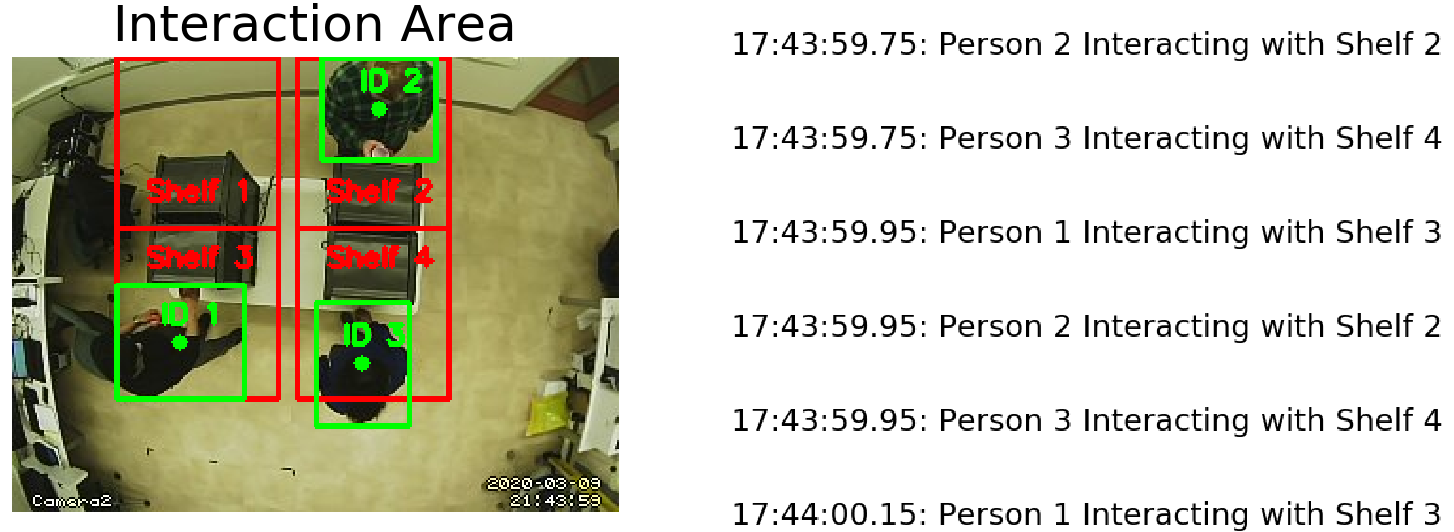} 
		\caption{\em The video-trackers track people and detect interactions with shelf instances from the video-feed of an overhead camera.}
		\label{fig:interact}
	\end{figure*}

	For experiments with real video trackers, we established a
	zone in our laboratory with an array of open shelves.  The
	zone was monitored with eleven cameras and each shelf has two racks 
	each having its
	own camera for recording the content of the shelf and any
	changes in the content.  The eleven area based cameras were
	used to monitor humans as they approached the shelves or
	retracted away from them.

	The video feeds from all the cameras are processed by two PC class client machines. The client machines allot separate processes for the video-trackers hooked to each camera. The VTDR module on the server side, shown in Fig \ref{fig:VTBlock} provides  a  Java  RMI  (Remote  Method  Invocation)  based  plug-n-play  interface  for  the  software clients running  the video-trackers. The video trackers directly call the E-API using the RMI stub classes for uploading the event information asynchronously and concurrently to the server memory, as mentioned in details in Section \ref{sec:vtdr}. 
	
	\begin{figure}[htbp]
		\centering
		\includegraphics[width=\textwidth]{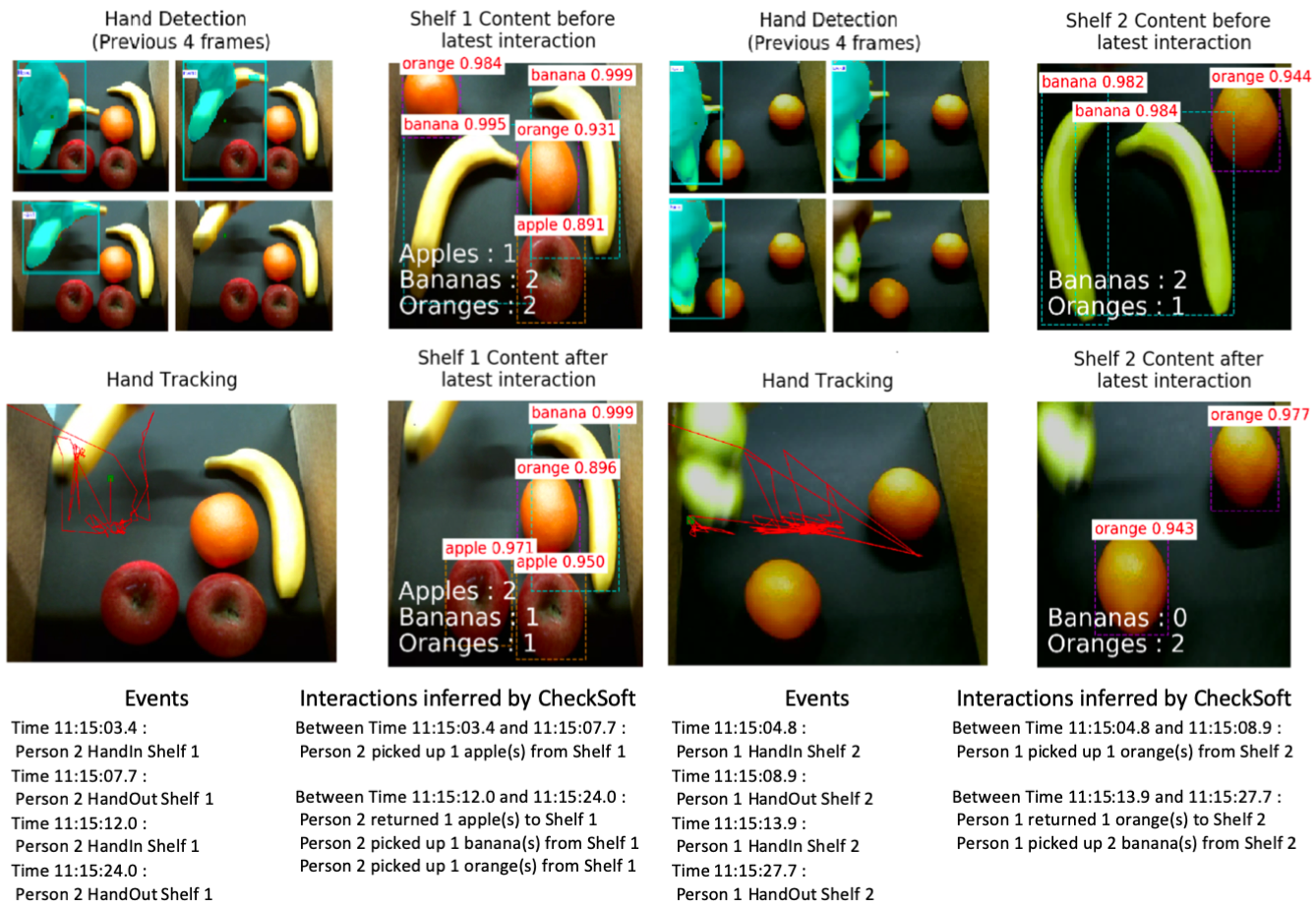} 
		\caption{\em The video-trackers track hand movement inside the shelf (detection for four previous time instants are shown in blue and tracking shown by red lines in the images on the left side) and recognize objects in different shelves before and after interactions from the video-feed of the shelf cameras (as shown by the corresponding labels in the images on the right side). The images to the right also show a count of the number of instances of apples, oranges and/or bananas detected before and after the latest interaction. The news-feed shows the inferences made by CheckSoft in response to the hand-based events detected. }
		\label{fig:action}
	\end{figure}

	As can be seen in Fig. \ref{fig:VTBlock}, each process on the client machines runs a {\tt EventDetector.py} program that detects the various events from the video-feed of the corresponding camera and 
	calls the function {\em recordEventSequence(Event)} made available through the Blackboard interface. The event recording subroutines are implemented in the {\tt VTDRServer.java} program
	that encodes the event information and records it in the {\tt EventSeq} queue. The encoded event information is 
	then dispatched to the different event handler modules of CheckSoft for further processing.


	The video-trackers shown in Fig. \ref{fig:VTBlock} are responsible for tracking person instances and
	their hand movements as well as shelves and the objects 
	present within each shelf to detect events which trigger the different software modules 
	in {\tt CheckSoft}. 
	{
	Fig. \ref{fig:entryexit} shows the video-trackers detecting human entry and exit events from the video-feed of the overhead cameras installed at the entrance and exit.}
	Fig \ref{fig:interact} illustrates the results of person and shelf tracking from the video feed of an overhead camera and the news-feed shows which person is interacting with which shelf.  Fig \ref{fig:action} illustrates hand-based events being detected and video trackers detecting and recognizing objects in storage units before and after an interaction from the video feed of the shelf cameras. 
	
	The shelves shown in Fig \ref{fig:action} contains 3 types of objects -- apples, bananas and oranges. In this simple application, the video-trackers keep track of the number of products of each type present in each of the shelves at any time. The news-feed shows the hand-based events and the application-specific interaction information between person and objects in the shelves as inferred by {\tt CheckSoft}. Based on the content information shown before and after the interaction, the reader can verify the operation of {\tt CheckSoft} for the latest interaction in each of the shelves. 
	\section{Conclusion}
	\label{sec:conclusion}
	CheckSoft is based on several time-honored principles of object-oriented software design [25]. For example, one of the most venerable such principles is that the clients of a software system should only have to program to the public interfaces in the software system. CheckSoft subscribes to this principle by requiring the video tracker clients to only have to be aware of the declaration of the method headers in the Blackboard interface.
	
	Our main goal in this paper was to present a scalable software architecture that can run asynchronously vis-a-vis the video trackers, and that incorporates a finite-state machine based reasoning framework for keeping track of concurrent people-object interactions in people-centric spaces. {\tt CheckSoft} is designed to handle concurrent events simultaneously using a multi-processed event-driven architecture. It is also designed to provide a significant measure of immunity to errors in the event data generated
	by the video trackers.  This is done by enforcing consistency in the finite-state logic between
	the different events related to the same overall person-object interaction.
	
	CheckSoft has so far been tested with both the simulated video trackers and some simple scenarios involving actual video trackers. That was intentional since all we wanted to accomplish at this stage was to formulate the basic architectural design of the software. Our future work would involve testing {\tt CheckSoft} in large-scale applications specifically with regards to scalability and tolerance to noise in real-world applications. 












\typeout{}
\printbibliography

@misc{sarkar2020CheckSoft,
  title={Scalable event-driven software architecture for the automation of people-centric systems},
  author={Sarkar, Rohan and Kak, Avinash},
  year={2020},
  month=dec # "~10",
  publisher={Google Patents},
  note={US Patent App. 16/436,164}
}

@InProceedings{Roda2016,
author="Roda, Cristina
and Rodr{\'i}guez, Arturo
and Navarro, Elena
and L{\'o}pez-Jaquero, V{\'i}ctor
and Gonz{\'a}lez, Pascual",
editor="de la Prieta, Fernando
and Escalona, Mar{\'i}a J.
and Corchuelo, Rafael
and Mathieu, Philippe
and Vale, Zita
and Campbell, Andrew T.
and Rossi, Silvia
and Adam, Emmanuel
and Jim{\'e}nez-L{\'o}pez, Mar{\'i}a D.
and Navarro, Elena M.
and Moreno, Mar{\'i}a N.",
title="Towards an Architecture for a Scalable and Collaborative AmI Environment",
booktitle="Trends in Practical Applications of Scalable Multi-Agent Systems, the PAAMS Collection",
year="2016",
publisher="Springer International Publishing",
address="Cham",
pages="311--323",
isbn="978-3-319-40159-1"
}

@InProceedings{Greaves2018,
author="Greaves, Brian
and Coetzee, Marijke
and Leung, Wai Sze",
editor="Furnell, Steven
and Mouratidis, Haralambos
and Pernul, G{\"u}nther",
title="Access Control Requirements for Physical Spaces Protected by Virtual Perimeters",
booktitle="Trust, Privacy and Security in Digital Business",
year="2018",
publisher="Springer International Publishing",
address="Cham",
pages="182--197",
isbn="978-3-319-98385-1"
}

@ARTICLE{Tsigkanos2018,
author={C. {Tsigkanos} and L. {Pasquale} and C. {Ghezzi} and B. {Nuseibeh}},
journal={IEEE Transactions on Dependable and Secure Computing},
title={On the Interplay Between Cyber and Physical Spaces for Adaptive Security},
year={2018},
volume={15},
number={3},
pages={466-480},
doi={10.1109/TDSC.2016.2599880},
ISSN={2160-9209},
month={May},}

@article{Almeida2019,
title = "A distributed event-driven architectural model based on situational awareness applied on internet of things",
journal = "Information and Software Technology",
volume = "111",
pages = "144 - 158",
year = "2019",
issn = "0950-5849",
doi = "https://doi.org/10.1016/j.infsof.2019.04.001",
url = "http://www.sciencedirect.com/science/article/pii/S0950584918301137",
author = "Ricardo Borges Almeida and Victor Renan Covalski Junes and Roger da Silva Machado and Diórgenes Yuri Leal da Rosa and Lucas Medeiros Donato and Adenauer Corrêa Yamin and Ana Marilza Pernas"
}

@article{Sciphor2019,
title = "Euphoria: A Scalable, event-driven architecture for designing interactions across heterogeneous devices in smart environments",
journal = "Information and Software Technology",
volume = "109",
pages = "43 - 59",
year = "2019",
issn = "0950-5849",
doi = "https://doi.org/10.1016/j.infsof.2019.01.006",
url = "http://www.sciencedirect.com/science/article/pii/S0950584919300096",
author = "Ovidiu-Andrei Schipor and Radu-Daniel Vatavu and Jean Vanderdonckt"
}

@book{Wagner2006,
 author = {Wagner, Ferdinand and Schmuki, Ruedi and Wagner, Thomas and Wolstenholme, Peter},
 title = {Modeling Software with Finite State Machines},
 year = {2006},
 isbn = {0849380863},
 publisher = {Auerbach Publications},
 address = {Boston, MA, USA},
}

@ARTICLE{Anvur1990, 
author={A. {Avnur}}, 
journal={Computing Control Engineering Journal}, 
title={Finite state machines for real-time software engineering}, 
year={1990}, 
volume={1}, 
number={6}, 
pages={275-278}, 
doi={10.1049/cce:19900076}, 
ISSN={0956-3385}, 
month={Nov},}

@article{IBM2008,
author = {Li, Ying and Brown, Lisa and Hampapur, Arun and Lu, Max and Senior, Andrew and Shu, Chiao-Fe},
year = {2008},
month = {10},
pages = {315-327},
title = {IBM smart surveillance system (S3): Event based video surveillance system with an open and extensible framework},
volume = {19},
journal = {Mach. Vis. Appl.},
doi = {10.1007/s00138-008-0153-z}
}

@INPROCEEDINGS{Vezzani2010, 
author={R. {Vezzani} and R. {Cucchiara}}, 
booktitle={2010 IEEE Computer Society Conference on Computer Vision and Pattern Recognition - Workshops}, 
title={Event driven software architecture for multi-camera and distributed surveillance research systems}, 
year={2010}, 
volume={}, 
number={}, 
pages={1-8}, 
doi={10.1109/CVPRW.2010.5543825}, 
ISSN={2160-7516}, 
month={June},}

@book{Ben-Ari1990,
 author = {Ben-Ari, M.},
 title = {Principles of Concurrent and Distributed Programming},
 year = {1990},
 isbn = {0-13-711821-X},
 publisher = {Prentice-Hall, Inc.},
 address = {Upper Saddle River, NJ, USA},
}

@article{Mustanar2019,
  author    = {Mustansar Fiaz and
               Arif Mahmood and
               Soon Ki Jung},
  title     = {Tracking Noisy Targets: {A} Review of Recent Object Tracking Approaches},
  journal   = {CoRR},
  volume    = {abs/1802.03098},
  year      = {2018},
  url       = {http://arxiv.org/abs/1802.03098},
  archivePrefix = {arXiv},
  eprint    = {1802.03098},
  bibsource = {dblp computer science bibliography, https://dblp.org}
}

@misc{Gyori2018,
  title={Shelf with Integrated electronics},
  author={B. Gyori and I. Medrano and A. Frenkel and P. Java},
  year={2018},
  month={Sep},
  publisher={Utility Patent Grant(B1)},
  note={US Patent 10064502}
}

@InProceedings{Frontoni2013,
author="Frontoni, Emanuele
and Raspa, Paolo
and Mancini, Adriano
and Zingaretti, Primo
and Placidi, Valerio",
editor="Petrosino, Alfredo
and Maddalena, Lucia
and Pala, Pietro",
title="Customers' Activity Recognition in Intelligent Retail Environments",
booktitle="New Trends in Image Analysis and Processing -- ICIAP 2013",
year="2013",
publisher="Springer Berlin Heidelberg",
address="Berlin, Heidelberg",
pages="509--516",
isbn="978-3-642-41190-8"
}

@INPROCEEDINGS{Singh2016,
author={B. {Singh} and T. K. {Marks} and M. {Jones} and O. {Tuzel} and M. {Shao}},
booktitle={2016 IEEE Conference on Computer Vision and Pattern Recognition (CVPR)},
title={A Multi-stream Bi-directional Recurrent Neural Network for Fine-Grained Action Detection},
year={2016},
volume={},
number={},
pages={1961-1970},
doi={10.1109/CVPR.2016.216},
ISSN={1063-6919},
month={June},}

@INPROCEEDINGS{Bhargava2007,
author={M. Bhargava and Chia-Chih Chen and M. S. Ryoo and J. K. Aggarwal},
booktitle={2007 IEEE Conference on Advanced Video and Signal Based Surveillance},
title={Detection of abandoned objects in crowded environments},
year={2007},
volume={},
number={},
pages={271-276},
doi={10.1109/AVSS.2007.4425322},
ISSN={},
month={Sept},}

@ARTICLE{JavaRMI, 
title={Java Remote Method Invocation - Distributed Computing for Java, \url{http://www.oracle.com/technetwork/java/javase/tech/index-jsp-138781.html}}}

@INPROCEEDINGS{Radke2011,
author={Z. Wu and R. J. Radke},
booktitle={2011 IEEE Computer Society Conference on Computer Vision and Pattern Recognition - Workshops},
title={Real-time airport security checkpoint surveillance using a camera network},
year={2011},
volume={},
number={},
pages={25-32},
doi={10.1109/CVPRW.2011.5981718},
ISSN={2160-7508},
month={June},}

@inproceedings{Hinze2009,
 author = {Hinze, Annika and Sachs, Kai and Buchmann, Alejandro},
 title = {Event-based Applications and Enabling Technologies},
 booktitle = {Proceedings of the Third ACM International Conference on Distributed Event-Based Systems},
 series = {DEBS '09},
 year = {2009},
 isbn = {978-1-60558-665-6},
 location = {Nashville, Tennessee},
 pages = {1:1--1:15},
 articleno = {1},
 numpages = {15},
 url = {http://doi.acm.org/10.1145/1619258.1619260},
 doi = {10.1145/1619258.1619260},
 acmid = {1619260},
 publisher = {ACM},
 address = {New York, NY, USA},
}

@misc{Chandy2006,
  author = {Chandy, Mani K.},
  description = {events},
  publisher = {Gartner Application Integration and Web Services Summit 2006},
  title = {Event-Driven Applications: Costs, Benefits and Design Approaches},
  year = 2006
}

@InProceedings{Etzion2005,
author="Etzion, Opher",
editor="Adi, Asaf
and Stoutenburg, Suzette
and Tabet, Said",
title="Towards an Event-Driven Architecture: An Infrastructure for Event Processing Position Paper",
booktitle="Rules and Rule Markup Languages for the Semantic Web",
year="2005",
publisher="Springer Berlin Heidelberg",
address="Berlin, Heidelberg",
pages="1--7",
isbn="978-3-540-32270-2"
}

@INPROCEEDINGS{Ning1997, 
author={J. Q. Ning}, 
booktitle={Proceedings Fifth International Symposium on Assessment of Software Tools and Technologies}, 
title={Component-based software engineering (CBSE)}, 
year={1997}, 
volume={}, 
number={}, 
pages={34-43}, 
doi={10.1109/AST.1997.599909}, 
ISSN={}, 
month={Jun},}

@article{Parnas1972,
 author = {Parnas, D. L.},
 title = {On the Criteria to Be Used in Decomposing Systems into Modules},
 journal = {Commun. ACM},
 issue_date = {Dec. 1972},
 volume = {15},
 number = {12},
 month = dec,
 year = {1972},
 issn = {0001-0782},
 pages = {1053--1058},
 numpages = {6},
 url = {http://doi.acm.org/10.1145/361598.361623},
 doi = {10.1145/361598.361623},
 acmid = {361623},
 publisher = {ACM},
 address = {New York, NY, USA},
}
\newpage 
\appendix
\section{| Tables}
\FloatBarrier
\begin{table*}[h!]
	\centering
	\arrayrulecolor{black}
	\resizebox{\textwidth}{!}{%
		\begin{tabular}{|l|l|l|} 
			\hline
			\rowcolor[rgb]{0.792,0.792,0.792} \textbf{Class~} & \textbf{Attributes} & \textbf{Description} \\ 
			\hline
			{\cellcolor[rgb]{1,0.906,0.741}} & {\cellcolor[rgb]{1,0.906,0.741}}\textbf{ID} & This stores an unique identifier for each instance of the different entities. \\ 
			\hhline{|>{\arrayrulecolor[rgb]{1,0.906,0.741}}->{\arrayrulecolor{black}}|-|-|}
			{\cellcolor[rgb]{1,0.906,0.741}}  \textbf{Entity} & {\cellcolor[rgb]{1,0.906,0.741}}\textbf{PhysicalState} & \begin{tabular}[c]{@{}l@{}}This stores data related to the positional co-ordinates and other physical attributes of the real-world entity.\\Storing these values is optional in the current implementation , but might be useful in actual systems.~\end{tabular} \\ 
			\hhline{|>{\arrayrulecolor[rgb]{1,0.906,0.741}}->{\arrayrulecolor{black}}|-|-|}
			{\cellcolor[rgb]{1,0.906,0.741}} & {\cellcolor[rgb]{1,0.906,0.741}}\textbf{Ownership} & \begin{tabular}[c]{@{}l@{}}This stores the ownership relations between the different entities, based on the application. It has two lists :\\ {\tt Owns} : where it stores the information for the instances that this instance is the owner of and\\ {\tt OwnedBy} : where it stores the information for the instances that owns this instance.\end{tabular} \\ 
			\hline
			\hline 
			{\cellcolor[rgb]{0.835,1,0.835}} & {\cellcolor[rgb]{0.835,1,0.835}}\textbf{ActionQ} & \begin{tabular}[c]{@{}l@{}}This data attribute records all the interaction information for each of the different objects that the human\\ interacted with. It is a hashtable of queues which uses the object instances that the {\tt HumEnt} interacted with\\ as keys and stores the history of interaction between the {\tt HumEnt} and each of the objects in separate\\ queues. The interaction information is stored in the temporal order of occurrence. ~\end{tabular} \\ 
			\hhline{|>{\arrayrulecolor[rgb]{0.835,1,0.835}}->{\arrayrulecolor{black}}|-|-|}
			\multirow{-1}{*}{{\cellcolor[rgb]{0.835,1,0.835}}} & {\cellcolor[rgb]{0.835,1,0.835}}\begin{tabular}[c]{@{}>{\cellcolor[rgb]{0.835,1,0.835}}l@{}}\textbf{BufferBefore} \\ \textbf{BufferAfter}\end{tabular} & \begin{tabular}[c]{@{}l@{}}
				This data attribute is primarily used for temporarily storing the storage content 
				before and after multiple \\concurrent interactions that the {\tt HumEnt} instance was involved 
				in, at any particular moment. Multiple such\\ entries are indexed using unique key pairs.~\end{tabular} \\ 
			\hhline{|>{\arrayrulecolor[rgb]{0.835,1,0.835}}->{\arrayrulecolor{black}}|-|-|}
			\multirow{-3}{*}{{\cellcolor[rgb]{0.835,1,0.835}}\textbf{HumEnt}} & {\cellcolor[rgb]{0.835,1,0.835}}\begin{tabular}[c]{@{}>{\cellcolor[rgb]{0.835,1,0.835}}l@{}}\textbf{ContentBefore} \\ \textbf{ContentAfter}\end{tabular} & \begin{tabular}[c]{@{}l@{}}This data attribute is primarily used for computation and stores the storage content
				before and after a\\ particular interaction that the HumEnt instance was involved in.\end{tabular} \\ 
			\hhline{|>{\arrayrulecolor[rgb]{0.835,1,0.835}}->{\arrayrulecolor{black}}|-|-|}
			{\cellcolor[rgb]{0.835,1,0.835}} & {\cellcolor[rgb]{0.835,1,0.835}}\textbf{InferenceQ} & \begin{tabular}[c]{@{}l@{}}This data attribute records the inferences derived from the history of interaction stored in the ActionQ,\\ between the {\tt HumEnt} and each of the objects. The inferences are stored for each of the objects in separate\\ queues, in the temporal order of occurrence. ~\end{tabular} \\ 
			\hhline{|>{\arrayrulecolor[rgb]{0.835,1,0.835}}->{\arrayrulecolor{black}}|-|-|}
			\multirow{-3}{*}{{\cellcolor[rgb]{0.835,1,0.835}}} & {\cellcolor[rgb]{0.835,1,0.835}}\begin{tabular}[c]{@{}>{\cellcolor[rgb]{0.835,1,0.835}}l@{}}\textbf{Actions}\end{tabular} & \begin{tabular}[c]{@{}l@{}}This data attribute stores a list of different types of actions that the {\tt HumEnt} instance can perform, specific\\ to the application. \end{tabular} \\ 
			\hline
				{\cellcolor[rgb]{1,1,0.675}} & {\cellcolor[rgb]{1,1,0.675}}\textbf{Content} & \begin{tabular}[c]{@{}l@{}}Each storage entity maintains its own individual content information with the corresponding time-stamp\\ in this data attribute.~\end{tabular} \\ 
				\hhline{|>{\arrayrulecolor[rgb]{1,1,0.675}}->{\arrayrulecolor{black}}|-|-|}
				{\cellcolor[rgb]{1,1,0.675}}\textbf{StoEnt} & {\cellcolor[rgb]{1,1,0.675}}\textbf{Update} & \begin{tabular}[c]{@{}l@{}}This is a Boolean variable that indicates whether the Content data attribute should be updated or not.\\ This is set to True only if the content information reported by the video trackers is reliable. ~\end{tabular} \\ 
				\hhline{|>{\arrayrulecolor[rgb]{1,1,0.675}}->{\arrayrulecolor{black}}|-|-|}
				{\cellcolor[rgb]{1,1,0.675}} & {\cellcolor[rgb]{1,1,0.675}}\textbf{InUse} & \begin{tabular}[c]{@{}l@{}}This is a Boolean variable that indicates whether the storage entity is currently involved in an interaction.~\end{tabular} \\ 
			\hline
			{\cellcolor[rgb]{1,1,0.675}}\textbf{OBlob} & {\cellcolor[rgb]{1,1,0.675}}\textbf{Characteristics} & \begin{tabular}[c]{@{}l@{}} This stores the information related to the characteristics of the object specific
				to the application. For\\ example, in an  airport checkpoint security this could be 
				the classification defining the security threat\\ posed and in a retail store this could
				store the price of the object for automated billing.~\end{tabular} \\
			\hline
			\multicolumn{1}{l}{} & \multicolumn{1}{l}{} & \multicolumn{1}{l}{}
	\end{tabular}}
	\vspace{-0.15in}
	\caption{\em Attributes of the different classes associated with each Entity in CheckSoft}
	\label{tbl:ds}
\end{table*}
\FloatBarrier
\begin{table*}[h!]	
	\centering \arrayrulecolor{black}
	\resizebox{\textwidth}{!}{%
		\begin{tabular}{|l|l|l|} 
			\hline
			\rowcolor[rgb]{0.792,0.792,0.792} \textbf{Class} & \textbf{Attributes} & \textbf{Description} \\ 
			\hline
			{\cellcolor[rgb]{1,0.906,0.741}}\textbf{Event} & {\cellcolor[rgb]{1,0.906,0.741}}\begin{tabular}[c]{@{}l@{}} \textbf{String Type}\\\textbf{String Time} \end{tabular}& \begin{tabular}[c]{@{}l@{}} This is the root class from which child classes for different events are derived. The time-stamp\\
				indicating when the event occurred is stored in the attribute {\tt Time} and the specific event type\\
				is stored in the attribute {\tt Type}. \\
				This class has a member function {\tt List\textless String\textgreater encodeEventInfo()} that can encode the\\
				information related to the type, time-stamp and the information regarding 
				the entities involved\\
				with the specific event in the following format : \\
				{\tt [Type, Time, Information regarding Entities involved in the event]}
			\end{tabular} \\
			\hline
			\hline 
			{\cellcolor[rgb]{1,1,0.675}}\textbf{HumanEnter} & {\cellcolor[rgb]{1,1,0.675}}\textbf{HumEnt H} & \begin{tabular}[c]{@{}l@{}} This class is associated with the event that a new Human Entity enters the monitored space.\end{tabular} \\ 
			\hline
			{\cellcolor[rgb]{1,1,0.675}}\textbf{HumanExit} & {\cellcolor[rgb]{1,1,0.675}}\textbf{HumEnt H} & \begin{tabular}[c]{@{}l@{}} This class is associated with the event that an existing Human Entity exits the monitored space.\end{tabular} \\ 
			\hline
			{\cellcolor[rgb]{1,1,0.675}}\textbf{HandIn} & {\cellcolor[rgb]{1,1,0.675}}\begin{tabular}[c]{@{}l@{}} \textbf{HumEnt H}\\\textbf{StoEnt S} \end{tabular}& \begin{tabular}[c]{@{}l@{}} This class is associated with the event that any of the hands of an existing Human Entity \\ reaches inside an existing Storage Entity.\end{tabular} \\
			\hline
			{\cellcolor[rgb]{1,1,0.675}}\textbf{HandOut} & {\cellcolor[rgb]{1,1,0.675}}\begin{tabular}[c]{@{}l@{}} \textbf{HumEnt H}\\\textbf{StoEnt S} \end{tabular}& \begin{tabular}[c]{@{}l@{}} This class is associated with the event that all the hands of an existing Human Entity comes\\ outside an existing Storage Entity.\end{tabular} \\ 
			\hline
			{\cellcolor[rgb]{1,1,0.675}}\textbf{StorageInstantiate} & {\cellcolor[rgb]{1,1,0.675}}\begin{tabular}[c]{@{}l@{}} \textbf{StoEnt S}\\\textbf{HumEnt H} \end{tabular}& \begin{tabular}[c]{@{}l@{}} This class is associated with the event that a new Storage Entity is instantiated by an existing\\ Human Entity.\end{tabular} \\ 
			\hline
			{\cellcolor[rgb]{1,1,0.675}}\textbf{StorageReturn} & {\cellcolor[rgb]{1,1,0.675}}\begin{tabular}[c]{@{}l@{}} \textbf{StoEnt S}\\\textbf{HumEnt H} \end{tabular}& \begin{tabular}[c]{@{}l@{}} This class is associated with the event that an existing Storage Entity is returned by an existing\\ Human Entity.\end{tabular} \\
			\hline
			{\cellcolor[rgb]{1,1,0.675}}\textbf{StorageUpdate} & {\cellcolor[rgb]{1,1,0.675}}\begin{tabular}[c]{@{}l@{}} \textbf{StoEnt S}\\\textbf{List\textless OBlob\textgreater O} \end{tabular}& \begin{tabular}[c]{@{}l@{}} This class is associated with the event that new content information is available for an existing\\
				Storage Entity. This class has a member function {\tt List\textless String\textgreater  parseContentInfo()} \\
				that can parse the information related to the different Object Blobs present inside the Storage\\ Entity at any particular moment of time. \end{tabular} \\
			\hline
			\multicolumn{1}{l}{} & \multicolumn{1}{l}{} & \multicolumn{1}{l}{}
	\end{tabular}}
	\vspace{-0.15in}
	\caption{\em Attributes and description of the different classes associated with each Event in CheckSoft}
	\label{tbl:events}
\end{table*}
\FloatBarrier
\vspace{-2in}
\FloatBarrier
\begin{table*}[h!]
	\begin{center}
		\resizebox{\textwidth}{!}{%
			\begin{tabular}{ |c|l|l| } 
				\hline
				\rowcolor[rgb]{0.792,0.792,0.792}{\bf \large Event} & {\bf \large Event Handler} & \multicolumn{1}{|c|}{{\bf \large Functionality performed by CheckSoft}} \\
				\hline
				{\tt Human} & {\bf HandleEntry} & Assign an available worker process from the {\tt HumEnt\_Group}. Instantiate an instance derived\\
				{\tt Enter} & & from the {\tt HumEnt} class, depending on the application and initialize the instance with an unique \\
				& & {\sc id}, update the time of entry {\tt t$_{entry}$} and initialize the other data attributes. \\
				\hline
				{\tt Storage} & {\bf HandleInstantiate} & Assign an available worker process from the {\tt StoEnt\_Group}. Instantiate an instance derived\\
				{\tt Instantiate} & & from the {\tt StoEnt} class, depending on the application and initialize the instance with an unique\\
				& &  {\tt ID}, update the time of instantiation {\tt t$_{instantiate}$} and initialize the other data attributes. If the \\
				& & {\tt StoEnt} $S_k$ was instantiated by a {\tt HumEnt} instance $H_i$ then update the ownership information \\
				& & in the {\tt OwnedBy} list in $S_k$. {\tt Ownership} indicating the owner of $S_k$ is $H_i$ and the {\tt Owns}\\
				& & list in $H_i$. {\tt Ownership} indicating $H_i$ owns $S_k$.\\
				\hline
				{\tt Storage} & {\bf HandleReturn} & The  time of return {\tt t$_{return}$} is updated and the entity information is permanently recorded. The \\
				{\tt Return} & & {\tt StoEnt} instance $S_k$ is deleted and the allotted worker process is then freed and made \\
				& & available to  the {\tt StoEnt\_Group}.\\
				\hline
				{\tt Storage} & {\bf HandleUpdate} & This event notifies the worker process for {\tt StoEnt} $S_k$ that new content information is available \\ 
				{\tt Update} & & from the video trackers and consequently fetches the latest {\tt OBlob} content from the \\
				& & corresponding {\tt Storage$_Y$.csv} file and updates it in $S_k$.{\tt Content} only if $S_k$.{\tt Update = True}. \\
				\hline
				{\tt HandIn}  & {\bf Interaction Module} & This event triggers the Interaction module (Section \ref{sec:IntM})  and the process for {\tt HumEnt} $H_i$ fetches\\
				& {\em -- Association Submodule}&  the noise-filtered content just before the event, from the process for {\tt StoEnt} $S_k$ using a gptwc() \\
				& & function call over the {\tt Inter\_comm} and stores it in $H_i$.{\tt BufferBefore} . The process for \\
				& & StoEnt $S_k$ sets $S_k$.{\tt Update = False} to prevent updating the state of $S_k$ during the interaction.\\
				\hline
				{\tt HandOut} & {\bf Interaction Module} &  This event triggers the Interaction module (Section \ref{sec:IntM})  and the process for {\tt HumEnt} $H_i$ fetches \\
				& {\em -- Association Submodule} & the noise-filtered content just after the event, from the process for {\tt StoEnt} $S_k$ using a gptwc() \\
				& {\em -- Interaction State Machine}& function call over the {\tt Inter\_comm} and stores it in $H_i$.{\tt BufferAfter}. It then checks for a\\
				&  {\bf Inference Module} & matching {\tt HandIn} event and if a matching entry is found in $H_i$.{\tt BufferBefore}, the elementary \\
				& {\em -- Data Extractor} &  interaction information is extracted and then recorded in $H_i$.{\tt ActionQ}.  Subsequently, the \\ 
				&{\em -- Inference State Machine} & Inference module (Section \ref{sec:InfM}) is triggered, that analyses the latest interaction history of the \\
				& {\em -- Anomaly Detector} & {\tt OBlob} involved in the interaction and records the inferences made in the $H_i$.{\tt InferenceQ}.  \\
				& &  It raises alerts if {\tt HumEnt} $H_i$ was involved in any anomalous interactions. The process for \\
				& & StoEnt $S_k$ sets $S_k$.{\tt Update = True} to allow updates to the state of $S_k$ after the interaction.\\
				\hline
				{\tt Human} & {\bf Inference Module}& This event also triggers the Inference module (Section \ref{sec:InfM}) that analyses the interaction history \\
				{\tt Exit} & {\em -- Data Extractor} &  with each {\tt OBlob} in $H_i$.{\tt ActionQ} and records the inferences made in the $H_i$.{\tt InferenceQ}. \\
				& {\em -- Inference State Machine}& It raises alerts if {\tt HumEnt} $H_i$ was involved in any anomalous interactions.  Once all inferences \\
				& {\em -- Anomaly Detector} & are made, the time of exit {\tt t$_{exit}$} is updated and the entity information is permanently recorded. \\
				& {\bf HandleExit}& The instance is then deleted and the allotted worker process is freed and made available to the\\
				& & {\tt HumEnt\_Group}.\\
				\hline
.		\end{tabular}}
		\caption{\em Functionality of the different Event Handlers}
		\label{tbl:evhandler}
	\end{center}
	\vspace{-0.1in}
\end{table*}
\FloatBarrier
\section{| General Purpose Two Way Communication}
\label{appendix:GPTWC}
\begin{figure}[htb]
	\begin{center}
		\includegraphics[width = 0.45\textwidth]{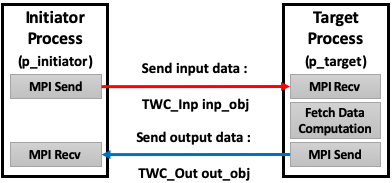}
		\caption{\em General Purpose Two Way Communication function}
		\label{fig:gptwc}
	\end{center}
\end{figure}
The signature of the {\tt gptwc()} function as follows:

\fontsize{9}{4}
\begin{verbatim}
TWC_Out gptwc(Comm comm_name, String oper_type, int p_initiator, int p_target, TWC_Inp inp_obj)
\end{verbatim}
\normalsize

In order to understand the signature of {\tt gptwc()}, the
reader would find it helpful to also look at the depiction
in Figure \ref{fig:gptwc} that shows an initiator process
and a target process.  The former would also like to send a
data object {\tt inp\_obj}, of type {\tt TWC\_Inp}, to the
latter and receive from the latter a result in the form of a
data object denoted {\tt out\_obj} of type {\tt TWC\_Out}.

Revisiting the function signature shown above, the first
parameter in the function signature is {\tt comm\_name}
which must be set to the communicator needed for the two-way
communication link. The parameter {\tt oper\_type} indicates
the specific type of functionality that needs to be executed
by the target process. The parameter {\tt p\_initiator} is
the initiator process rank that requested the two-way
communication link. By the same token, the parameter {\tt
	p\_target} is the target process rank that is the other
endpoint of the communication link. The parameter {\tt
	inp\_obj} is an instance of the class {\tt TWC\_Inp} that
has two attributes: (1) {\tt d\_input} that is set to the
data block being sent by the initiator process to the target
process; and (2) {\tt mode} that is used to notify the
target process the type of operation it needs to perform as
follows:
\[
{\tt mode}=
\begin{cases}
0: & \text{Fetch Data using {\tt d\_input}} \\
1: & \text{Compute operation on {\tt d\_input}} \\
2: & \text{Fetch using {\tt d\_input} followed} \\
& \text{by a compute operation } \\
\end{cases}
\]

The {\tt gptwc()} function returns an instance {\tt
	out\_obj} of the class {\tt TWC\_Out} which also has two
attributes: (1) {\tt d\_output} attribute which is set to
the result that the initiator process is expecting from the
target process; and (2) {\tt status} that indicates if any
errors were encountered when processing the data object sent
over by the initiator process.  When {\tt status} has a
value of 1, that signifies success; other values denote
specific types of errors. 
\section{| Inference Logic for different Applications}
\label{appendix:InfLogic}
\subsection{Airport Checkpoint Security}
{ The airport checkpoint security deals with distinct and unique objects and each of them are assigned an unique {\sc ID}. For interactions with each of these objects, there is a temporal relationship between consecutive interactions involving the same person and the same object. To elaborate further, the higher level interactions such as -- {\em Divest object, Move object from bin, Move object to bin, Leave behind object and Collect object} depends on the sequence of two consecutive elementary actions as shown in Table \ref{tbl:InfCheck}.\footnote{The symbol $\phi$ in the table represents an empty value (meaning that the next elementary action in the sequence is the first interaction with the object), A and R represent the elementary actions Add and Remove respectively and DNC represents Do Not Care and so it could be either $\phi$, A or R. }
	
	The {\tt HumEnt} instances own {\tt OBlob} instances and {\tt StoEnt} instances and hence these are the two ownership relationships that are tested by the Anomaly Detector. The first interaction determines whether the {\tt HumEnt} instance divested an item or not and if so, he is {\em Set} as the owner of the {\tt OBlob}. The ownership information of the {\tt StoEnt} instances are set before at the time of their instantiation. It can be seen in Table \ref{tbl:InfCheck}, how the Inference module can detect different types of anomalies and can either issue a warning or raise alarms, whenever a non-owner {\tt HumEnt} instance interacts with a {\tt OBlob} and {\tt StoEnt} instance that does not belong to him/her/them. }

\subsection{Automated Retail Store}
{ On the other hand, the automated retail store application deals with multiple objects of the same product type and hence have the same {\sc ID}. Interactions with a specific type of product can happen in any arbitrary temporal sequence. For example, a {\tt CustEnt} might pick-up 5 items and then return back 2 items and another {\tt CustEnt} can pick up an item one by one and return it one by one. In this application, the higher level interactions such as -- {\em Pick-up item, Return item to correct shelf and Misplace item in wrong shelf} depend only on the latest interaction with an item. This is why the inference module logic for an automated retail store analyzes only the latest interaction with an item instead of a sequence of elementary actions, as shown in Table \ref{tbl:InfStore}.\footnote{The symbols A and R represent the elementary actions Add and Remove respectively and DNC represents Do Not Care and so it could be either A or R. }
	
	The {\tt HumEnt} instances only become owners when they pay at the time of exit. So as long as a {\tt HumEnt} instance is within the store, {\tt StoEnt} instances {\em own} the {\tt OBlob} instances and this is the only ownership relationship that is tested by the Anomaly Detector. The ownership information is assumed to be set at the time of installation as it is expected the store will have this information before. 
	
	It can be also be seen in Table \ref{tbl:InfStore}, how the Inference Module can keep track of how many objects were purchased and automatically bill the customer for each product. Further, it can identify when a product is misplaced and notify support staff. Besides, it keeps track of how many times a particular type of product was inspected and returned and how many times a product was actually purchased, and this data can be used by businesses to optimize their operations and build a more efficient inventory management system. }

\begin{table*}[t]
	\centering
	\setlength{\extrarowheight}{0pt}
	\addtolength{\extrarowheight}{\aboverulesep}
	\addtolength{\extrarowheight}{\belowrulesep}  		\setlength{\aboverulesep}{0pt}
	\setlength{\belowrulesep}{0pt}
	\arrayrulecolor{black}
	\resizebox{\textwidth}{!}{%
		\begin{tabular}{|c!{\vrule width \heavyrulewidth}c!{\vrule width \heavyrulewidth}c!{\vrule width \heavyrulewidth}l!{\vrule width \heavyrulewidth}c!{\vrule width \heavyrulewidth}c!{\vrule width \heavyrulewidth}c!{\vrule width \heavyrulewidth}c!{\vrule width \heavyrulewidth}} 
			\bottomrule
			\multicolumn{4}{!{\vrule width \heavyrulewidth}c!{\vrule width \heavyrulewidth}}{{\cellcolor[rgb]{1,0.882,0.675}} \large \textbf{Inference State Machine } } & \multicolumn{2}{c|}{{\cellcolor[rgb]{1,0.675,0.675}}\large \textbf{Anomaly Detector} }   & \multicolumn{1}{c|}{{\cellcolor[rgb]{1,0.675,0.675}}\large \textbf{Tasks} }                                        \\ 
			\hhline{--------}
			\rowcolor[rgb]{1,0.882,0.675}  & {\cellcolor[rgb]{1,0.882,0.675}} &  {\cellcolor[rgb]{1,0.882,0.675}}  & {\cellcolor[rgb]{1,0.882,0.675}} & \multicolumn{2}{l!{\vrule width  \heavyrulewidth}}{{\cellcolor[rgb]{1,0.675,0.675}}\textbf{HumEnt} $H_i$ \textbf{owns}}  & {\cellcolor[rgb]{1,0.675,0.675}}                                  \\ 
			\rowcolor[rgb]{1,0.882,0.675} \multirow{-2}{*}{{\cellcolor[rgb]{1,0.882,0.675}}\begin{tabular}[l]{@{}>{\cellcolor[rgb]{1,0.882,0.675}}l@{}}\textbf{Triggered}\\ \textbf{by Event }\end{tabular}}                                                          & \multirow{-2}{*}{{\cellcolor[rgb]{1,0.882,0.675}}\begin{tabular}[l]{@{}>{\cellcolor[rgb]{1,0.882,0.675}}l@{}}\textbf{Latest Sequence of}\\ \textbf{Elementary Actions }\end{tabular}}   & \multirow{-2}{*}{{\cellcolor[rgb]{1,0.882,0.675}}\begin{tabular}[l]{@{}>{\cellcolor[rgb]{1,0.882,0.675}}l@{}}\textbf{Higher Level Interaction }\end{tabular}} & \multirow{-2}{*}{{\cellcolor[rgb]{1,0.882,0.675}}\begin{tabular}[c]{@{}>{\cellcolor[rgb]{1,0.882,0.675}}l@{}}\textbf{Set/Test} \\
					\textbf{Ownership}\end{tabular}} & {\cellcolor[rgb]{1,0.675,0.675}}\textbf{OBlob} & {\cellcolor[rgb]{1,0.675,0.675}}\textbf{StoEnt } & \multirow{-2}{*}{{\cellcolor[rgb]{1,0.675,0.675}}}  \\ 
			\bottomrule
			\multirow{5}{*}{HandOut} & \multirow{5}{*}{$\phi$, A} & \multirow{2}{*}{Divest own object $O_j$ in own Bin $S_k$} &   & \multirow{2}{*}{$O_j$-Yes}  & \multirow{2}{*}{$S_k$-Yes}  & Append $O_j$ in \\ 
			&  & & Set (OBlob)&   & & $H_i$.Ownership\\ 
			\cmidrule[\heavyrulewidth]{3-3}\cmidrule[\heavyrulewidth]{5-7} 
			&  & \multirow{2}{*}{Divest own object $O_j$ in other's Bin $S_k$} & Test (StoEnt) & \multirow{3}{*}{$O_j$-Yes}  & \multirow{3}{*}{$S_k$-No}  & Append $O_j$ in \\ 
			&  & & &   & & $H_i$.Ownership\\
			&  & & &   & & Raise Alarm\\
			\bottomrule
			HandOut & $\phi$, R & Taking other's object $O_j$ from other's Bin $S_k$& Test & $O_j$-No  & $S_k$-No  & Raise Alarm \\ 
			\bottomrule
			
			\multicolumn{1}{!{\vrule width \heavyrulewidth}c!{\vrule width \heavyrulewidth}}{\multirow{10}{*}{HandOut}} & \multirow{10}{*}{R, A} & \begin{tabular}[c]{@{}l@{}}Move own object $O_j$ from own Bin $S_f$ to own Bin $S_t$\end{tabular} & \multirow{10}{*}{Test} & $O_j$-Yes & $S_f$-Yes, $S_t$-Yes &  \\ 
			\cmidrule[\heavyrulewidth]{3-3}\cmidrule[\heavyrulewidth]{5-7} 
			\multicolumn{1}{!{\vrule width\heavyrulewidth}l!{\vrule width \heavyrulewidth}}{} &  &  \begin{tabular}[c]{@{}l@{}}Move own object $O_j$ from own Bin $S_f$ to other's Bin $S_t$\end{tabular}  & &  $O_j$-Yes & $S_f$-Yes, $S_t$-No & Raise Alarm \\
			\cmidrule[\heavyrulewidth]{3-3}\cmidrule[\heavyrulewidth]{5-7} 
			\multicolumn{1}{!{\vrule width\heavyrulewidth}l!{\vrule width \heavyrulewidth}}{} &  &  \begin{tabular}[c]{@{}l@{}}Move own object $O_j$ from other's Bin $S_f$ to own Bin $S_t$\end{tabular}  & & $O_j$-Yes & $S_f$-No, $S_t$-Yes & Raise Alarm \\
			\cmidrule[\heavyrulewidth]{3-3}\cmidrule[\heavyrulewidth]{5-7} 
			\multicolumn{1}{!{\vrule width\heavyrulewidth}l!{\vrule width \heavyrulewidth}}{} &  &  \begin{tabular}[c]{@{}l@{}}Move own object $O_j$ from other's Bin $S_f$ to other's Bin $S_t$\end{tabular}  & & $O_j$-Yes & $S_f$-No, $S_t$-No & Raise Alarm \\
			\cmidrule[\heavyrulewidth]{3-3}\cmidrule[\heavyrulewidth]{5-7} 
			
			\multicolumn{1}{!{\vrule width \heavyrulewidth}c!{\vrule width \heavyrulewidth}}{} & & \begin{tabular}[c]{@{}l@{}}Move other's object $O_j$ from own Bin $S_f$ to own Bin $S_t$\end{tabular} &  & $O_j$-No & $S_f$-Yes, $S_t$-Yes & Raise Alarm \\ 
			\cmidrule[\heavyrulewidth]{3-3}\cmidrule[\heavyrulewidth]{5-7} 
			\multicolumn{1}{!{\vrule width\heavyrulewidth}l!{\vrule width \heavyrulewidth}}{} &  &  \begin{tabular}[c]{@{}l@{}}Move other's object $O_j$ from own Bin $S_f$ to other's Bin $S_t$\end{tabular}  & &  $O_j$-No & $S_f$-Yes, $S_t$-No & Raise Alarm \\
			\cmidrule[\heavyrulewidth]{3-3}\cmidrule[\heavyrulewidth]{5-7} 
			\multicolumn{1}{!{\vrule width\heavyrulewidth}l!{\vrule width \heavyrulewidth}}{} &  &  \begin{tabular}[c]{@{}l@{}}Move other's object $O_j$ from other's Bin $S_f$ to own Bin $S_t$\end{tabular}  & & $O_j$-No & $S_f$-No, $S_t$-Yes & Raise Alarm \\
			\cmidrule[\heavyrulewidth]{3-3}\cmidrule[\heavyrulewidth]{5-7} 
			\multicolumn{1}{!{\vrule width\heavyrulewidth}l!{\vrule width \heavyrulewidth}}{} &  &  \begin{tabular}[c]{@{}l@{}}Move other's object $O_j$ from other's Bin $S_f$ to other's Bin $S_t$\end{tabular}  & & $O_j$-No & $S_f$-No, $S_t$-No & Raise Alarm \\
			\bottomrule
			\multicolumn{1}{!{\vrule width \heavyrulewidth}l!{\vrule width \heavyrulewidth}}{\multirow{2}{*}{HumanExit}} & \multirow{2}{*}{DNC, A} & \begin{tabular}[c]{@{}l@{}}Left own object $O_j$ in own Bin $S_k$\end{tabular} & \multirow{2}{*}{Test} & $O_j$-Yes & $S_k$-Yes & Warn~$H_i$ \\ 
			\cmidrule[\heavyrulewidth]{3-3}\cmidrule[\heavyrulewidth]{5-7} \multicolumn{1}{!{\vrule width\heavyrulewidth}l!{\vrule width \heavyrulewidth}}{} &  &  \begin{tabular}[c]{@{}l@{}}Left own object $O_j$ in other's Bin $S_k$\end{tabular}  & & $O_j$-Yes & $S_k$-No & Raise Alarm \\
			\bottomrule
			
			\multicolumn{1}{!{\vrule width \heavyrulewidth}c!{\vrule width \heavyrulewidth}}{\multirow{4}{*}{HumanExit}} & \multirow{4}{*}{DNC, R} & \begin{tabular}[c]{@{}l@{}}Collect own object $O_j$ from own Bin $S_k$\end{tabular} & \multirow{4}{*}{Test} & $O_j$-Yes & $S_k$-Yes &  \\ 
			\cmidrule[\heavyrulewidth]{3-3}\cmidrule[\heavyrulewidth]{5-7} \multicolumn{1}{!{\vrule width\heavyrulewidth}l!{\vrule width \heavyrulewidth}}{} &  & \begin{tabular}[c]{@{}l@{}}Collect other's object $O_j$ from other's Bin $S_k$\end{tabular}  & & $O_j$-No & $S_k$-No & Raise Alarm \\
			\cmidrule[\heavyrulewidth]{3-3}\cmidrule[\heavyrulewidth]{5-7} \multicolumn{1}{!{\vrule width\heavyrulewidth}l!{\vrule width \heavyrulewidth}}{} &  & \begin{tabular}[c]{@{}l@{}}Collect other's object $O_j$ from own Bin $S_k$\end{tabular}  & & $O_j$-No & $S_k$-Yes & Raise Alarm \\
			\bottomrule
			
	\end{tabular}}
	
	\caption{Inference module logic for Airport Checkpoint Security Application.}
	\label{tbl:InfCheck}
\end{table*}

\begin{table*}
	\centering
	\setlength{\extrarowheight}{0pt}
	\addtolength{\extrarowheight}{\aboverulesep}
	\addtolength{\extrarowheight}{\belowrulesep}
	\setlength{\aboverulesep}{0pt}
	\setlength{\belowrulesep}{0pt}
	\arrayrulecolor{black}
	\resizebox{0.95\textwidth}{!}{%
		\begin{tabular}{|c!{\vrule width \heavyrulewidth}c!{\vrule width \heavyrulewidth}c!{\vrule width \heavyrulewidth}l!{\vrule width \heavyrulewidth}c!{\vrule width \heavyrulewidth}c!{\vrule width \heavyrulewidth}c!{\vrule width \heavyrulewidth}c!{\vrule width \heavyrulewidth}} 
			\bottomrule
			\multicolumn{4}{!{\vrule width \heavyrulewidth}c!{\vrule width \heavyrulewidth}}{{\cellcolor[rgb]{1,0.882,0.675}} \large \textbf{Inference State Machine } } & \multicolumn{1}{c|}{{\cellcolor[rgb]{1,0.675,0.675}}\large \textbf{Anomaly Detector} }   & \multicolumn{1}{c|}{{\cellcolor[rgb]{1,0.675,0.675}}\large \textbf{Tasks} }                                        \\ 
			\hhline{--------}
			\rowcolor[rgb]{1,0.882,0.675}  & {\cellcolor[rgb]{1,0.882,0.675}} &  {\cellcolor[rgb]{1,0.882,0.675}}  & {\cellcolor[rgb]{1,0.882,0.675}} & \multicolumn{1}{l!{\vrule width  \heavyrulewidth}}{{\cellcolor[rgb]{1,0.675,0.675}}\textbf{StoEnt} $S_k$ \textbf{owns OBlob} $O_j$}  & {\cellcolor[rgb]{1,0.675,0.675}}                                  \\ 
			\rowcolor[rgb]{1,0.882,0.675} \multirow{-2}{*}{{\cellcolor[rgb]{1,0.882,0.675}}\begin{tabular}[l]{@{}>{\cellcolor[rgb]{1,0.882,0.675}}l@{}}\textbf{Triggered}\\ \textbf{by Event }\end{tabular}}                                                          & \multirow{-2}{*}{{\cellcolor[rgb]{1,0.882,0.675}}\begin{tabular}[l]{@{}>{\cellcolor[rgb]{1,0.882,0.675}}l@{}}\textbf{Latest Sequence of}\\ \textbf{Elementary Actions }\end{tabular}}   & \multirow{-2}{*}{{\cellcolor[rgb]{1,0.882,0.675}}\begin{tabular}[l]{@{}>{\cellcolor[rgb]{1,0.882,0.675}}l@{}}\textbf{Higher Level Interaction }\end{tabular}} & \multirow{-2}{*}{{\cellcolor[rgb]{1,0.882,0.675}}\begin{tabular}[c]{@{}>{\cellcolor[rgb]{1,0.882,0.675}}l@{}}\textbf{Set/Test} \\
					\textbf{Ownership}\end{tabular}} & {\cellcolor[rgb]{1,0.675,0.675}} & \multirow{-2}{*}{{\cellcolor[rgb]{1,0.675,0.675}}}  \\ 
			\bottomrule
			HandOut   & R   & Picked up item $O_j$ from Shelf $S_k$ & Test & Yes & nPurchase++; nInspect++;\\ 
			\bottomrule
			\multicolumn{1}{!{\vrule width \heavyrulewidth}l!{\vrule width \heavyrulewidth}}{\multirow{3}{*}{HandOut}} & \multirow{3}{*}{A} & \begin{tabular}[c]{@{}l@{}}Returned item $O_j$ to correct shelf $S_k$\end{tabular}  & \multirow{3}{*}{Test} & Yes & nReturn++; nPurchase--;\\ 
			\cmidrule[\heavyrulewidth]{3-3}\cmidrule[\heavyrulewidth]{5-6}
			\multicolumn{1}{!{\vrule width \heavyrulewidth}l!{\vrule width \heavyrulewidth}}{}                     &                    &                       \begin{tabular}[c]{@{}l@{}}Misplaced item $O_j$ in wrong shelf $S_k$\end{tabular}                                                                                                                                                                                                       &                                                                                                                                                                    & No                                                                                                                                                                                & \begin{tabular}[c]{@{}l@{}}nMisplace++; nPurchase--;\\Notify CustEnt and StaffEnt\end{tabular}  \\ 
			\bottomrule
			\multicolumn{1}{!{\vrule width \heavyrulewidth}l!{\vrule width \heavyrulewidth}}{HumanExit}                    & DNC                & -      &-&-         & \begin{tabular}[c]{@{}l@{}}Add amount for all OBlobs $O_j$ \\
				in $H_i.ActionQ$ to total bill as : \\ $H_i$.Amount+=[nPurchase*$O_j$.Price]\end{tabular}        \\
			\bottomrule
	\end{tabular}}
	
	\caption{Inference Module Logic for Automated Retail Store Application.}
	\label{tbl:InfStore}
\end{table*}


\end{document}